\documentclass[showpacs, preprintnumbers, nofootinbib, aps, prd, superscriptaddress,10pt, twocolumn, showkeys, notitlepage]{revtex4-2}

\usepackage{graphicx,amssymb,amsmath,amsthm,amsfonts,epsfig, setspace}

\usepackage[linktocpage]{hyperref}
\usepackage[usenames,dvipsnames]{color}
\usepackage{epstopdf}
\usepackage{pifont}
\definecolor{darkred}{rgb}{0.5,0,0}
\definecolor{darkgreen}{rgb}{0,0.5,0}
\definecolor{darkblue}{rgb}{0,0,0.5}
\definecolor{prussian}{rgb}{0.0, 0.19, 0.33}
\definecolor{richelectricblue}{rgb}{0.03, 0.57, 0.82}
\definecolor{teal}{rgb}{0.0, 0.5, 0.5}
\definecolor{mediumseagreen}{rgb}{0.24, 0.7, 0.44}
\definecolor{lust}{rgb}{0.9, 0.13, 0.13}
\definecolor{ballblue}{rgb}{0.13, 0.67, 0.8}
\definecolor{darkcyan}{rgb}{0.0, 0.55, 0.55}
\definecolor{mountainmeadow}{rgb}{0.19, 0.73, 0.56}
\definecolor{palecarmine}{rgb}{0.69, 0.25, 0.21}
\definecolor{richcarmine}{rgb}{0.84, 0.0, 0.25}
\definecolor{tangelo}{rgb}{0.98, 0.3, 0.0}
\definecolor{venetian}{rgb}{0.784,0.031,0.082}
\definecolor{bdfrance}{rgb}{0.192,0.549,0.906}

\hypersetup{colorlinks=true, citecolor=venetian,
linkcolor=bdfrance, urlcolor=lust}
\usepackage{tensor}
\usepackage{mathtools}
\usepackage{amsbsy}
\usepackage{bm}

\usepackage{cleveref}

\usepackage{appendix}

\newcommand{\be}{\begin{equation}}
\newcommand{\ee}{\end{equation}}
\newcommand{\bear}{\begin{eqnarray}}
\newcommand{\eear}{\end{eqnarray}}

\newcommand{\nn}{\nonumber}

\begin{document}

\title{SMBH shadows: gravity fingerprints revealed by spectral line
background radiation}

\author{Konstantinos Kostaros}
\email{kkostaro@auth.gr}
\affiliation{Department of Physics, Aristotle University of Thessaloniki, Thessaloniki 54124, Greece}

\author{Padelis Papadopoulos}
\email{padelis@auth.gr}
\affiliation{Department of Physics, Aristotle University of Thessaloniki, Thessaloniki 54124, Greece}
\affiliation{Research Center for Astronomy, Academy of Athens, Soranou Efesiou 4, GR-11527 Athens, Greece}

\author{Wing-Fai Thi}
\affiliation{Karpfenstrasse 18, 81825 Munich, Germany}

\author{George Pappas}
\email{gpappas@auth.gr}
\affiliation{Department of Physics, Aristotle University of Thessaloniki, Thessaloniki 54124, Greece}

\begin{abstract} Supermassive Black Holes (SMBHs) at the centers of galaxies, illuminated by the radiation fields of their accretion disks, can reveal valuable information on BH spacetime geometry via their shadows, which can be used for strong-gravity tests. However, the continuum emission from their highly turbulent hot plasma, while a powerful SMBH illuminator, is expected to be strongly time-varying and with very inhomogeneous brightness. This, in turn, can mask, or make rather difficult to discern, important SMBH-related effects, like the appearance of the light-ring, rendering them ineffective as probes of strong gravitational lensing physics. Besides being an inhomogeneous and strongly time-varying $''$illuminator$''$, the hot plasma accretion disk emission extends all the way to the SMBH's last stable circular orbit. This then leads to the superposition of the strongly-lensed radiation from the area of the light-ring to the continuum emission coming directly from the inner hot accretion disk, effectively making gravitational lensing physics hard to separate from accretion disk Astrophysics. These problems could be overcome if one utilizes the spectral line radiation field emanating from the cooler parts of the extended accretion disk, the so-called Broad Line Region (BLR), and especially its expected neutral phase, as a more distant, but still adequately strong SMBH illuminator, typically found at $r\sim (10^2-10^4)\, R_s$ ($R_s=2GM/c^2$). This kind of illumination can provide a cleaner image of the light-ring region, and thus allow more information on the spacetime geometry around the SMBH to be obtained. Here, we examine some of the benefits of such a spectral line rather than continuum emission illumination in discerning strong-gravity physics near SMBHs and their observability. We expand on the fact that such emission can provide a smoking gun signal of lensing, in the form of an Einstein ring, caused by the gravitational field of the BH. To first order, the imaging of the Einstein ring and its spectroscopic signature can facilitate the measurement of the SMBH mass, while the second order effects associated with the light-ring can constrain the SMBH spin, and even identify deviations from the Kerr spacetime.
\end{abstract} 
  
\maketitle

\section{Introduction}

Directly imaging the vicinity of a black hole's (BH) event horizon has been a long-standing dream of astrophysicists, realized only in the imagination of science fiction. Very Long Baseline Interferometry (VLBI) has recently fulfilled this dream producing the first images of the shadows of supermassive black holes (SMBHs) \cite{Akiyama_2019,EHT_M87_2,EHT_M87_3,EHT_M87_4,EHT_M87_5,EHT_M87_6,EHT_Sgr1,EHT_Sgr2,EHT_Sgr3,EHT_Sgr4,EHT_Sgr5,EHT_Sgr6,EHT_M87_7}, through the Event Horizon Telescope collaboration (EHT). These unique observations are paving the road for new paths of exploration on BH physics. So far, the EHT has imaged the shadows of the SMBHs M87*, at the center of the M87 galaxy, and Sgr A*, at the center of our Milky Way galaxy \cite{Akiyama_2019,EHT_Sgr1}. These images encode information for the structure of the spacetime near the location of the light-ring, i.e., the vicinity of the unstable photon orbit, but also for the astrophysical environment of the BH, i.e., the accretion region near the innermost stable circular orbit (ISCO) \cite{Volkel:2020xlc,Glampedakis:2021oie,Glampedakis:2023eek,Gralla:2020srx,Younsi:2021dxe,Bauer:2021atk,Glampedakis:2023eek,Gralla:2020nwp,Lima:2021las,Medeiros:2019cde,Gralla:2019xty,Johannsen_2010,Olmo:2023lil}. More recently, the EHT collaboration also extracted information on the polarization of the emission near the BH, which carries additional information about this region of spacetime and the properties of the emitting plasma \cite{EventHorizonTelescope:2021bee,EventHorizonTelescope:2021srq,EventHorizonTelescope:2023gtd}. 

Plasma orbits around the BH forming an accretion disk, converting potential energy into heat and radiation until it reaches the ISCO, where it plunges into the BH and the disk ends. The shadow of a BH forms as hot accretion disk plasma emits electromagnetic continuum radiation that gets scattered by the BH's gravitational potential, and in particular, the peak of that potential for the motion of photons, which defines the unstable photon orbit, creating the light-ring. The scattering of this radiation at the light-ring is an extreme form of gravitational lensing that results in a characteristic bright and thin ring-like image (hence the name). For BHs, particularly those with rapid rotation, the ISCO is located near the light-ring and forms a region where brightness reaches a maximum and then drops abruptly, which could look like a bright ring. Therefore, direct emission from the part of the inner accretion disk near the ISCO could appear as a $''$light-ring$''$, resulting in confusion issues, masking the properties of the true light-ring and intertwining the information on strong-gravity physics with that of accretion disk physics \cite{Gralla:2019xty,Glampedakis:2021oie,Ozel:2021ayr,Lara:2021zth}. 

One characteristic of particular interest is the higher-order structure of the light-ring, which is comprised of many sub-rings produced by the photons that circle the unstable photon orbit many times (zoom-whirl orbits). These photons, depending on the number of turns they perform, produce rings of increasing order. The observation of these individual rings and the comparison of their relative brightnesses can reveal the encoded strong-gravity information \cite{Olmo:2023lil}, assuming ideal conditions of illumination. Unfortunately, such an observation will suffer from the complexities of the accretion physics for two main reasons. Firstly it will be because direct emission from the accretion will contaminate with strong radiation the very faint signal from the higher order light-rings \cite{Gralla:2019xty}, and secondly because the higher order light-rings receive their illumination in a rather complicated way from the accretion disk, with different order rings illuminated by radiation coming from different parts of a disk, which is far from having a uniform brightness. Moreover, any such higher-order ring illumination will also inadvertently incorporate the strong temporal variability expected in the strongly turbulent magnetohydrodynamic (MHD) environments of the hot inner accretion disks around SMBHs. Since some of these tests of General Relativity (GR) and the Kerr solution rely on comparing the relative brightnesses of the light-rings of different orders, the aforementioned non-uniform and time-varying illumination renders such comparisons meaningless for the extraction of strong-gravity information \cite{Olmo:2023lil}. Furthermore, similar tests of strong gravity that rely on the precise determination of the size and shape of the shadow of the BH may also suffer from the expected strong variability of the accretion disk emission \cite{Volkel:2020xlc,Gralla:2020srx,Younsi:2021dxe,Fernandes:2024ztk,Gyulchev:2024iel,DeliyskiPhysRevD.111.064068,BenAchour:2025uzp}.  

Thus, in order to use the light-ring to test the Kerr nature of BHs \cite{Kostaros_2022,Kostaros:2024vbn} or perform precise tests of GR \cite{Olmo:2023lil,Younsi:2021dxe}, one must find a way around these accretion-physics-related problems. In the previous discussion, we pointed out the two main issues: the confusion from the emission coming from a region too close to the light-ring of the SMBH, and the rapid variability of the emission from the inner part of the accretion disk, where the dynamical time-scales are very short. One idea recently proposed to solve these problems is to illuminate the BH with radiation coming from a source located farther away from the central region \cite{Kostaros:2024vbn}. Illuminating the SMBH, and the associated light-ring, from a larger distance (provided that enough radiation flux density still reaches the SMBH \cite{Astro_paper}) will keep the emission coming directly from the disk well-separated from that scattered by the light-ring, removing any ISCO-related confusion and producing a clean signal. This will likely also resolve the issues of non-uniform illumination of the higher order light-rings, making these types of tests of GR and the Kerr metric available. This is not the only advantage of a more distant SMBH illuminator, however. Indeed, for the plasma orbiting the BH, the dynamical timescales are a function of the BH mass (which defines both a length scale and a time scale) and the distance from it, with higher mass or larger distances yielding longer timescales. Hence, having the source of illumination farther away from the BH reduces the problem of rapid variability of the associated radiation field. This also means steadier illumination conditions during observations, and thus more robust images, especially when VLBI imaging is involved. Should the new, more distant, SMBH illuminator also be much less turbulent than the hot MHD-turbulent plasma of the inner accretion disk, it will add to both aforementioned benefits, especially regarding the expected time-variability of the associated radiation field.

Recent work has shown that such a more distant SMBH illuminator may actually exist, and, even more advantageously, it is a spectral line rather than a continuum emission source, which, as we shall see, enables additional classes of GR tests. This illuminator comes in the form of the so-called Broad Line Region (BLR), a cooler, more extended gas disk that exists around the SMBHs that power (via accretion) the Active Galactic Nuclei (AGN) \cite{Thi:2024dny, Astro_paper, Chem_paper}. The existence of the BLR is well established, taking its name from its wide Doppler-broadened emission lines, the widest observed within galaxies, a telltale mark of deep gravitational potentials, and the key observational characteristic that got the BLR discovered in the first place \cite{Seyf43, Bald03, Pet06, Gas09, Czer11, Czer19, Bask14, Bask18, Fer20, Czer23}. It could even be claimed that, along with the tremendous luminosities of the AGN, it was the wide BLR lines (Full Width Zero Intensity $\sim (2000-10^4)\,km\,s^{-1}$) that gave the first evidence of the deep gravitational wells attributed to SMBHs in galactic centers. The strong UV radiation emanating from the hot inner accretion disk ionizes its regions further out, producing an $\rm H^+$-rich thin disk layer which then (along with other ionized atoms) emits the recombination lines that revealed the BLR, and helped study its physical conditions.

The optical/UV recombination lines from the ionized BLR gas, henceforth BLR$^{+}$, could in principle provide effective SMBH illumination from the distances outlining the BLR$^{+}$ disk ($r$$\sim $$(100-10^4) R_s$, $R_s=2GM/c^2$). However any SMBH-lensed emission may then   be strongly absorbed by the dust-rich interstellar medium of the host galaxy, especially along lines of sight on the SMBH+(disk) equatorial plane, where the relevant lensing effects will be the strongest (this interstellar absorption is the reason why the BLR$^{+}$ lines are not visible in the so called Type 1 AGN). However it was recently proposed that the warm ($\sim 5000\,K$) BLR$^{+}$ ionized gas is only a small fraction of a  more massive,  neutral,  and cooler ($\sim (1500-2000)\,K$) gas phase (henceforth BLR$^{0}$) in a disk configuration, with the BLR$^{+}$ being only a thin layer. The thermodynamic and chemical conditions of BLR$^{0}$ allow for neutral atoms and even molecules to be abundant, whose radio and IR lines can then be an effective distant SMBH illuminator \cite{Thi:2024dny, Astro_paper, Chem_paper}, unaffected by interstellar dust absorption. Future, perhaps space-borne, IR interferometers can then image the spectral-line radiation distributions near the SMBH equatorial plane where the relevant GR signatures are the strongest, unhindered by interstellar medium absorption within the host galaxy. This capability, enabled by the unique chemical and thermal BLR$^{0}$ conditions \cite{Thi:2024dny}, make it better than BLR$^{+}$ as a possible SMBH spectral line illuminator\footnote{Also, interferometers are less technically challenging in IR/high-radio than in the optical/UV frequency domain}.

Details for the BLR$^{0}$ conditions are presented in companion papers in preparation (\cite{Astro_paper, Chem_paper}). Here we only briefly mention the attributes making it better than BLR$^{+}$ as a SMBH spectral line illuminator: 1) it is far less turbulent than the BLR$^{+}$, which often contains strong outflows (complicating the SMBH illumination geometry and the resulting line profiles), 2) it is dominated by Keplerian velocity fields (simplifying lensing analysis and the construction of the emergent line profiles), 3) its lines are {\it effectively} optically thin (allowing much simpler radiative transfer computations),  4) its disk is much thinner than that of the BLR$^{+}$ (simplifying geometric assumptions), and 5) the  BLR$^{0}$ can form even when BLR$^{+}$ cannot. This can occur for the largest SMBHs whose colder accretion disks emit very weak ionizing UV radiation (this is broadly expected from accretion disk theory), which may be the best objects for future observational studies of the GR effects examined here.

The combination of far-away illumination with spectral line emission at wavelengths for which the disk is optically thin allows for many interesting possibilities. As has been reported in \cite{PRL_companion}, and we expand here, illuminating the BH with a source farther away results in a prominent Einstein ring that forms around the central BH. Furthermore, the lensing that causes the Einstein ring leaves a distinctive signature in the broadened spectrum of the accretion disk. These are telltale signals of gravitational lensing from the BH. In addition, the light-ring takes the emission from the disk that is directed towards the BH and redirects it towards the observer, forming a relatively bright image of the light-ring in the form of a relatively thin emission line that encodes additional information about the spacetime around the BH. These are the effects that we will investigate here.

This paper is organized as follows: Section \ref{sec:setup} gives the setup of the accretion disk we use and describes our integration scheme. Section \ref{sec:Ering} presents the results of the ray-tracing and the radiative transfer and details the formation of an Einstein ring, where we discuss also its use in measuring the mass of the SMBH. In Section \ref{sec:light-ring}, we turn our attention to the light-ring, and in Section \ref{sec:testsKerr}, we discuss how it could be used to get information about the properties of the spacetime near SMBHs. Finally, in Section \ref{sec:concl} we present our Conclusions. We generally use geometric units, where $G=c=1$, but for some equations, we have these physical constants restored. Our unit of length (scale) is given by the mass of the central BH, $M$, which can be taken to be equal to 1.

\section{Disk setup and radiative transfer}
\label{sec:setup}
%

For our investigation, the SMBH is illuminated by a disk structure  located at a radial distance between $r_{in}=500M$ and $r_{out}=1000M$. In cylindrical coordinates ($\varpi,\varphi,z)$, our disk  extends between $500M<\varpi<1000M$ and $-15M<z<15M$, so as to have a radius-height relation of $H(\varpi)/\varpi<10\%$ \cite{Astro_paper}. In some cases, we vary the height of the disk, as well as its outer radius, in order to explore the effect on the image and spectra.  We assume a density profile for the gas, which also plays the part of the local rest-frame emissivity $j_0$, that follows a simple law. In fact, we will use two such profiles, i.e., 
\be \label{eq:emis1}
j_0(r,\theta)=j_0\exp\left(-\frac{r-500M}{r_{\text{scale}}}\right)\exp\left(-\kappa\lvert \cos\theta\rvert\right)
,
\ee
with $r_{\text{scale}}=2000M$, and $\kappa=10$, which is the same profile used in \cite{PRL_companion},
and 
\be \label{eq:emis2}
j_0(r,\theta)=j_0 \left(\frac{r}{R_{in}}\right)^{-p}\exp\left(-\kappa\lvert \cos\theta\rvert\right)
,
\ee
with $R_{in}=500M$, $p=2.5$, and $\kappa=10$ \cite{Storchi_Bergmann_2017}. The $(r,\theta)$ coordinates here are the spherical coordinates. These two profiles essentially differ in how rapidly the density drops with the cylindrical radius $\varpi$, where the second profile drops at the outer edge of the disk down to $20\%$ of the density of the inner edge, while the first profile drops a little less rapidly, i.e., down to $80\%$. Both profiles have an exponential decrease with the height $z$. We use these two different profiles in order to test how strongly such changes affect the images and the spectra. 

For the disk material in orbit around the central BH, we assume a velocity profile that follows the relation, 
\begin{equation}
v^{\phi}=\left(\frac{\sqrt{M}}{(r\sin\theta)^{3/2}+a\sqrt{M}}\right),
\label{eq:velocityprofSL}
\end{equation}
where $a=J/M$ is the spin parameter of the BH, which describes an almost Keplerian disk. These are choices usually made in the literature \cite{Younsi2012,Fuerst_2004,Fuerst_2007}. 

At this point, we also need to make an assumption on the BH illumination. As  already discussed, we will assume  spectral line emission at some rest frame frequency $\nu_0$. The emission frequency will be arbitrary, and everything is normalized in terms of  $E/E_0$. Furthermore, we assume an optically thin spectral line (see \cite{Astro_paper}).

We can now perform the radiative transfer for our disk in the BH background and calculate the image of the disk and the shadow, as well as the various spectra. We first perform ray-tracing from the observer to the target, following the algorithm detailed in previous work \cite{Kostaros_2022,Kostaros:2024vbn}. Along the null rays, we then integrate the equation of radiative transfer, expressed in terms of the Lorentz Invariant intensity $\mathcal{I}_\nu=I_\nu/\nu^3$,  
\cite{Younsi2012,Fuerst_2004,Fuerst_2007},
\begin{equation}
    \frac{d\mathcal{I}_\nu}{d\lambda}=\gamma^{-1}\left(\frac{j_0}{\nu^3_{0}}\right),
\end{equation}
where $\gamma=\nu/\nu_0$, with $\nu$ being the frequency and $"0"$ indicates quantities in the local rest frame. The equation in this form assumes zero absorption along the geodesic. We calculate along the geodesic the specific intensity
\begin{equation}
\begin{split}
    dI_\nu&=d\mathcal{I}_\nu \nu_{obs}^3=\gamma^{-1}\left(\frac{j_0}{\nu_0^3}\right)\nu^3_{obs}d\lambda\\
    &=\gamma^2 j_0 d\lambda,
\end{split}
\end{equation}
which gives the contribution to the specific intensity from every gas  volume element  along the geodesic \cite{Fuerst_2004}. Alternatively to the specific intensity $I_\nu$, one can use the specific intensity in terms of the photon energy $I_E$, which is more convenient for presenting the emitted spectra.

A spectral line given by a delta function of the form $\delta(E-E_0)$, centered on the energy $E_0$ is of course non-physical. Since we assume that the gas is turbulent with $\Delta V_{\rm turb}=150 km/s$, the emission line will be Doppler-broadened; and for this we assume a Gaussian line profile: 
\be 
j(r,\theta) = \frac{j_0(r,\theta)}{4\pi} \frac{e^{-(E-E_0)^2/2\sigma^2}}{\sqrt{2\pi}\sigma},
\ee
with $\sigma=5\times10^{-4}E_0$. From the specific intensity, $I_E$, one can calculate the bolometric intensity by integrating 
\begin{equation}
    dI=I_E dE.
\end{equation}
One can thus have the integrated intensity images of the disk and the BH. 
From the specific intensity, one can calculate the specific flux by integrating the solid angle over which the source is viewed, 
\begin{equation}
    dF_E = I_Ed\Omega.
\end{equation}
For our configuration, the solid angle on the observer's screen is 
\begin{equation}
    d\Omega=\frac{db d\alpha}{D_L^2},
\end{equation}
where $D_L$ is the distance from the observer to the BH, while $b$ and $\alpha$ are the impact parameters on the observer's screen with respect to the BH, which are related to the impact parameters defined by Bardeen \cite{Bardeen:1973tla} ($b$ is related to the $p_\phi$ momentum of the photon, while $\alpha$ is related to the $p_\theta$ momentum). We express these impact parameters in units of  BH mass, and $db d\alpha$ is the size of the observer screen's pixels. 

We use Kerr BHs of various rotation rates, from non-rotating (Schwarzschild BH) up to the Thorne limit at $a=0.998M$ \cite{1974ApJThorne} (where we remind that the mass is our length scale and can be set equal to 1), while for some applications we will use the Johannsen-Psaltis (JP) spacetime \cite{Johannsen:2011dh} as a non-Kerr BH, following earlier work \cite{Kostaros:2024vbn}. The disk is placed on the equatorial plane of the BH, while the observer is placed at various angles with respect to the axis of rotation of the BH. We will use either the angle $\theta_0$ between the observer's line of sight and the axis of rotation of the BH, or the inclination $\iota$ between the observer's line of sight and the equatorial plane, which is the disk plane. 

\section{The disk image, the black hole shadow and the total spectra}
\label{sec:Ering}
%
\begin{figure*}
\begin{center}
\includegraphics[width=0.22\textwidth]{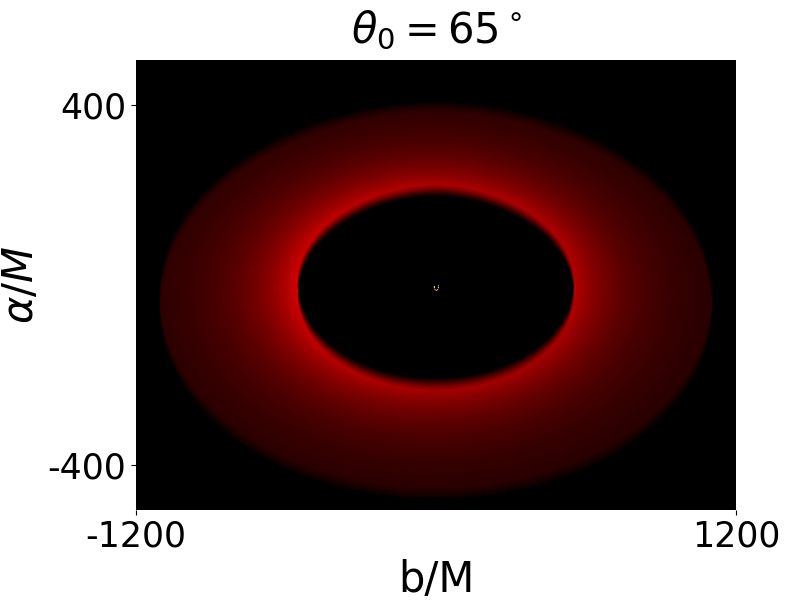}
\includegraphics[width=0.22\textwidth]{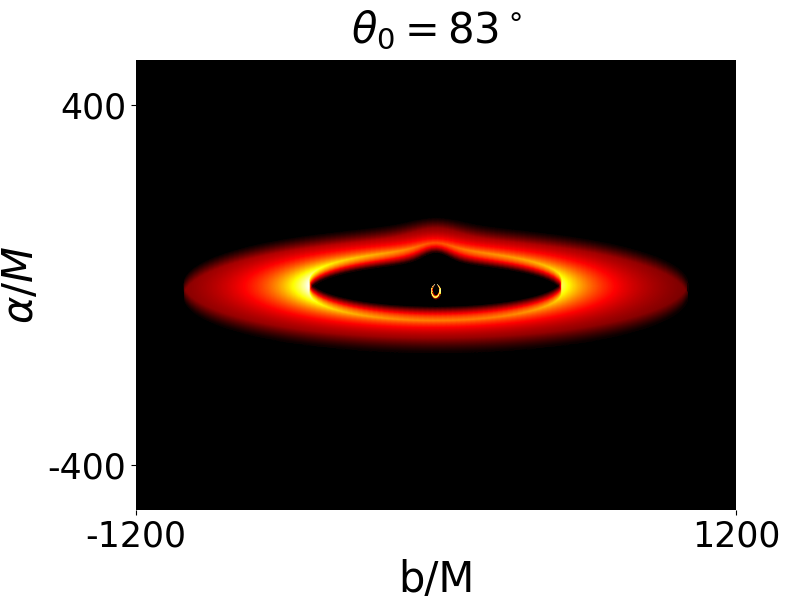}
\includegraphics[width=0.22\textwidth]{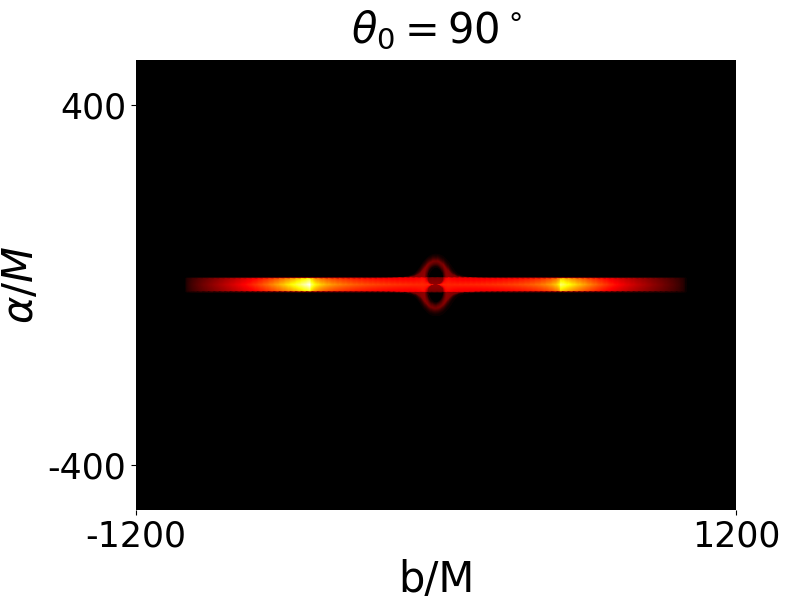}
\includegraphics[width=0.22\textwidth]{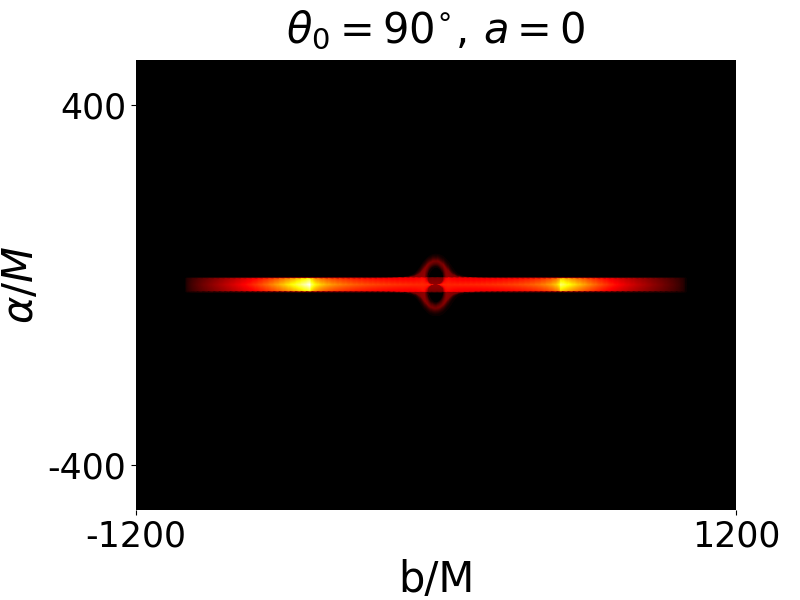}

\includegraphics[width=0.22\textwidth]{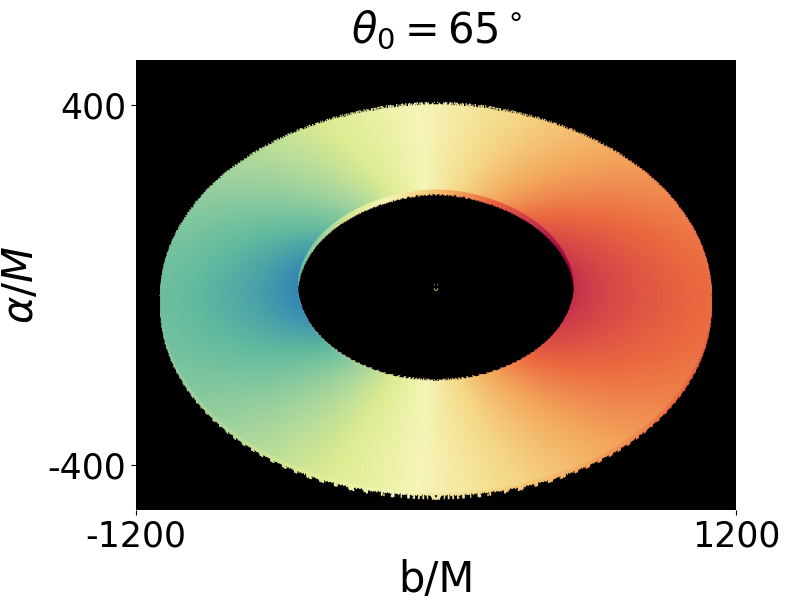}
\includegraphics[width=0.22\textwidth]{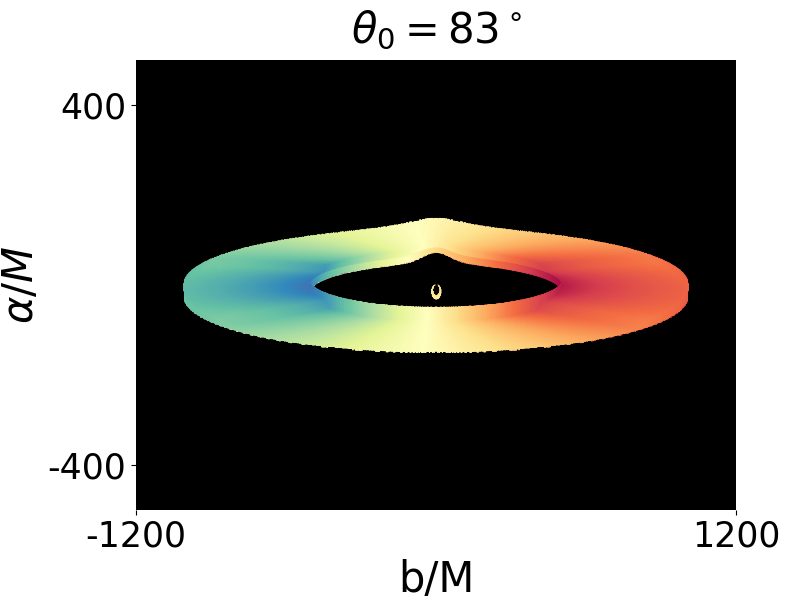}
\includegraphics[width=0.22\textwidth]{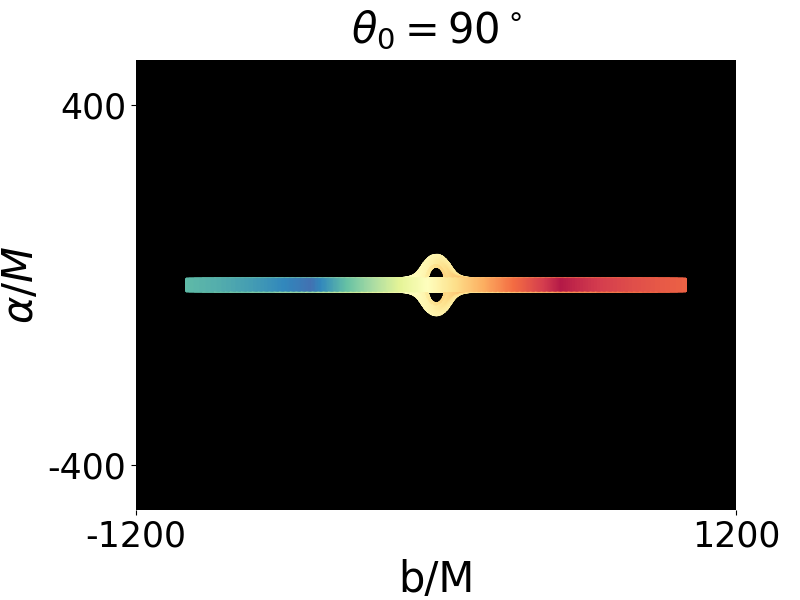}
\includegraphics[width=0.22\textwidth]{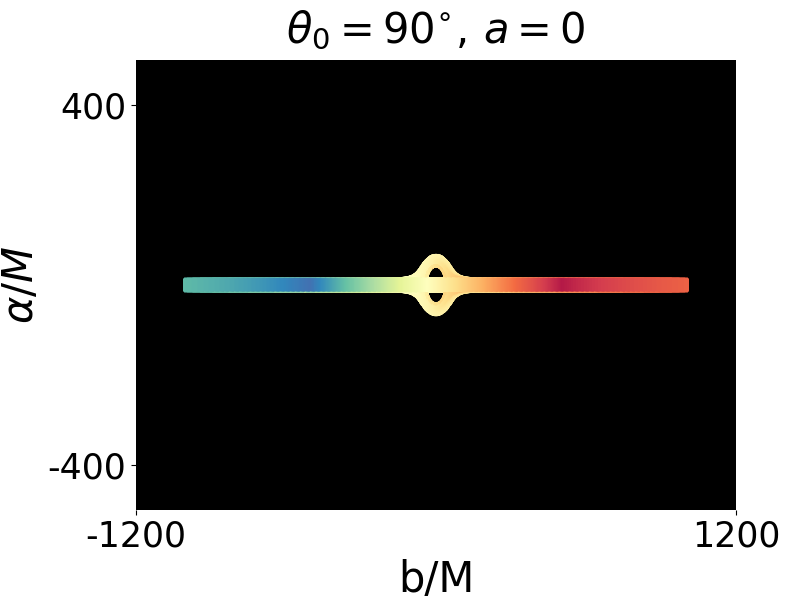}

\includegraphics[width=0.22\textwidth]{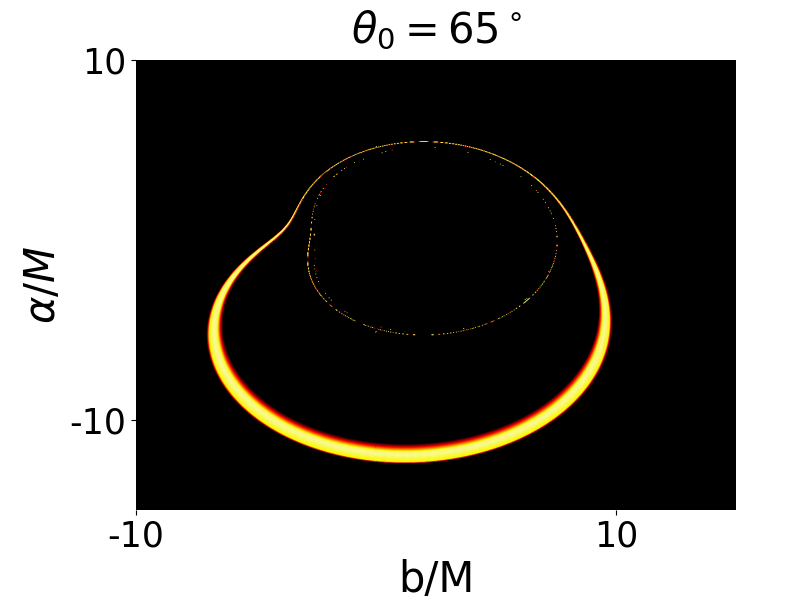}
\includegraphics[width=0.22\textwidth]{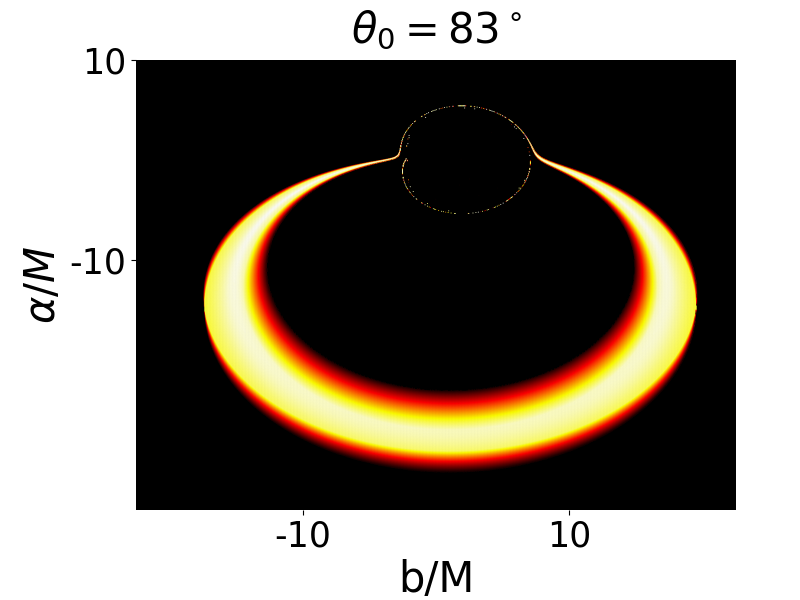}
\includegraphics[width=0.22\textwidth]{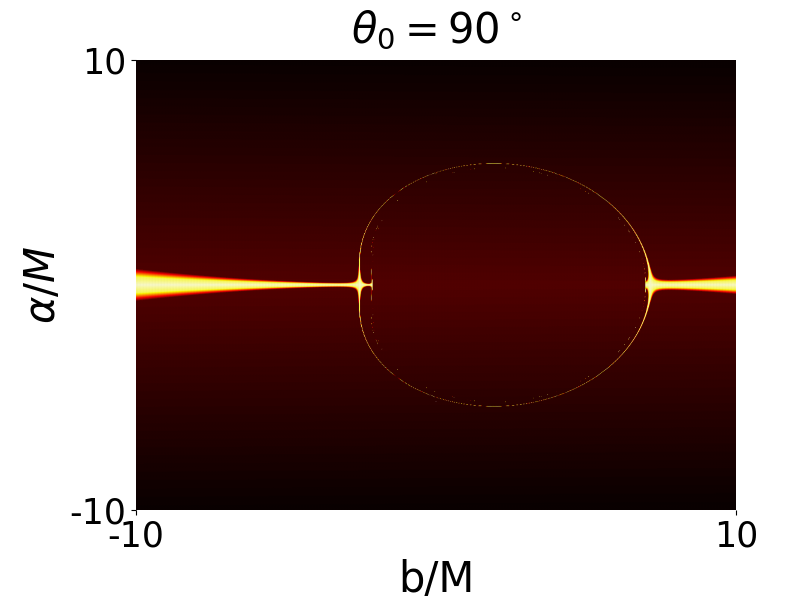}
\includegraphics[width=0.22\textwidth]{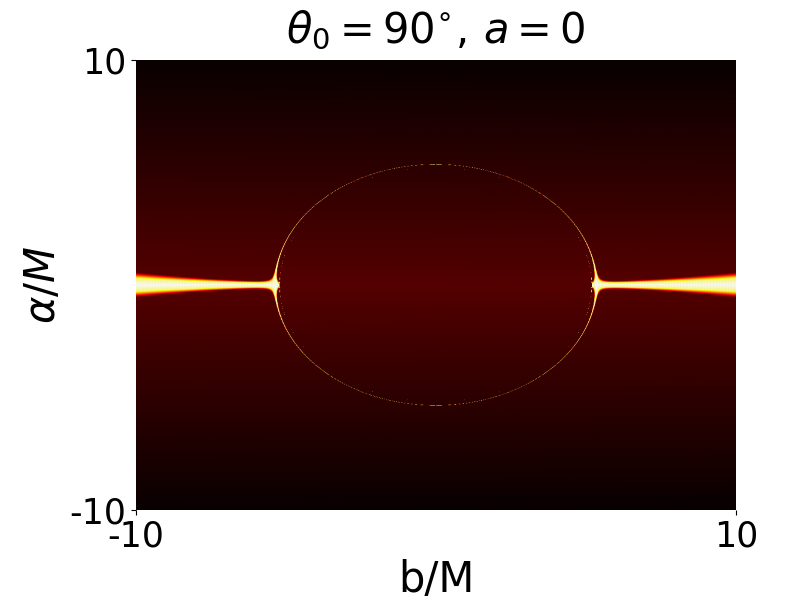}

\includegraphics[width=0.22\textwidth]{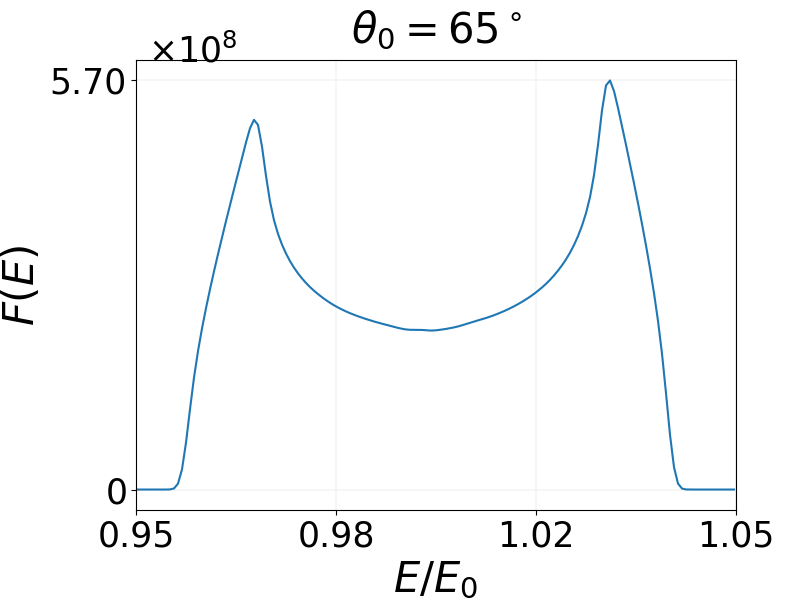}
\includegraphics[width=0.22\textwidth]{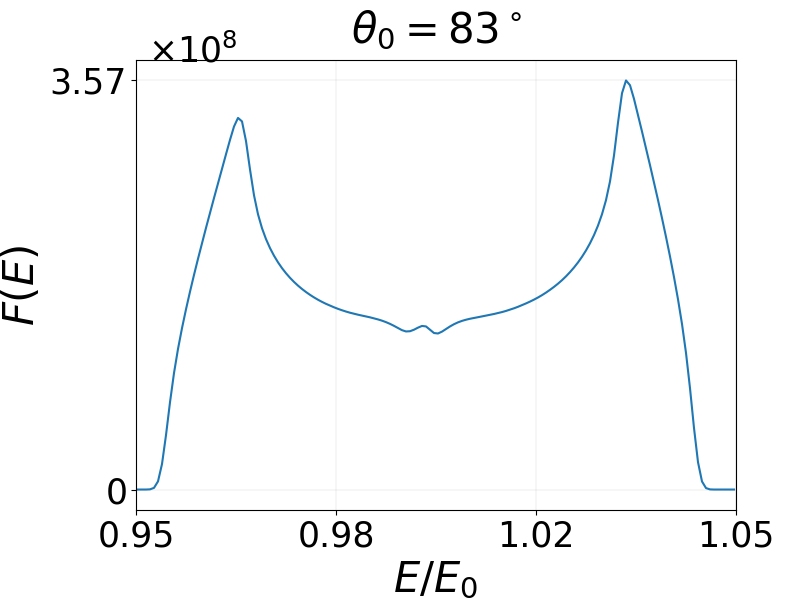}
\includegraphics[width=0.22\textwidth]{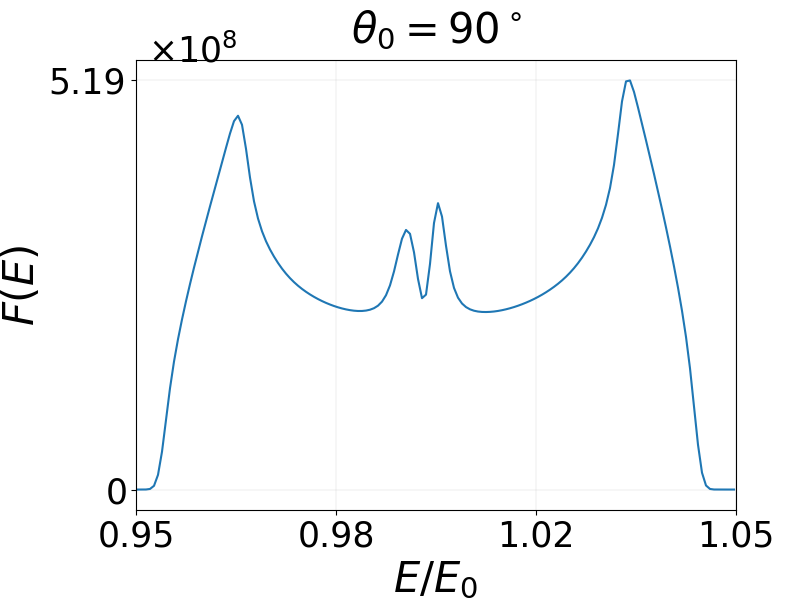}
\includegraphics[width=0.22\textwidth]{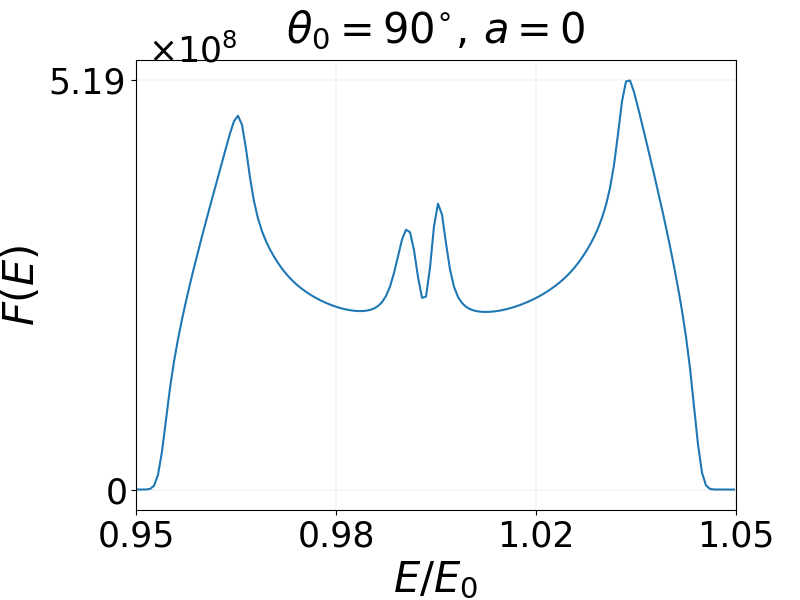}
\caption{Integrated intensity images, redshift maps, and observed spectra of an optically thin disk located at $500M<\varpi<1000M$ and $-15M<z<15M$. The first three columns correspond to a Kerr black hole with spin $a=0.998M$ for viewing inclinations $\theta_0=65,85^{\circ}$, and $90^{\circ}$ (left to right), while the last column on the right corresponds to a Schwarzschild BH viewed at an inclination of $90^{\circ}$. The figures have been produced using the emissivity profile given by eq. \eqref{eq:emis2}.}
\label{fig:TT_bigscale1}
\end{center}
\end{figure*}
%

We now present the results of the radiative transfer, which can be seen in \cref{fig:TT_bigscale1}. From top to bottom \cref{fig:TT_bigscale1} shows (see also \cref{fig:TT_bigscale} in the Appendix) the integrated intensity image of the disk, the redshift map, the strongly lensed image of the disk coming from the central region of the BH near the light-ring (shadow area), and the spectra produced by both the direct emission from the disk and the strongly lensed emission that the BH redirects towards the observer. These are shown for viewing angles of $\theta_0=65^{\circ}, 83^{\circ}$, and $90^{\circ}$ from the axis of symmetry of the BH (from left to right). In addition to the Kerr BHs, we also present at the rightmost column of the figure a Schwarzschild BH at $90^{\circ}$ for comparison. For \cref{fig:TT_bigscale1} we have used the emissivity profile given by eq. \eqref{eq:emis2} (while \cref{fig:TT_bigscale} is produced using eq. \eqref{eq:emis1}, the same as in \cite{PRL_companion}).

In the images for $\theta_0=83^{\circ}$ and $\theta_0=90^{\circ}$, one can clearly see the lensing effect the BH produces to the image of the far side of the disk, something that is not present in the $\theta_0=65^{\circ}$ image. The feature we observe, i.e., the distortion of the disk image, is essentially the beginning of the formation of an Einstein ring around the BH due to the emission from the far side of the disk \cite{PRL_companion}. The details of the formation of the Einstein ring will be discussed in the following subsection.

On the third row of \cref{fig:TT_bigscale1} we have zoomed-in to the strongly lensed region around the BH. This image is produced by the light that the disk emits towards the BH, i.e., towards the center, and gets strongly scattered, being redirected towards the observer. The image has two characteristic rings; the larger one is a lensed image of the disk, and the smaller one is the light-ring around the object's shadow that is produced by the photons that orbit the BH several times near the unstable photon orbit. This is essentially formed from all the higher order images of the disk.

Finally, the bottom row of \cref{fig:TT_bigscale1} presents the spectra produced by the accretion disk. The spectra demonstrate the well known double-horn feature, observed from disks in BLRs \cite{Storchi_Bergmann_2017}, while on-top of that they present some additional interesting features, more prominent at edge-on orientations. Specifically, the observed novel feature presents as a double-peaked spectral line centered at around $E/E_0=1$ (modulo some small gravitational shift). The observability of the narrower spectral line feature depends on the inclination, and the line becomes more visible as the inclination becomes more edge-on.   
While the broad double-horn is a well-understood effect due to the kinematics of the disk and the related Doppler and beaming effects that alter the rest frame emitted spectrum, the superimposed narrower spectral line feature with the two peaks is, as far as we know, an effect that hasn't been considered before in the literature.

\begin{figure*}
\begin{center}  
    \includegraphics[width=0.45\textwidth]{90_fardisk_intensity_2_pl.png}  
    \includegraphics[width=0.45\textwidth]{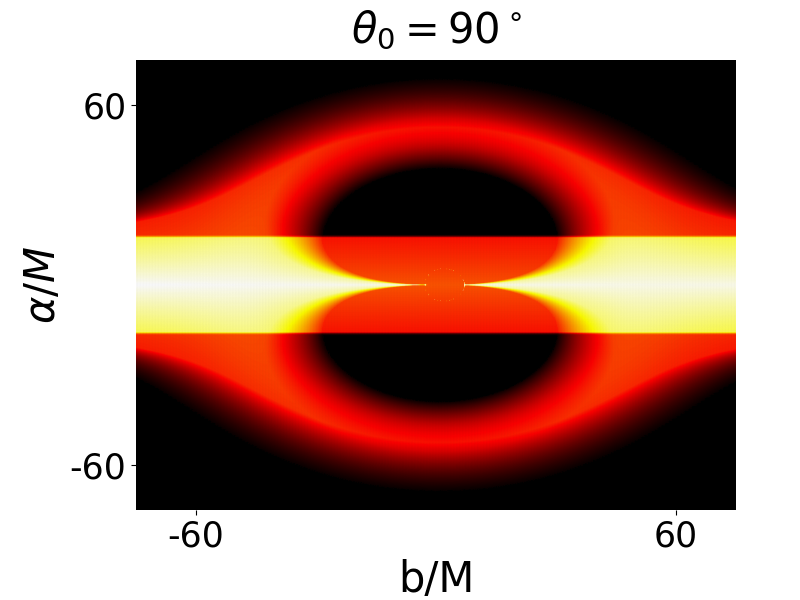} 
    
    \includegraphics[width=0.45\textwidth]{90_stronglylensed_intensity_2_pl.png}  
    \includegraphics[width=0.45\textwidth]{hist_theta90_pl.png} 
 \caption{Larger view of the images and spectrum of an edge-on BLR disk for a Kerr BH rotating at $a=0.998M$ (as in \cref{fig:TT_bigscale1}).}
 \label{fig:ring1}
\end{center}
\end{figure*}

This feature, seen in more detail in \cref{fig:ring1}, is a smoking gun signal of the formation of an Einstein ring around the central BH, illuminated by the far side of the disk.

\subsection{The Einstein ring: measuring the SMBH mass}
%
%
\begin{figure}[htbp]
\begin{center}
    \includegraphics[width=0.45\textwidth]{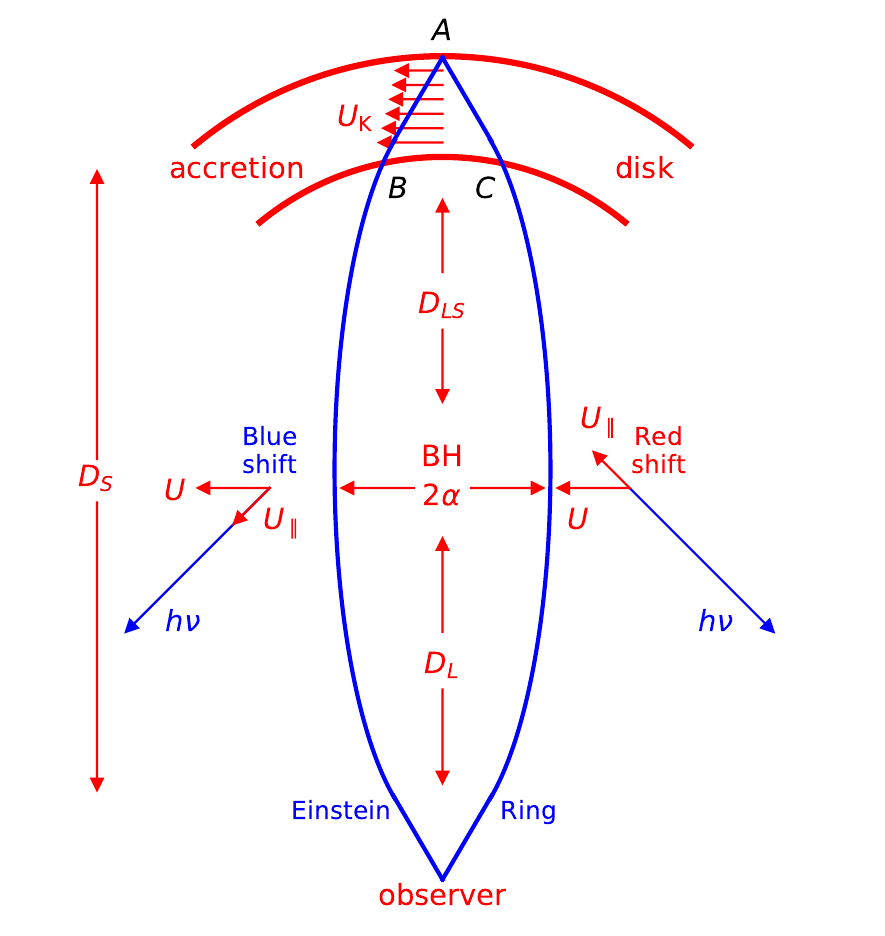}
    \includegraphics[width=0.45\textwidth]{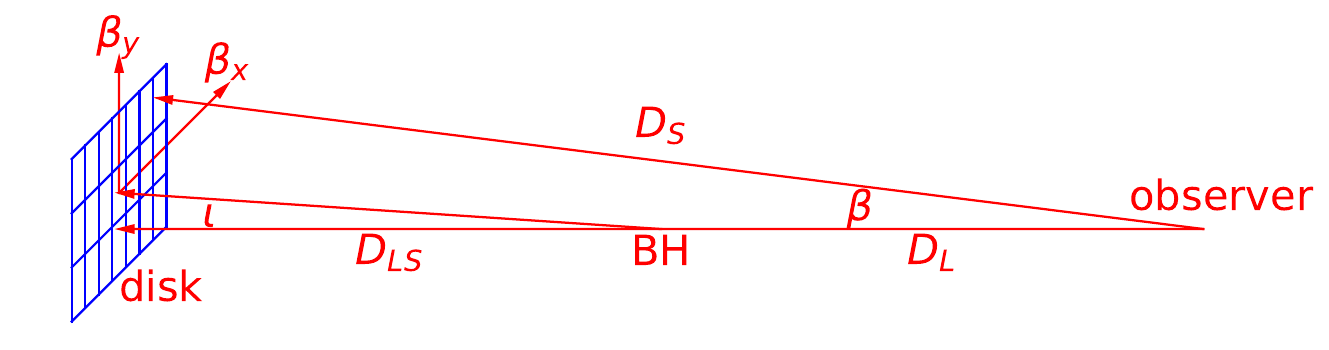}
 \caption{Cartoon of the geometry of the emission coming from behind the black hole and towards the observer, that will form the Einstein ring.}
 \label{fig:cartoon1}
\end{center}
\end{figure}

We will now give a detailed demonstration of how the Einstein ring forms and produces the central feature in the spectrum. We will start by making the assumption that the central BH is approximately a $''$point mass$''$ (which given the size of the disk, is a reasonable assumption). When the disk is viewed edge-on, the photons coming from the emitting gas directly behind the BH $''$point-mass$''$ lens would form an Einstein ring (see \cref{fig:ring1}). The angular size of the Einstein ring for a radiation source, a lens, and an observer, with a geometry like the one given in \cref{fig:cartoon1}, is given by the expression \cite{2018bookCongdon},
\be 
    \theta_E=\sqrt{\frac{D_{LS}}{D_L D_S}\frac{4GM}{c^2}},
\ee
where $D_S$ is the distance of the radiation source from the observer, $D_{LS}$ is the distance of the radiation source from the lens (related to the radial position of the disk in our case), and $D_L$, as we have said, is the distance of the lens (i.e., BH) from the observer. For the geometry of our observer-BH-disk system, where $D_S\simeq D_L\gg D_{LS}$, the size of the Einstein ring will be 
\be\label{eq:alpha_E}
\alpha_E= D_L \theta_E=2\sqrt{D_{LS} G M/c^2},
\ee
where the radius of the Einstein ring $\alpha_E$ is essentially the impact parameter of the photons with respect to the BH, and is the radius of the Einstein ring on the observer's image plane. Therefore, expressed in units of the BH mass $M$,  the image size of the Einstein ring will be, 
    \be \bar{\alpha}_E=\left(\frac{\alpha_E}{M}\right)=2\left(\frac{D_{LS}}{M}\right)^{1/2},
    \ee
which for a disc with $500M<r<1000M$ yields a size $45<\bar{\alpha}_E<63$ (just as seen in \cref{fig:ring1}). In this configuration, apart from the formation of the extended Einstein ring, the emission from behind the BH along the line of sight, is also magnified, forming in this way the central prominent peaks in the line profile  (see \cref{fig:TT_bigscale1} and \cref{fig:ring1}). 

We will now estimate the magnification for a simple setup and by adopting some simplifying assumptions. We start by assuming that at a distance $D_{LS}$ behind the BH, we have a flat surface emitting isotropically with uniform specific intensity $I_\nu'$ (the disk). The specific flux arriving at an observer far away, assuming that nothing intervenes between the observer and the source,  will be 
    \be 
    dF_\nu'= I_\nu' d\Omega',
    \ee
where $d\Omega'$ is the solid angle that the observer perceives and $I_\nu'$ is the emitted specific intensity. In this case, since the propagation takes place in empty and flat space, the specific intensity arriving at the observer is $I_\nu=I_\nu'$, but in general, it is not $I_\nu$ itself that is conserved along propagation, but the invariant specific intensity $\mathcal{I}_\nu$. Similarly, the perceived angular size by the observer is equal to the angular size perceived in flat space, i.e., $d\Omega'=d\Omega$. The total specific flux will then be given by the integral 
    \be 
F_\nu=\int I_\nu d\beta_x d\beta_y,
    \ee
where $(\beta_x, \beta_y)$ are the two angles associated with the angular size of the source in the observer's sky and are related to the two impact parameters, $\alpha$, and $b$.
    
Now, let us consider the case where a point-mass gravitational lens exists between the source and the observer. We will assume for simplicity that the gravitational or other shift to the frequency is small and therefore $$I_\nu'/\nu'^3=I_\nu/\nu^3\rightarrow I_\nu'=I_\nu,$$
which is an assumption that will not affect the result of our calculation. The specific flux arriving at the observer in this case will be 
 \be 
    dF_\nu'= I_\nu' d\Omega'=I_\nu \mu d\Omega,
\ee
with $\mu=d\Omega'/d\Omega$ being the total magnification of the radiating source due to a point mass lens, which is ~\cite{2018bookCongdon}
\be \mu = \frac{u^2+2}{u\sqrt{u^2+4}}, \ee
where $u=\beta/\theta_E$ is the normalised angular actual position of the source on the emission plane. The apparent position of the source will be given by the angle \cite{2018bookCongdon}
\be \label{eq:point_lens}
\theta_\pm= \frac{\beta\pm\sqrt{\beta^2+4\theta_E^2}}{2},
\ee
which corresponds to the two images of the source formed by the BH lens. The source magnification  will then be,
\be 
\mu_{\rm tot}=\frac{\int I_\nu \mu d\beta_x d\beta_y}{\int I_\nu d\beta_x d\beta_y}=\frac{\int \mu d\beta_x d\beta_y}{\int d\beta_x d\beta_y}.
\ee
%
\begin{figure}[htbp]
\begin{center}
   \includegraphics[width=0.15\textwidth]{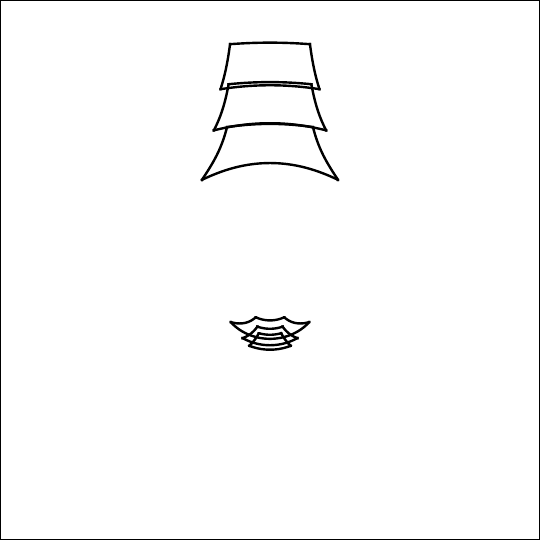}
   \includegraphics[width=0.15\textwidth]{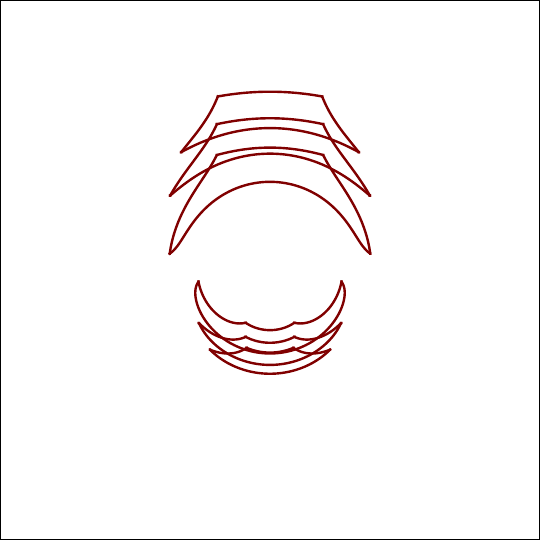}
   \includegraphics[width=0.15\textwidth]{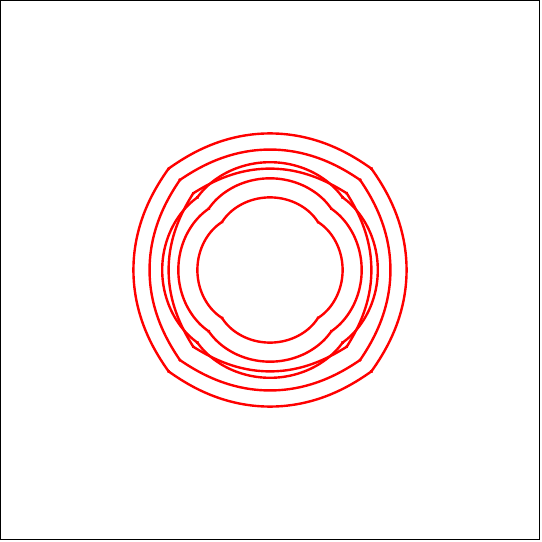}

\includegraphics[width=0.15\textwidth]{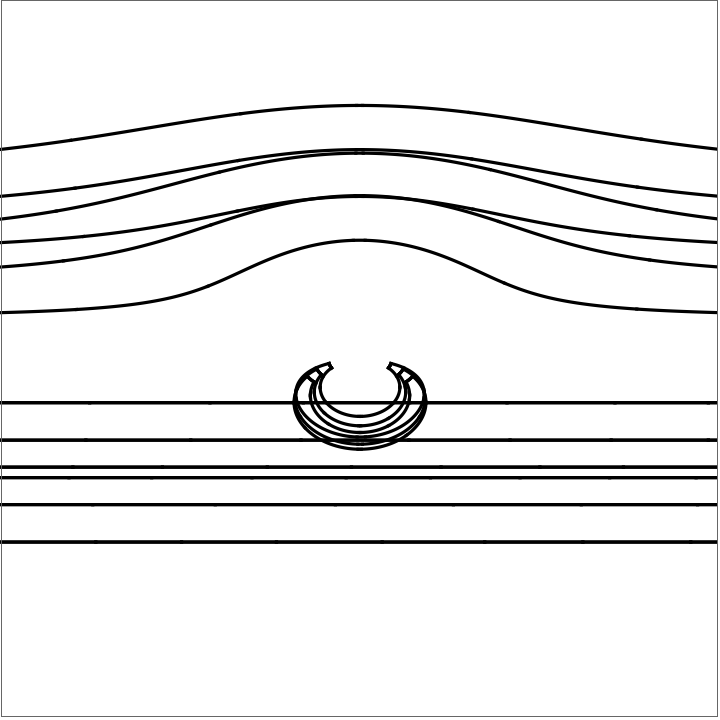}
   \includegraphics[width=0.15\textwidth]{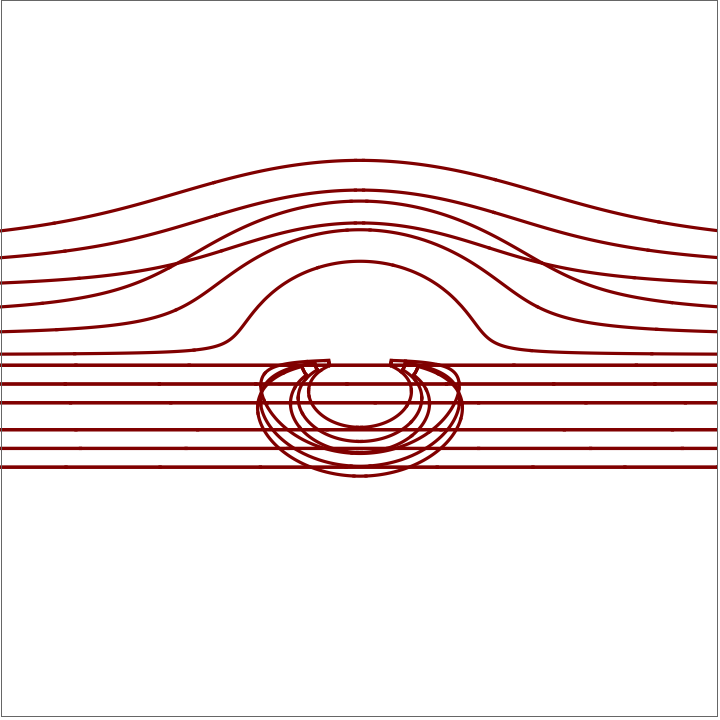}
   \includegraphics[width=0.15\textwidth]{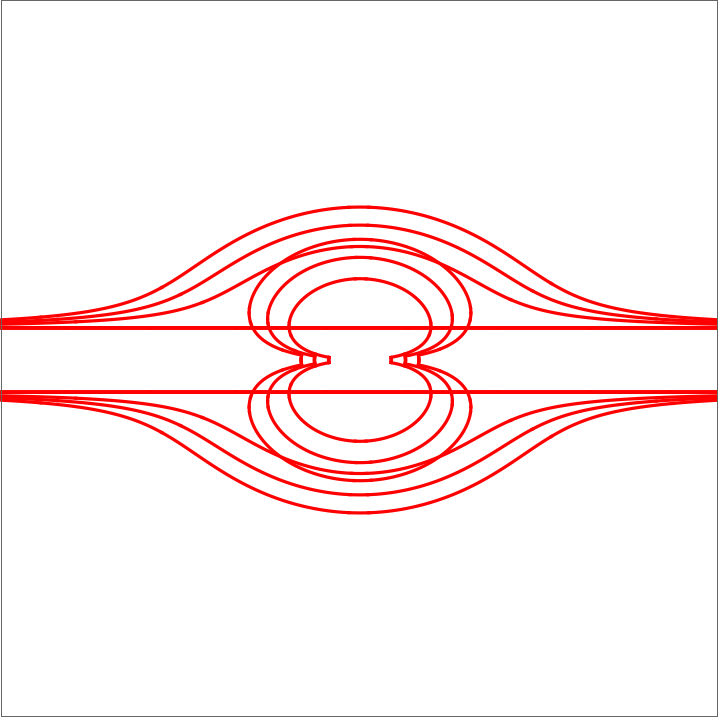}
 \caption{Depiction of the lensed image of the emission region for different viewing inclination angles of the disk. The top row shows the image from only one square forming the emission plane, for different inclinations, while the bottom row shows the image for an emission plane with a wider horizontal extension as to simulate a disk, also for three inclinations. The plots show three emission planes ($500M, 750M, 1000M$) for inclinations $\iota=4^o, 2^o, 0^o$
 (left to right). The two figures on the right column show the corresponding Einstein rings. The resemblance of the bottom row rightmost figure to the actual images from the ray-tracing is clear.}
 \label{fig:E_image}
\end{center}
\end{figure} 

\begin{figure}[htbp]
\begin{center}
    \includegraphics[width=0.4\textwidth]{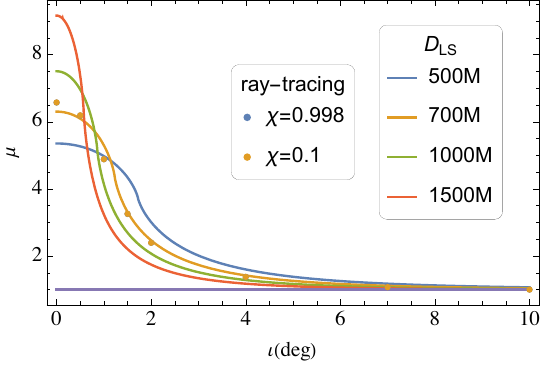}
 \caption{Magnification on the Einstein ring of a square area emission from behind the BH, theoretical (lines) and through ray-tracing (points). The points with the magnification from the ray-tracing for the two spin values practically coincide.}
 \label{fig:magnification}
\end{center}
\end{figure}

Assuming the geometry of \cref{fig:cartoon1} for the source plane and assuming the simplifying assumption that the inclination $\iota$ is small and that the emission is almost perpendicular to the emitting surface, one can calculate the shape of the source on the image plane of the observer, using the two images' positions given by eq. \eqref{eq:point_lens}, and also estimate the magnification for the various emission planes and inclinations. Using the approximation, $\beta^2=\beta_x^2+\beta_y^2$, and the fact that the inclination essentially causes only a shift of the location of the emission plane by, $\iota D_{LS}$, we have plotted the shape of the emission area as it is seen in the observers image plane in \cref{fig:E_image} as well as the magnification $\mu_{\rm tot}$ for various inclination angles, in \cref{fig:magnification}. The top row of \cref{fig:E_image} corresponds to a shape of the emission area that is a square with $-15M\leq \alpha \leq 15M$ and $-15M\leq b \leq15M$. Fig. \ref{fig:E_image} also shows at the bottom row the image for various inclinations of an emitting surface that is wider than the previous case (with the vertical width being $-15M\leq \alpha \leq 15M$, as before, but the horizontal width is larger than our observing window). This latter configuration is chosen so as to emulate the shape of a disk, and for this reason, we have both an emission band behind the BH as well as an emission band in front of the BH, which is not lensed. In all the images of \cref{fig:E_image} we show three emission planes, at distances $r=500M,750M,$ and $1000M$ that span the entire radial size of the corresponding disk. The magnification curves in \cref{fig:magnification} correspond to the first case of the emitting square areas and assume the emission planes at distances $r=500M,700M,1000M,$ and $r=1500M$.

%
Figure \ref{fig:E_image} essentially shows that the simplified theoretical calculation of the Einstein ring produced by the disk behind the BH using eq. \eqref{eq:point_lens}, describes the observed Einstein ring in the ray-tracing strikingly well, while \cref{fig:magnification} shows that the magnification is larger when the plane of emission is further away from the BH and therefore the maximum of the flux in the spectrum in \cref{fig:ring1} should correspond to the outer radius of the disk that is emitting. 

\begin{figure*}
\begin{center} \includegraphics[width=0.45\textwidth]{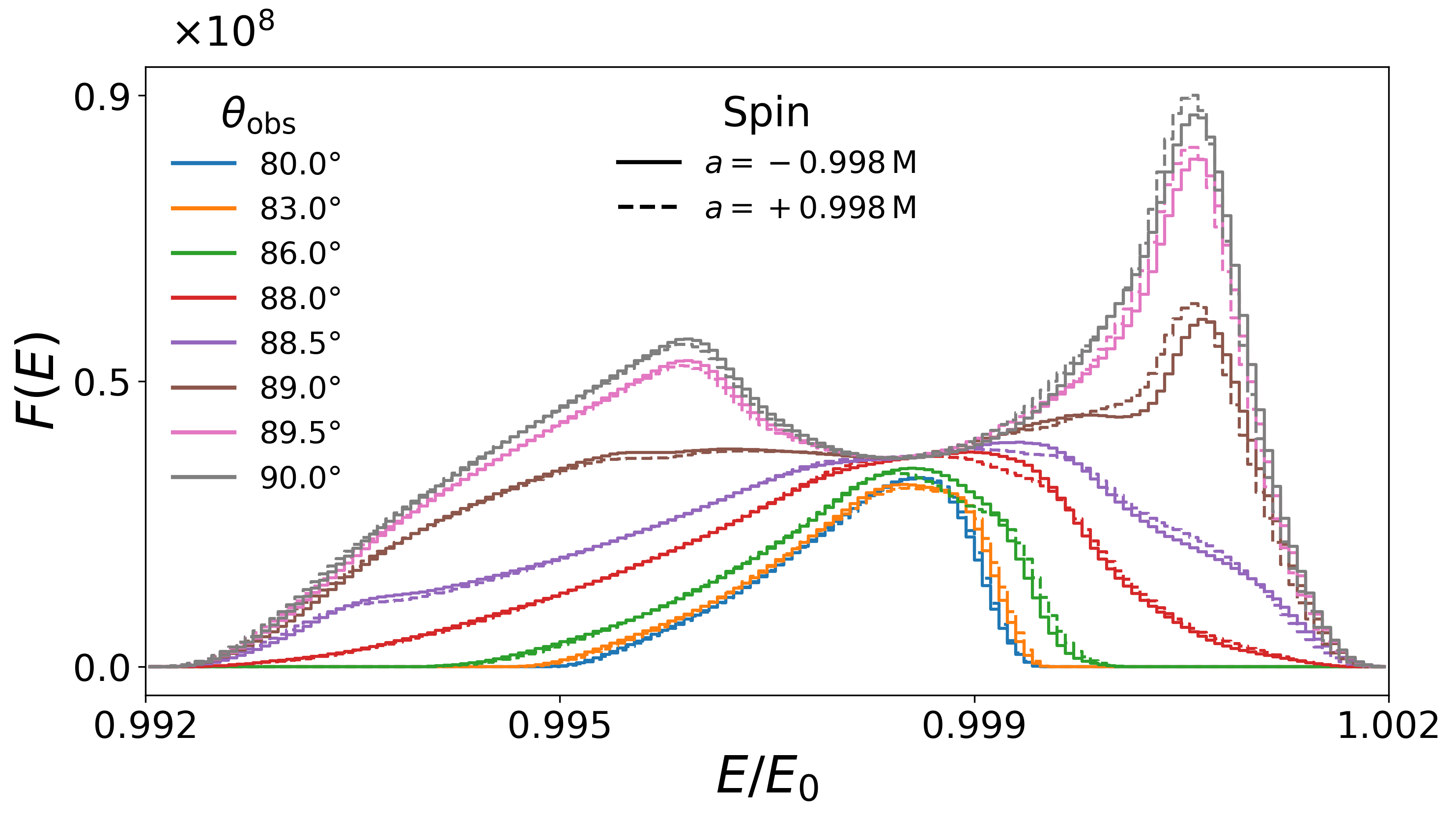}
    \includegraphics[width=0.45\textwidth]{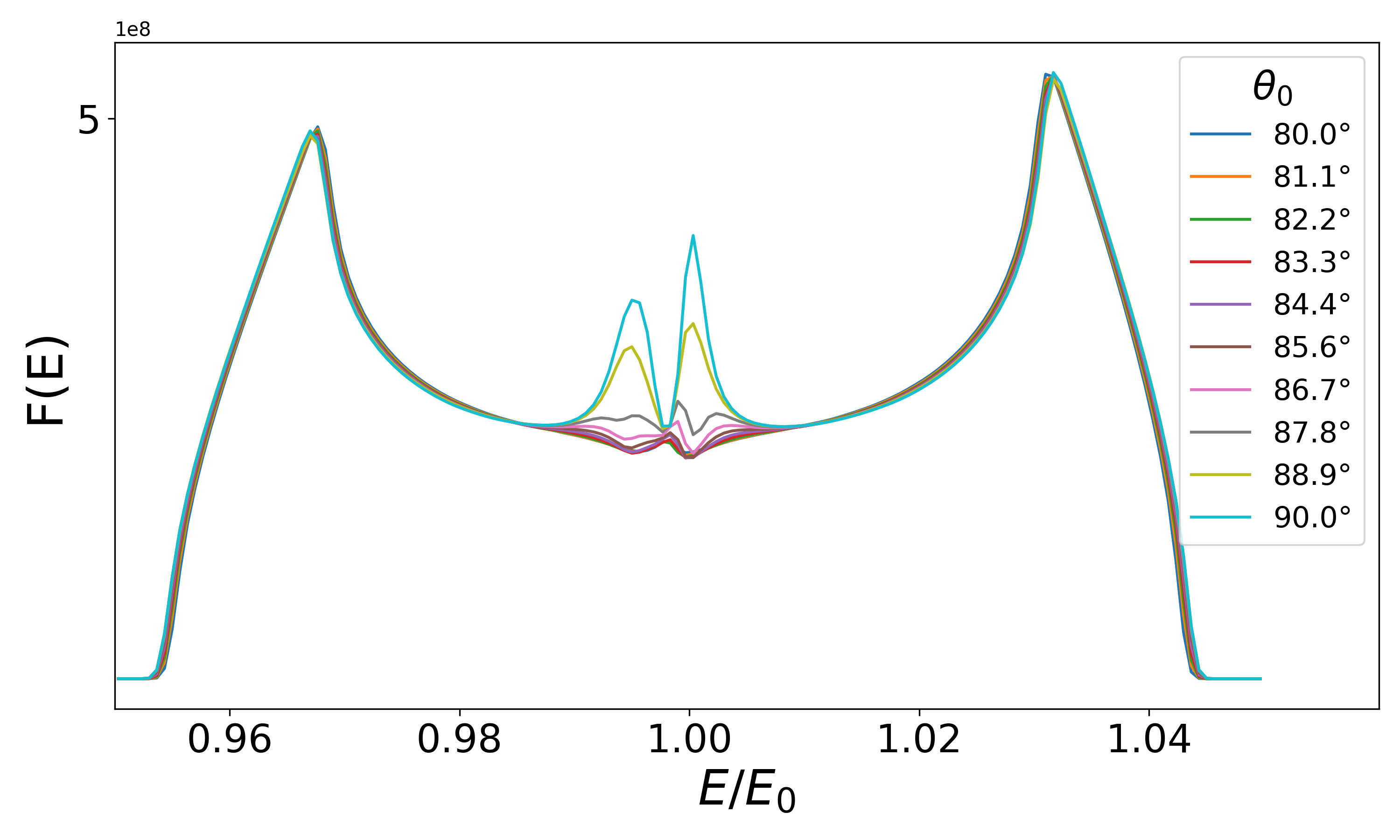}  \includegraphics[width=0.9\textwidth]{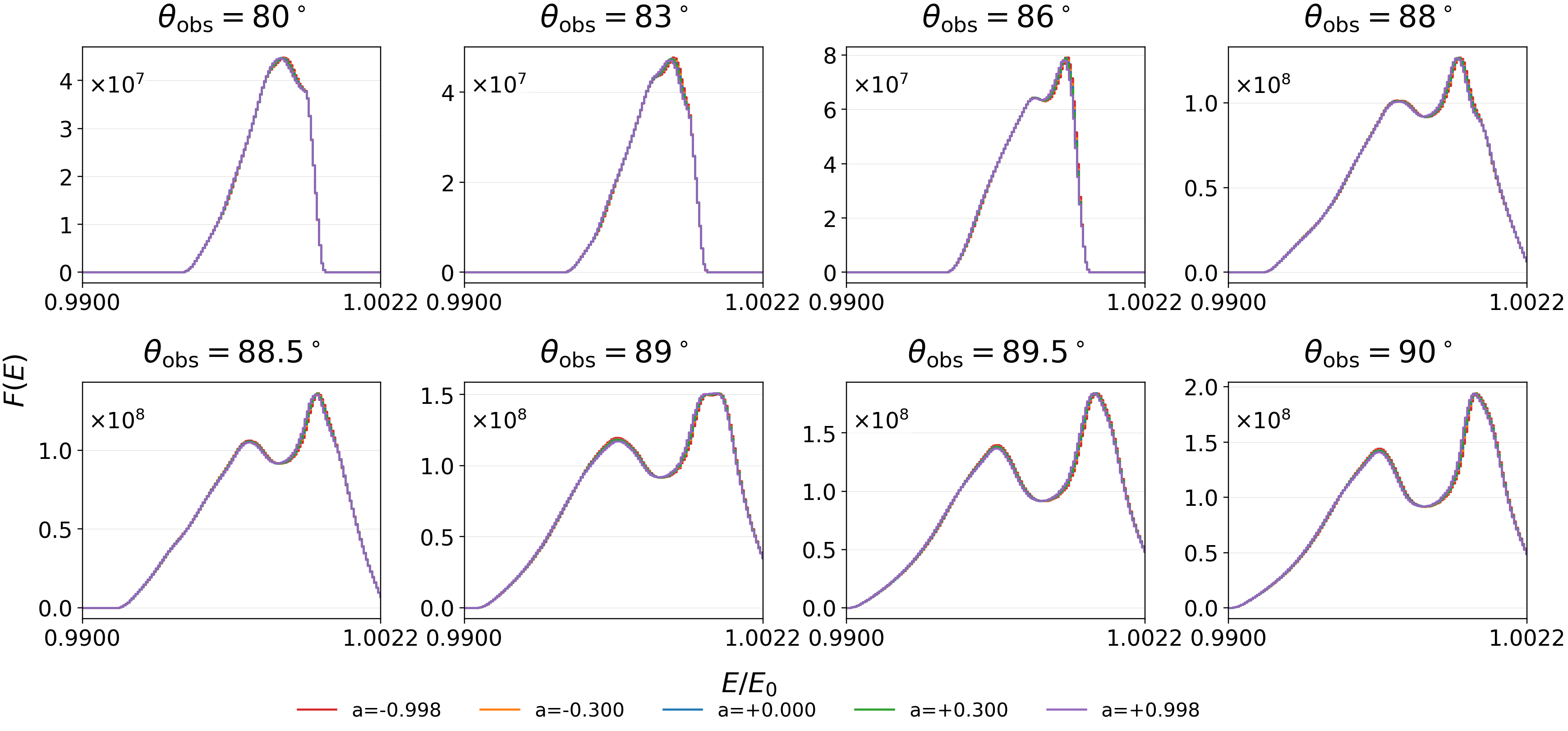}
 \caption{Demonstration of the emergence of the central magnified double peak of the spectral line with changing inclination, by isolating the emitting region to that behind the black hole as shown in \cref{fig:cartoon1}. The top-right plot shows how the combined total spectrum depends on the inclination. The plots also demonstrate the dependence of the spectrum on the BH spin. The figures have been produced using the emissivity profile given by eq. \eqref{eq:emis2}.}
 \label{fig:E_spectra}
\end{center}
\end{figure*} 
%
In order to further verify the emission and the line profile we would get from the ray-tracing, we have calculated the image and the spectrum using ray-tracing, assuming that only a block of material of square cross-section, located behind the black hole, is emitting. In this way, we can isolate the effect of the emission coming from that region of the disk and compare it to the theoretical calculation in order to test our model and interpretation. Furthermore, we have calculated the emission for various values of the BH's rotation from counter-rotating to co-rotating, all the way up to the Thorne limit. In addition, we have also calculated the full spectrum in the relevant range of inclinations between 80-90 degrees from the axis of rotation for a BH rotating with $a=0.998M$, to better see how the central feature emerges. The resulting spectra are shown in \cref{fig:E_spectra}, where we observe the emergence of two peaks under the various setups. In these spectra is apparent that the BH spin has negligible effects.

%
\begin{figure}[htbp]
\begin{center}
    \includegraphics[width=0.4\textwidth]{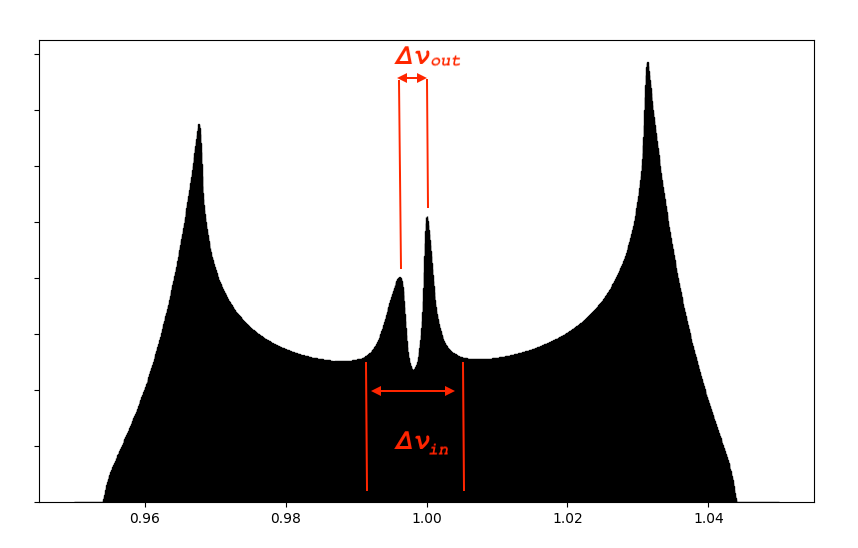}
 \caption{Illustration of the Doppler shifts relation to the observed spectrum.}
 \label{fig:cartoon2}
\end{center}
\end{figure} 
%
So far, we have focused our attention on the increase in the flux in the central region of the spectrum due to lensing, but there is also the effect of having two distinct peaks instead of one. The question then is, what is it that is causing the split into two peaks? In our previous calculations, we assumed that the rays are emitted perpendicular to the surface of emission. But as \cref{fig:cartoon1} shows, this is not exactly the case. The rays generally go through the volume of the gas at some angle. Therefore, the reason for having two peaks, one being red-shifted, while the other being blue-shifted, is because every ray intersecting the line of sight along the back side of the disc that contributes to the formation of the Einstein ring, will either move through the gas with the same sense as the gas is moving around the BH (left ray in \cref{fig:cartoon1}) or will move in the opposite sense (right ray in \cref{fig:cartoon1}). 

The rays following a polar orbit will be the only ones without any energy shift, but these orbits are only a small fraction of all the light rays, providing less flux in the middle between the two peaks. One can give an estimate of the relevant Doppler shifts, assuming a thin lens approximation, where the impact parameter of a light ray coming from behind the black hole will be equal to the distance of the observer from the black hole times the angular size of the Einstein ring, which will be $\theta_E D_L$. 
Since in this approximation, the light rays are straight lines, the ray intersects the fluid element of the disc (moving horizontally) at the point of emission, at an angle that will result in a projected velocity along the ray equal to 
\be u_\parallel = u_K \frac{\theta_E D_L}{D_{LS}}=\sqrt{\frac{GM}{D_{LS}}}\sqrt{\frac{4GM}{c^2D_{LS}}}=\frac{2GM}{c D_{LS}},
\ee
which will be either in the direction of the emission (for the rays on the left) or against the direction of emission (for the rays on the right). The total Doppler shift between left and right rays will then be,
\be \label{eq:Dnu}
     \frac{\Delta \nu}{\nu} = 2 \frac{u_\parallel}{c} =\frac{4GM}{c^2 D_{LS}}, 
    \ee
which for radii between $500M<r<1000M$ will give shifts between $\frac{4}{1000}<\frac{\Delta \nu}{\nu}<\frac{4}{500}$. This can be seen in \cref{fig:E_spectra} and within the wider context of the total emission line in \cref{fig:ring1} or \cref{fig:cartoon2} where the Doppler shift from the emission coming from the far side of the accretion disk (at $D_{LS}=r_{out}$) corresponds to the energy shift between the two central peaks, while the Doppler shift from the emission coming from the near side of the accretion disk (at $D_{LS}=r_{in}$) corresponds to the total width of the central magnified part of the line. 
The latter effect of the frequency splitting of the emitting spectral line $\Delta \nu$ together with the presence of an Einstein ring of size $\alpha_E$ provide for a way of measuring the mass of the central SMBH. Combining eq. \eqref{eq:Dnu} with eq. \eqref{eq:alpha_E}, we have for the mass of the BH \cite{PRL_companion}, 
\be
M=\sqrt{\frac{\alpha_E^2}{16}\frac{\Delta \nu}{\nu}},
\ee
in geometric units.

\subsection{The Einstein ring: robustness of the effect}
%
\begin{figure*}[h!t]
\begin{center}   \includegraphics[width=0.6\textwidth]{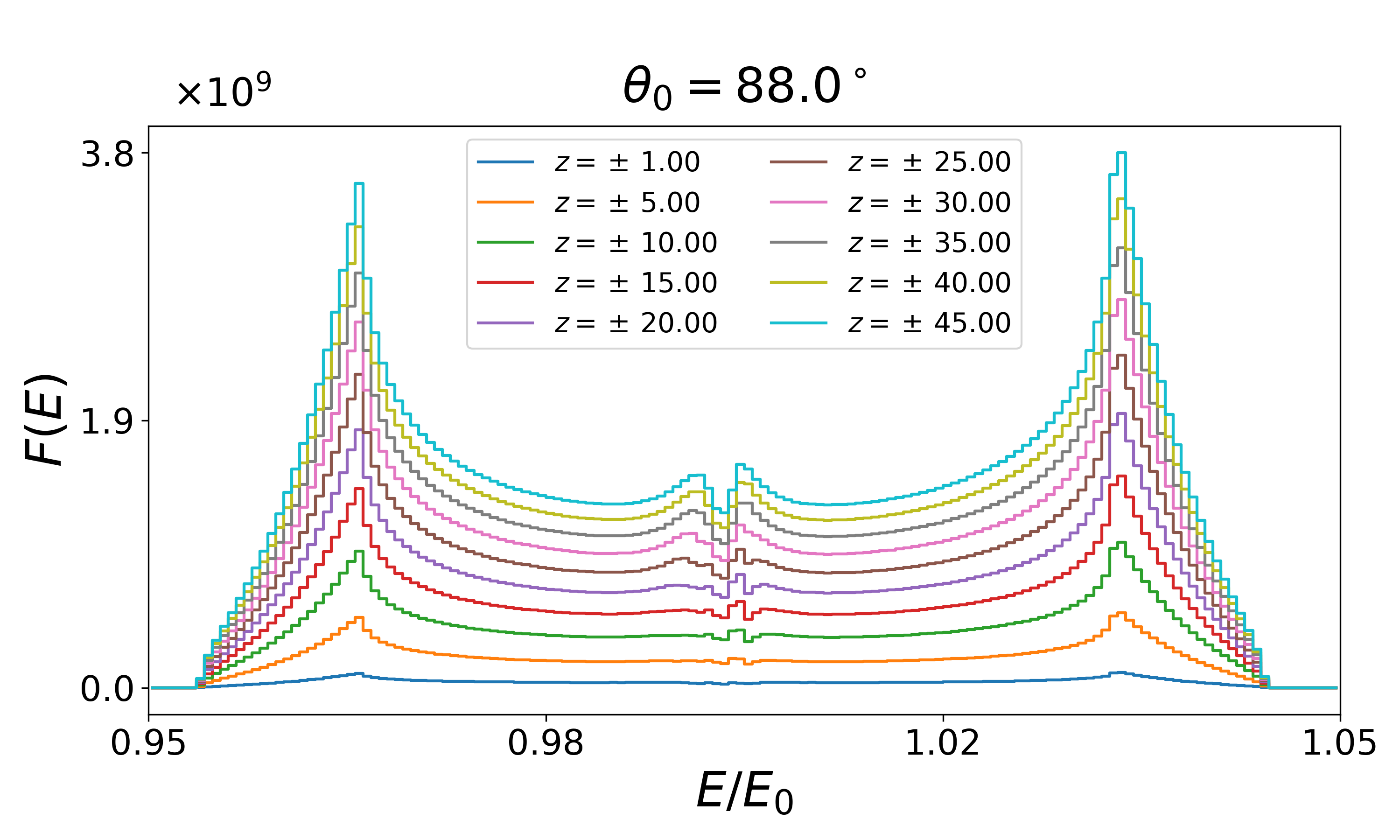}   \includegraphics[width=0.9\textwidth]{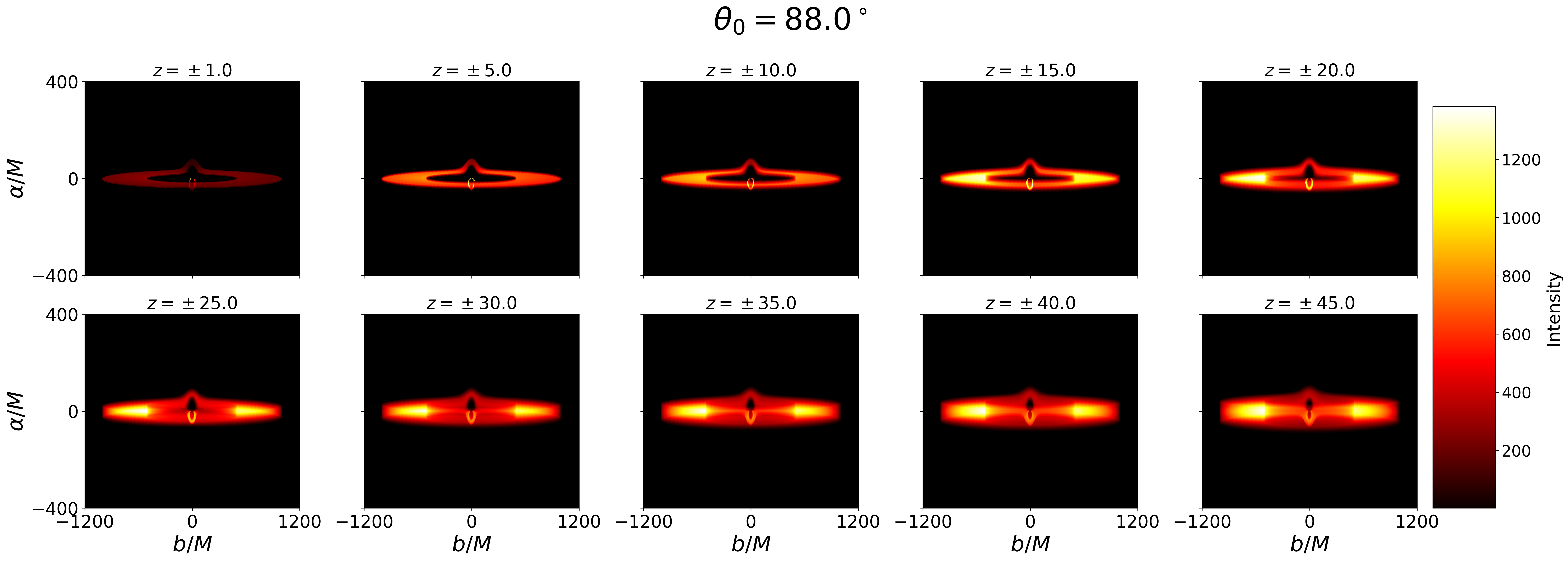}
 \caption{Dependence of the spectrum and the image of the accretion disk and the Einstein ring on the thickness of the disk. For this figure, we have used the intensity distribution given by eq \eqref{eq:emis1}.}
 \label{fig:robust}
\end{center}
\end{figure*} 
%
%

So far we have discussed the properties of the Einstein ring, how it forms, and its dependence on the inclination of the disk and the rotation rate of the BH. Here, we will explore how some additional parameters may affect the formation and the observability of the effect.  Specifically, we will explore how the thickness of the accretion disk affects the image and the spectrum. 

For this, we have considered disks with thicknesses ranging from $z=\pm1M$ to $z=\pm45M$. The main effect of changing the thickness of the disk is, as expected, on the total emitted and observed flux, as a thicker disk has more emitting material and therefore is expected to produce more radiation. This is clearly seen in the spectra shown in \cref{fig:robust} for a disk observed at an angle of $\theta_0=88^{\circ}$ (from the axis of rotation).  

The disk thickness also affects the formation of the Einstein ring and its fingerprints on the spectrum. As the thickness of the disk increases for a given inclination, the far side of the disk starts having some material located immediately behind the BH. As we have seen in the previous subsection, it is the material located directly behind the BH along the line of sight of the observer that is producing the emission that leads to the Einstein ring. We can see in \cref{fig:robust} in both the spectrum and the images, that indeed for increasing thicknesses, the characteristic image of an Einstein ring (see \cref{fig:ring1} and \cref{fig:E_image}) starts to form and the characteristic double-peak becomes more prominent. 

\begin{figure}[htbp]
\begin{center} \includegraphics[width=0.45\textwidth]{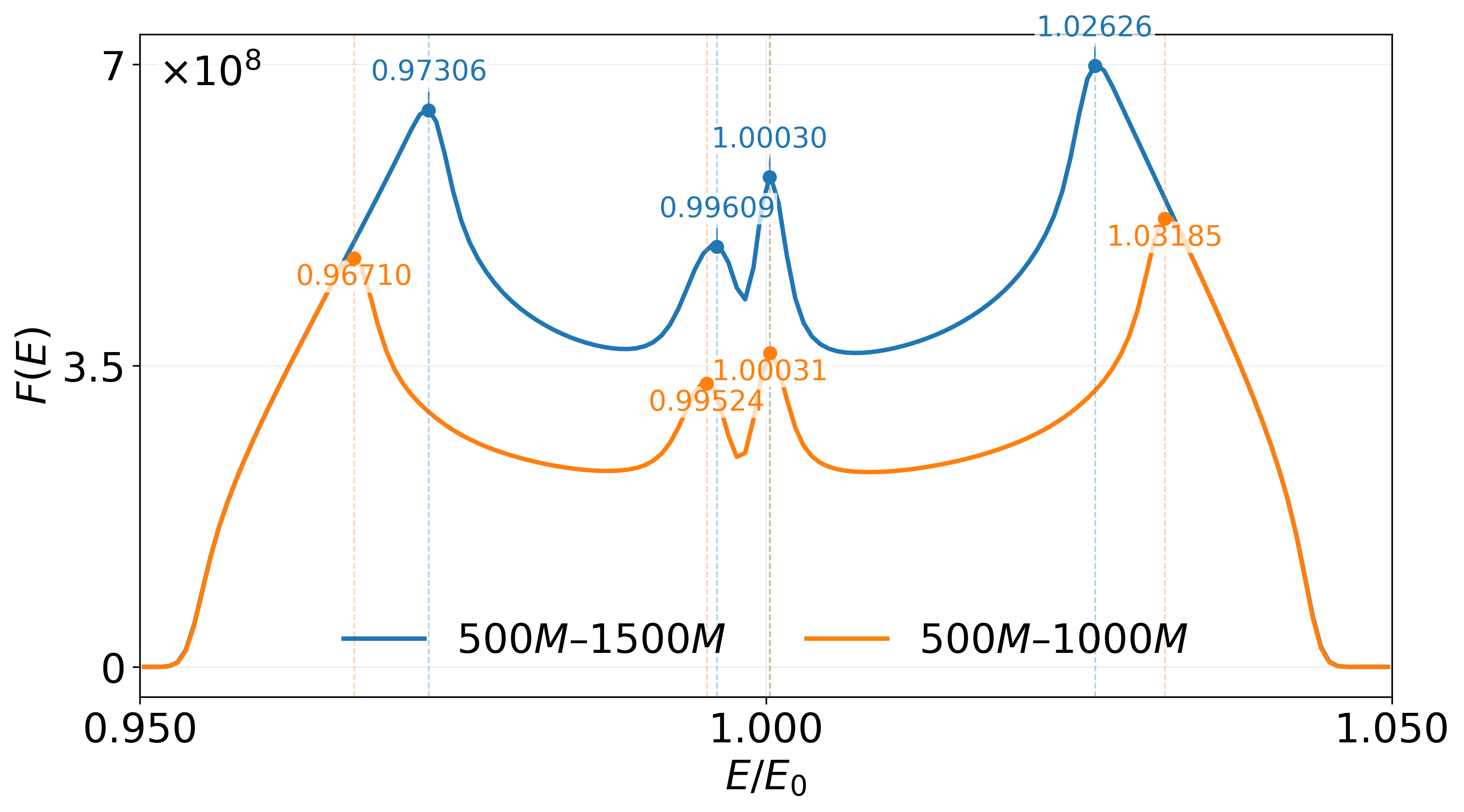}
\includegraphics[width=0.45\textwidth]{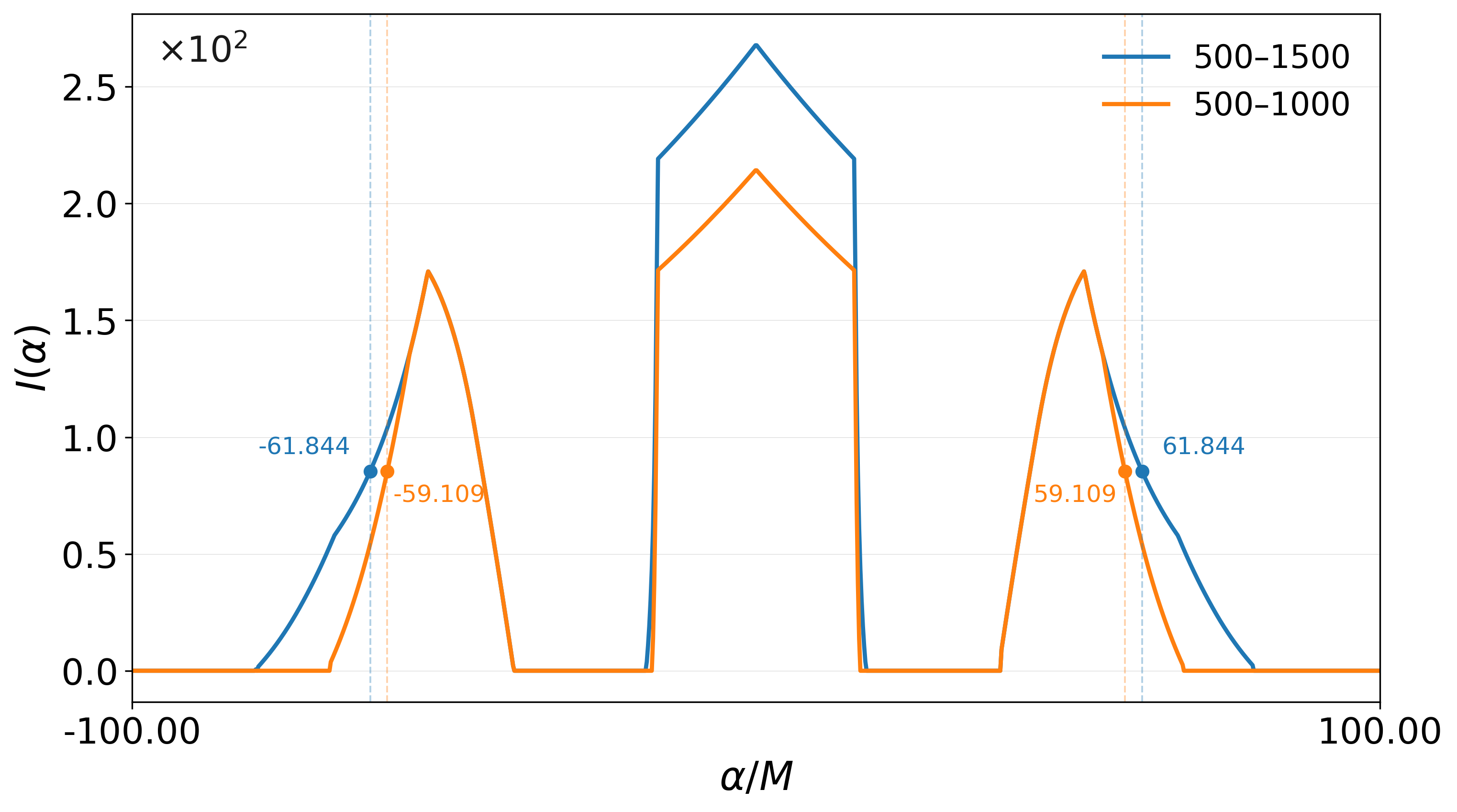}
 \caption{Effect on the spectrum and the Einstein ring of changes in the outer radius of the emission. For this figure, we have used the emissivity profile given by eq \eqref{eq:emis2}.}
 \label{fig:larger_disk}
\end{center}
\end{figure}

Another factor that can affect the appearance of the Einstein ring signature in the spectrum is that of how far out the emitting region extends. In a realistic setup, the result depends on a combination of the actual outer extent of the emitting region and the fraction of the emission that is coming from that outer region. Fig. \ref{fig:larger_disk} shows the spectra for a disk that extends out to a radius of $1000M$ and a disk that extends out to $1500M$. Under the spectra, we have a cross-section for a value of $b$ near zero, of the intensity of the disk as a function of the impact parameter $\alpha$. The two spikes, on the left and right sides, are the Einstein rings, and the central spike is the emission coming from the central horizontal band (corresponding to the near disk in front of the BH) and the light-ring. The intensity profiles from the two disks are, for the most part, the same, apart from the emission coming, for the second disk, from the region beyond the $1000M$. In the intensity profile, it is not clear where exactly should one consider that the Einstein ring starts and where it ends. We will assume the general qualitative criterion that the ring starts and ends at the location where the intensity is approximately half the maximum for the spike. This is going to give us an effective outer radius of the ring. Similarly we will assume that the frequency shift $\Delta \nu$ observed in the central twin-peak is related to that radius. Therefore, the frequencies of the central twin-peak should depend on that effective outer radius of emission $\varpi_{out}^{eff}$, which is related to the apparent outer size of the Einstein ring, $\alpha_E^{eff}$. The choice of where exactly the edge of the Einstein ring should be taken, is something that should be more thoroughly studied in order to produce a generic, robust relation between the mass of the BH and the observed $\Delta \nu/\nu$ and $\alpha_E$. For our purposes, though, we can see that for our setup, the result is quite robust. Using the aforementioned way for estimating the outer $\alpha_E^{eff}$, we have for the first case: $\alpha_E^{eff}\simeq 59M$, while $\Delta \nu/\nu \simeq 0.005$. This gives a mass of $M\simeq1.04$. For the second case, where the disk extends out to $1500M$, the estimate for the outer size of the Einstein ring is $\alpha_E^{eff}\simeq 62M$, while $\Delta \nu/\nu \simeq 0.0042$. This yields $M\simeq1.005$, in good agreement with the previous result. 

Therefore, thicker disks are expected to have a wider range of inclinations for which the Einstein ring and its spectral signatures will be observable, apart from the fact that they are expected to produce larger fluxes and have a stronger signal. Furthermore, the observed effects of the Einstein ring and the central twin-peak give robust mass estimates for the central BH even in terms of an effective outer radius and Einstein ring size. Nevertheless, a systematic algorithm for producing robust mass estimates in a general case needs to be developed, and we leave it for future work.

On the practical front, the spectral line emission from the BLR disk, which provides the radiation background to the BH for the Einstein ring to become visible, both geometrically and spectrally, must have no significant $''$contamination$''$ from non-disk structures or non-Keplerian velocity fields (e.g. strong gas outflows from the BLR disk). The former will complicate the Einstein ring geometry, and the latter will insert unrelated spectral features to the double-peak spectral features seen e.g. in \cref{fig:cartoon2}. In
future interferometric spectral line imaging of the Einstein ring around SMBHs, non-Keplerian velocity-field $''$contamination$''$ is more serious than non-disk emission structures (which may be resolved out if extended). This is because non-Keplerian velocity field contributions can be compact in nature (e.g. discrete outflowing gas clouds from the BLR disk, or random line-of-sight clouds from galaxy regions beyond the BLR), and can thus remain present in the final spectral line interferometric data. In that regard, the geometrically thin, much less turbulent,  neutral and dense BLR$^{0}$, with its unique neutral atomic or molecular lines at IR wavelengths, is the best possible SMBH spectral line illuminator that could deliver such data in the future \cite{Astro_paper}.

\subsection{Total spectrum and inclination}
\begin{figure}[htbp]
\begin{center} \includegraphics[width=0.45\textwidth]{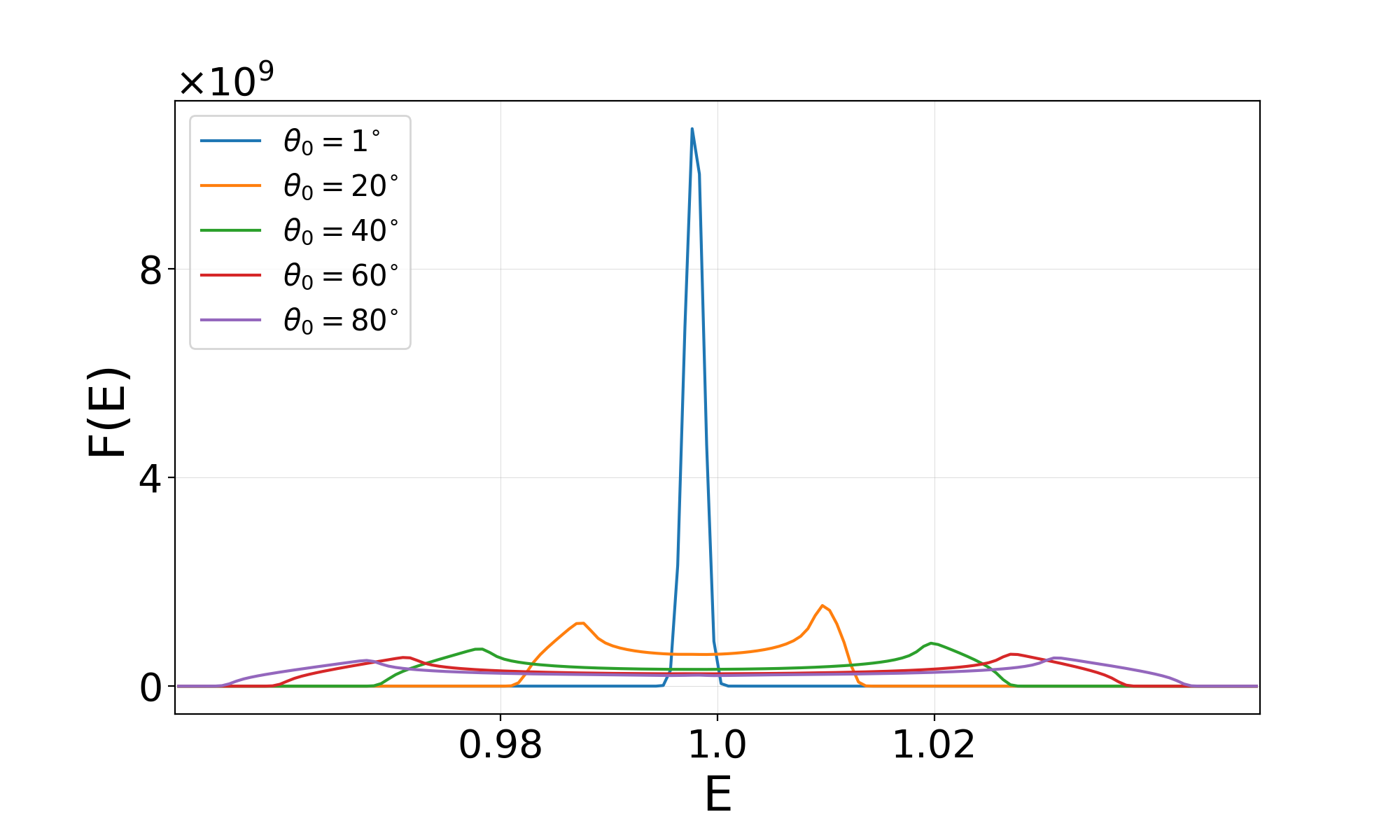}
 \caption{Effect of the inclination on the observed spectrum. For this figure, we used the emissivity profile given by eq \eqref{eq:emis2}.}
 \label{fig:spectra_i}
\end{center}
\end{figure}

%
A final thing worth examining, for completeness, about the total spectrum of the BLR disk is its dependence on the inclination. In \cref{fig:spectra_i} we present spectra for $\theta_0=1^{\circ},~20^{\circ},~40^{\circ},~60^{\circ},$ and $80^{\circ}$ (angle between rotation axis and line of sight to the observer). As expected, the line profile broadening depends on the viewing angle, with observers near the axis of rotation seeing a narrow spectral line, which has suffered some gravitational redshift but almost no Doppler broadening, while for observers away from the axis, the line starts having the familiar broadening and double-peak profile. Another notable feature is that the total frequency-integrated line flux received (the area under each profile) remains constant with observing angle. This is because of the assumed {\it effectively} optically thin BLR lines (\citet{Astro_paper}), which make the frequency-integrated line flux proportional to the total BLR gas mass, with the proportionality constant depending on line excitation factors and gas thermal conditions (which of course do not depend on the viewing angle).

A final thing to note is that, as the spectra demonstrate the well-known double-horn feature associated with BLRs, this also encodes information on the structure of the BLR disk, the viewing angle, and the mass of the SMBH \cite{Storchi_Bergmann_2017}. This information can provide an additional handle on the parameters of the system.

\section{Focusing on the light-ring:  measuring the spin}
\label{sec:light-ring}
%

The light-ring  encodes information on the BH mass (which sets the scale and  overall ring size), but in addition it also holds information about the BH spin. 
In particular, estimating the BH spin using the light-ring has been investigated extensively in the literature \cite{Takahashi_2004,Johannsen_2010,Johannsen:2013vgc,Broderick:2013rlq,Johannsen:2015mdd,Tsupko2017PhysRevD,Johnson:2019ljv,Gralla:2020srx,Younsi:2021dxe,Gralla:2020nwp,Cardenas-Avendano:2023dzo,Medeiros:2019cde,KumarWalia:2024omf}. The light-ring is affected in two ways by the rotation, at least in the case of Kerr BHs \cite{Bardeen:1973tla}. The first effect of rotation is that the ring (when the observer is not positioned near the axis of rotation) migrates from being centered at $(b,\alpha)=(0,0)$ (the position of the central BH) to being centered at some positive value of $b$ (when we have the BH rotating in the usual sense, i.e., the angular momentum pointing upwards). The second effect has to do with the fact that the shape of the light-ring changes from circular to that of a deformed circle with a depression on the left side (a sort of $D$-like shape), where the deformation becomes more significant for higher rotation rates that are closer to the maximal-rotation limit. In general, though, this deformation is small. We remind here that both the displacement and the deformation depend also on the inclination and are stronger for more edge-on inclinations, while for face-on inclinations the effects disappear. In general, the inclination is unknown and is thus a source of uncertainty, but in our case, the additional information from the spectrum could lift the inclination uncertainty. In any case, for the Einstein ring to be visible as well, the inclination must be close to edge-on, leaving little inclination dependence. 

Of the two effects, the second is generally easier to identify, since it is a change in the shape of the light-ring and is, in a sense, a modification of an invariant (coordinate-independent) property of the ring. The horizontal shift of the ring, on the other hand, is practically difficult to identify unless one has a very good frame of reference, which, when imaging the light-ring itself, is not available. For this reason, more work has been done on using the former effect for measuring the spin of the BH \cite{Broderick:2013rlq,Medeiros:2019cde} than using the latter effect \cite{Koutsantoniou:2022fev,Koutsantoniou:2023anr}. In the latter case, for example, it has been attempted to use the full accretion disk as a reference for measuring the shift of the light-ring with respect to the center of the $(b,\alpha)$ frame, as it is defined by the position of the BH. When the disk is close to the BH, then, as we have mentioned, it is difficult to separate the light-ring from the inner edge of the disk, while when the disk is far, the separation of scales can be too large for both the light-ring and the disk to be used to extract useful information.

The Einstein ring itself, as a purely geometrical and symmetric (i.e. $I(\vec{r})=I(-\vec{r})$) emission feature, can identify the BH center. This then allows re-setting the origin of the image coordinates on the BH center, more precisely than what the BLR disc emission could allow\footnote{Realistic BLR  disks will have
asymmetries in their line emission, making them less effective for such a role.}. Then determining the offset of the light-ring from the coordinate center can yield the BH spin. The aforementioned analysis,  and the use of Einstein-ring/light-ring offsets for spin determination, can be performed entirely using the primary measured quantities of interferometric observations, i.e., the visibilities \cite{Johnson:2019ljv,Gralla:2020srx,Gralla:2020nwp,Cardenas-Avendano:2023dzo}. This is an added
advantage, given the non-trivial complexity of interferometric image deconvolution techniques \cite{Corn99_N1, Corn99_N2}, especially when only a few baselines are involved\footnote{The case for VLBI and, initially at least, also in future space-borne IR interferometers imaging the BLR$^{0}$ lines.}. Moreover, positional offsets 
in the image plane translate to phase shifts in the visibility plane, {\it and phases are the most sensitively measured interferometer outputs} as their thermal noise probability distribution is strongly modified
even for signal/noise (S/N)$\sim $1-2, quite unlike visibility amplitudes whose probability distributions are
significantly modified away from that of thermal noise for S/N$\geq $5 \cite{Cra89}.

In interferometry, the measured visibilities $V(u,v)$ are samples of the $2D$ Fourier transform of the sky brightness $I(b,\alpha)$. Using the standard convention, with $(u,v)$ the baseline vector measured in wavelengths on the
Fourier plane, and $(b,\alpha)$ the image plane coordinates, it is 
 \begin{equation}
     V(u,v)=\iint I(b,\alpha)e^{-2\pi i \left(ub+v\alpha\right)}dbd\alpha,
 \end{equation}
(assuming the small field imaging approximation) with amplitude $A(u,v)$ and phase $\Phi(u,v)$:
 \begin{equation}
     A(u,v)\equiv |V(u,v)|,\quad \Phi(u,v)\equiv\arg V(u,v),
 \end{equation}
with the $(u,v)$ vector, of length $\rho=\sqrt{u^2+v^2}$, and the related visibility $V(u, v)$, probing angular scales of $D_L\theta\sim 1/\rho$, or scales of $r=\sqrt{b^2+\alpha^2}\sim 1/\rho$ in the image plane. Short baselines are sensitive to extended emission features (e.g. outflows), while long baselines record high spatial frequencies, i.e. small angular-size structures such as BH shadows, and their light-rings. A suitable choice of the short baseline $(u,v)$-range can then spatially filter out any extended emission from the corresponding interferometric images, no matter how bright it is. This is important if  the BLR  line emission  is superimposed on extended emission line distributions such as BLR winds or  background/foreground line-emitting regions of the BH host~galaxy\footnote{One can of course minimize such problems by selecting those spectral lines  unique to the BLR$^{0}$.}.

\begin{figure*}[h!t]
\begin{center} \includegraphics[width=0.32\textwidth]{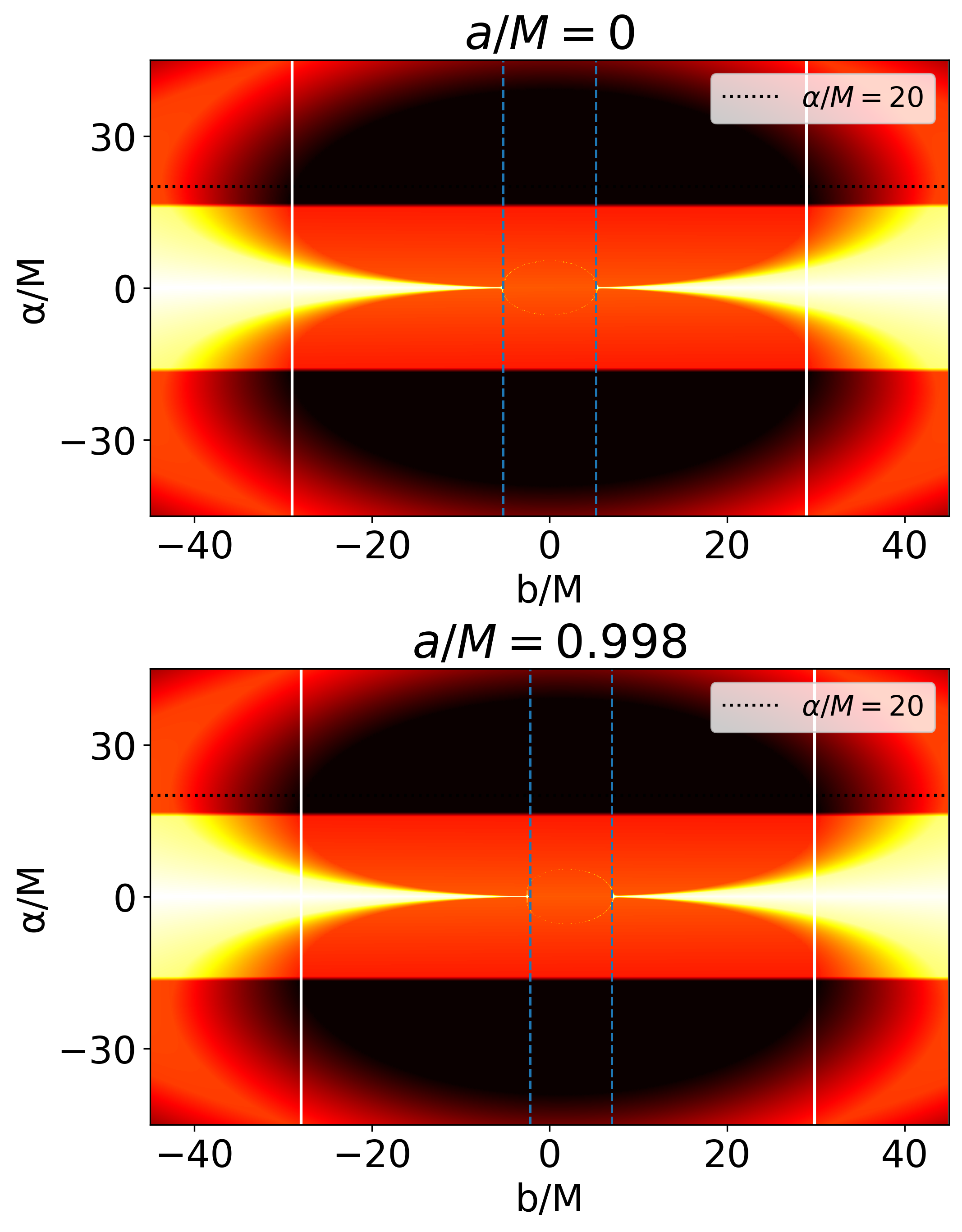}
\includegraphics[width=0.32\textwidth]{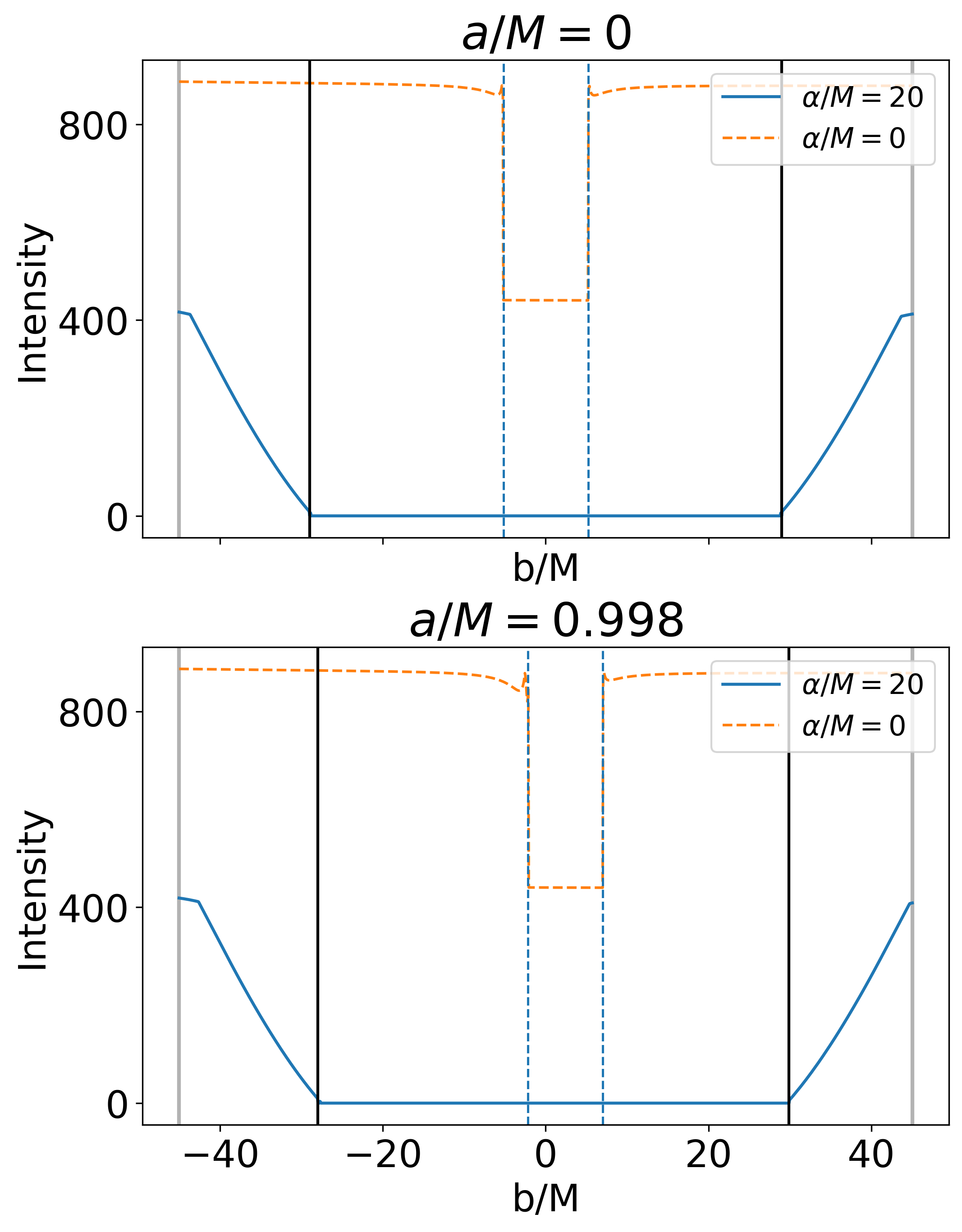}
\includegraphics[width=0.34\textwidth]{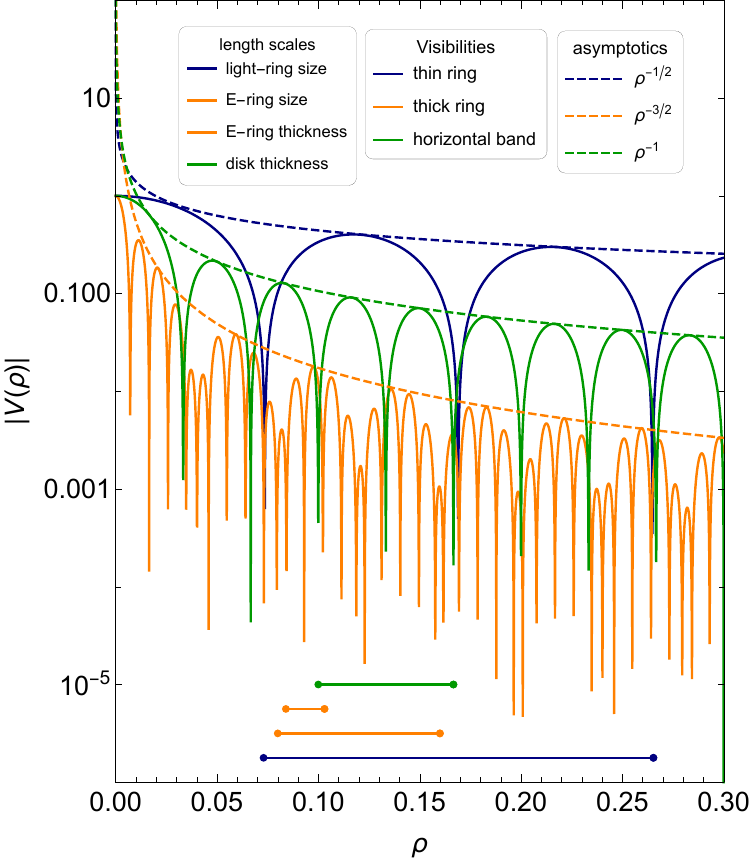}
 \caption{Comparison of the inner part of the Einstein ring and the light-ring in the image of the accretion disk around a non-rotating and a rotating SMBH at edge-on orientation with respect to the observer. The left image shows the intensity profiles on the image plane of the observer (top: non-rotating, bottom: rotating). The images show 3 characteristics of that region, i.e. the inner part of the Einstein ring, the light-ring, and a horizontal bright band, which is the near part of the disk that is located between the observer and the BH. The middle plots show cross-sections of the intensity images at impact parameters $\alpha=20M$ (solid) and $\alpha=0$ (dashed). The vertical lines in both cases indicate where the edges of the light-ring and the Einstein ring (for $\alpha=20M$) are located and are there to highlight the relative displacements of the rings. The plot on the right shows the visibilities for a thin ring, a thick ring (annulus), and a horizontal bright band of some vertical width much smaller than the horizontal (like the one in the intensity images). The plot also indicates the corresponding length scales in the $(u,v)$-plane. In the intensity images we note that the Einstein ring for the BH with spin $a=0.998M$ is displaced with respect to the non-rotating case by $\Delta b_E=0.96M$, while the light-ring is displaced by $\Delta b_{LR}=2.31M$, if we use the location of the half-width of the ring ($\left(b_{\textrm{left}}+b_{\textrm{right}}\right)/2$) as the center. The non-rotating rings have their centers at $(b,\alpha)=(0,0)$. These images have been produced with the emissivity profile given by eq. \eqref{eq:emis1}.}
 \label{fig:uv}
\end{center}
\end{figure*}

We can start examining now the signatures of the various structures of interest in the central region of the BH image on the measured $V(u,v)$. We  consider first a thin uniform ring of total unit intensity, i.e., 
\be 
I=\int_0^{2\pi}\int_0^{\infty}\frac{1}{2\pi r_0}\delta(r-r_0)rdrd\theta,
\ee
where $r_0$ is the radius of the thin ring in the image plane $(b,\alpha)$. 
For such a ring, the visibility would be given by, 
\bear
V(\rho,\phi)&=&\int_0^{2\pi}\int_0^{\infty}\frac{1}{2\pi r_0}\delta(r-r_0) e^{-2\pi i \rho r \cos(\theta-\phi)}rdrd\theta\nn\\
&=& J_0(2\pi r_0 \rho),
\eear 
where $J_m$ are the Bessel functions of the first kind. We can see that in the $(u,v)$-plane, the ring introduces features of scale $\rho=1/r_0$, i.e., inversely proportional to the radius of the ring, that are related to the periodicity of $V$ and its amplitude $A(u,v)=|V|$. This can be seen in \cref{fig:uv}. 

\begin{figure}[htbp]
\begin{center} \includegraphics[width=0.23\textwidth]{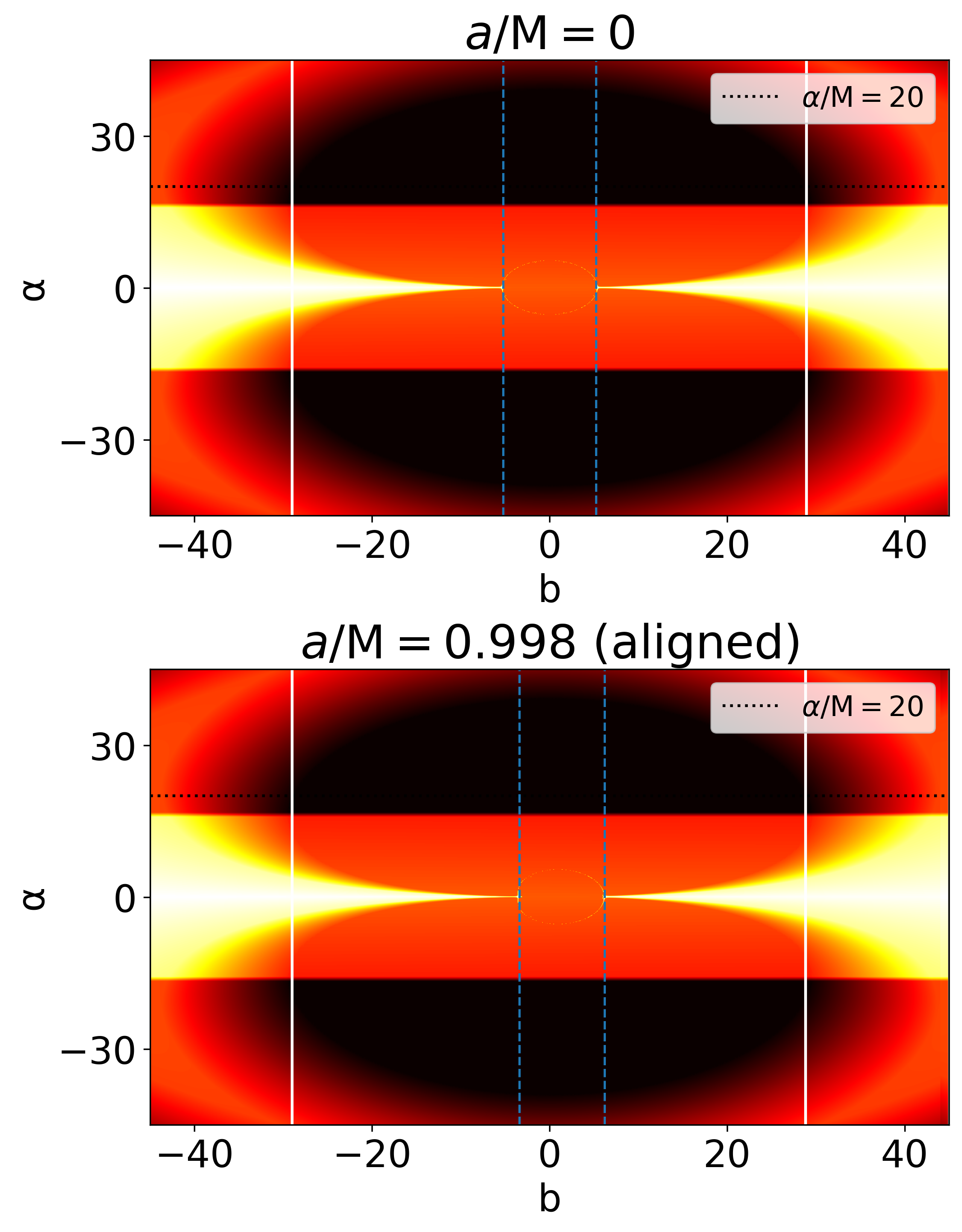}
\includegraphics[width=0.23\textwidth]{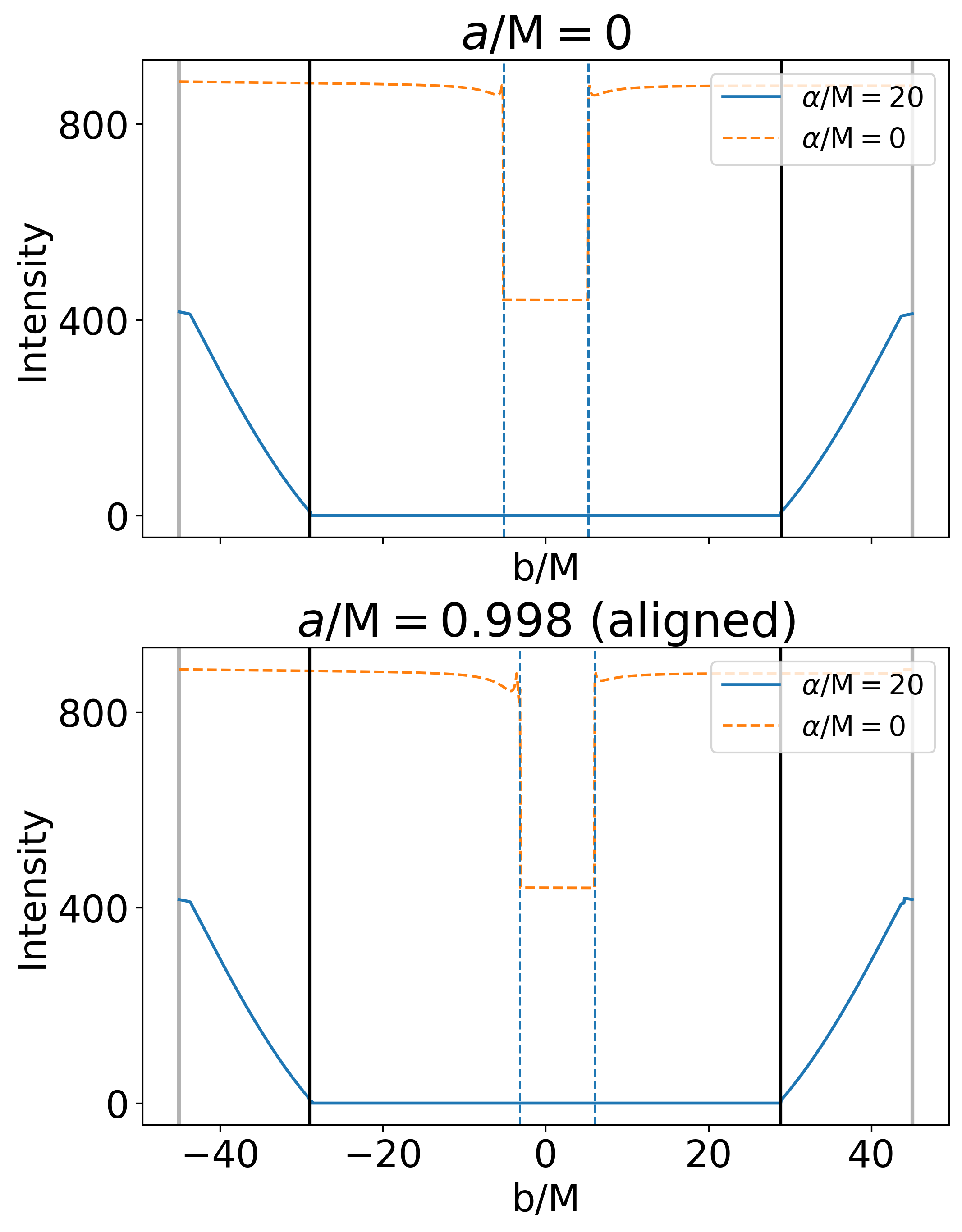}
 \caption{Aligned Einstein rings. Here, the light-ring shift is $\Delta b_{LR}=1.35M$.}
 \label{fig:uv2}
\end{center}
\end{figure}

 Let's now consider the images of two BLR disks, the one around a rotating BH and the other around a non-rotating BH (see \cref{fig:uv} and \cref{fig:uv2}).  The bright and symmetric Einstein ring emission in both cases can be used to set the image center (which is the phase center $\Phi =0$ in the $(u,v)$-plane) as to coincide with the SMBH center (this assumes an Einstein ring minimally displaced from the SMBH center, even in the case of  maximal spin which, as we will see, is the case). This takes care of any initial coordinate-center offsets\footnote{These are expected in any realistic interferometric imaging.}  by using solely the bright Einstein ring line emission geometry. After such a (0,0)-coordinate re-centering, the much smaller light-ring distribution of a non-rotating BH will be circular and centered on the (new) (0,0) image center, {\it but it will be offset and non-circular for a rotating one.} Models of the light-ring's shape and position can then robustly deliver the BH spin $a$, which, for a Kerr BH, controls both. Carrying out this analysis in the image plane is of course much more intuitive, yet it can be done entirely in the $(u,v)$-plane using the most reliable and most sensitive outputs of interferometric observations (especially when the $(u,v)$-plane is only sparsely covered), namely visibility phases.

 A $\Phi(u,v)$-based analysis of all the above in the $(u,v)$-plane will be the subject of future work. For now, we continue analyzing the effects of spin in the image plane by considering two intensity images of the shadow, one for zero and one for maximum ($a=0.998M$) rotation, each of them sampled on a uniform $(b,\alpha)$ grid. These images have been produced using the emissivity profile from eq. \eqref{eq:emis1}. At a chosen impact parameter of $\alpha=20M$ - slice, where the intensity drops sharply, and we are outside the horizontal bright band, we locate the first zeros in either direction in $b$ of each intensity profile and use the resulting $\Delta b$ offset to align the two Einstein rings, as we show in \cref{fig:uv2}. This alignment procedure, based on the intensity images, could be done in the $(u,v)$-plane directly, taking advantage of the symmetry and the geometry of the Einstein ring, circumventing the image altogether. We should note at this point some interesting features of the Einstein ring and the light-ring. We can see in \cref{fig:uv} that the Einstein ring in the non-rotating case is centered at the $(b,\alpha)=(0,0)$ point on the image plane. However, for the Kerr BH with $a=0.998M$, the center of the Einstein ring is slightly shifted by $\Delta b_E=0.96M$, where we use the location of the half-width of the ring $\left(b_{\textrm{left}}+b_{\textrm{right}}\right)/2$ to define the center. Performing the same procedure for the light-rings, while the light-ring in the non-rotating case is centered at $(0,0)$, the rotating BH light-ring is shifted by $\Delta b_{LR}=2.31M$. This is the shift of the light-ring before the alignment, after the alignment, it becomes $\Delta b_{LR}=1.35M$. These results are in agreement with \cite{Johannsen_2010,Johannsen:2013vgc}  

 While the thin ring approximation can be considered quite accurate for the light-ring, the Einstein ring, on the other hand, is geometrically relatively thick. So, we will consider a thick uniform ring with inner radius $r_{in}$ and outer radius $r_{out}$ of total unit intensity 
\be 
I=\int_0^{2\pi}\int_{r_{in}}^{r_{out}}\frac{1}{\pi \left(r_{out}^2-r_{in}^2\right)}rdrd\theta.
\ee
The visibility for such a ring will be
 \begin{equation}
     V_{\text{ann}}(\rho,\phi)=\frac{r_{out}J_1(2\pi\rho r_{out})-r_{in}J_1(2\pi\rho r_{in})}{\pi \left(r_{out}^2-r_{in}^2\right)\rho},
 \end{equation}
the amplitude of which can be seen in \cref{fig:uv}. This visibility reduces to the thin ring visibility in the limit where $r_{in}\rightarrow r_{out}$. While in the case of the thin ring, we had one characteristic scale, the annulus has two such scales. The first scale is related to the average radius of the ring, $(r_{out}+r_{in})/2$ and introduces a short-scale periodicity in the visibility, while the other scale is related to $(r_{out}-r_{in})/2$ (essentially the annulus thickness) that introduces a longer periodicity in the visibility. These can be seen in \cref{fig:uv} as a high frequency periodicity of $|V(\rho)|$ (Einstein ring size) and a shorter frequency modulation of $|V(\rho)|$
(Einstein ring thickness). 

Another significant feature in the intensity images is the bright band passing in front of the light-ring with a large horizontal size and a relatively smaller vertical width, which is due to the emission of the part of the disk that is between the observer and the BH. Assuming a uniformly emitting band of width $2b_{band}$ and height $2H_{disk}$, and of unit intensity,
\be
I=\int_{-b_{band}}^{+b_{band}}\int_{-H_{disk}}^{+H_{disk}}\frac{1}{4H_{disk}b_{band}}dbd\alpha,
\ee
the visibility function for the band will be 
\bear 
V(u,v)&=&\int_{-b_{band}}^{+b_{band}}\int_{-H_{disk}}^{+H_{disk}}\frac{e^{-2\pi i \left(ub+v\alpha\right)}}{4H_{disk}b_{band}}dbd\alpha\nn\\
&=&\frac{\sin\left(2\pi u b_{band}\right) \sin\left(2\pi v H_{disk}\right)}{4\pi^2 u v H_{disk}b_{band}},
\eear
which is similar to the form that the diffraction pattern of a thin slit has. For the band, the visibility has a periodicity in the $v$ direction that is related to the length scale of $H_{disk}$ and is essentially $1/H_{disk}$. This can be seen in \cref{fig:uv}, where we have plotted $V(u=0,v)=\frac{ \sin\left(2\pi v H_{disk}\right)}{2\pi v H_{disk}}$, with $v\rightarrow\rho$. 

The symmetric emission features considered so far yield only real visibilities with periodicities that depend on the different characteristic scales. One last characteristic of interest is the displacement of the ring, which can be either present for the light-ring or, to a significantly lesser extent, for the Einstein ring. In general, a displacement by $\Delta b$ is equivalent to a phase shift to the visibility in the $(u,v)$-plane of the form $e^{-2\pi i \Delta b u}$ \cite{Johnson:2019ljv}.  One can repeat the calculation for the thin ring, assuming that the ring is displaced by some small $\Delta b\ll r_0$. In this case, the ring will be given by an equation of the form, 
$r_{LR}(r,\theta)=r_0+\Delta b \cos\theta$. For such a ring, the visibility is:
\be
V(\rho,\phi) = J_0(2\pi r_0 \rho)\left(1-2i\pi\Delta b \rho \cos\phi\right),
\ee
where $\rho\cos\phi=u$, with the displacement of the ring due to the rotation of the BH  essentially encoded in
the visibility phase: $\Delta \Phi=-2\pi \Delta b \rho \cos\phi $. 

Such phase shifts can be very effectively modeled in VLBI and future space-VLBI $(u,v)$-data \cite{Johnson:2019ljv,Gralla:2020nwp,Cardenas-Avendano:2023dzo,Gralla:2020srx,Zineb:2024gwx,Fromm:2021flr,Gurvits:2019ioq}, where relative phases rather than amplitudes would be those most sensitively measured against noise, and remain unaffected by any absolute phase uncertainties.

However we must remember that it is the BLR disk now serving as a SMBH spectral line illuminator rather than any radio continuum  from the inner accretion disk. Thus, any space-borne interferometers of the future {\it must operate at the near and far IR rather than radio wavelengths in order to detect the effects examined here} (\cite{Astro_paper}).

\section{More tests of the Kerr metric using the light-ring properties}
\label{sec:testsKerr}
%

We will now turn our attention to the light-ring structure and morphology, and how it can provide information on the Kerr or the non-Kerr nature of the spacetime near SMBHs. As we have mentioned, the light-rings and their properties have been extensively studied. In part these investigations have used light-rings to probe possible deviations from the Kerr solution and from GR \cite{Johannsen_2010,Johannsen:2011dh,Johannsen:2015mdd,Johannsen:2013vgc,Broderick:2013rlq,Medeiros:2019cde,Gralla:2020srx,Kostaros_2022,Glampedakis:2021oie,Glampedakis:2023eek,Younsi:2021dxe,Kostaros:2024vbn}. These deviations have been explored in terms of the size and shape of the light-ring in the emission plane, but also in terms of the visibilities in the $(u,v)$-plane \cite{Gralla:2020srx,Johnson:2019ljv}. We can now consider a different probe of the light-ring, its corresponding spectrum, since we have the benefit of a BH illuminated by spectral line rather than continuum emission from the accretion disk.

For our investigation, we will use Kerr BHs rotating at different  rates, while for non-Kerr BHs we will use as a toy model the JP metric with the same spacetime setup as the one used in \cite{Kostaros:2024vbn}, where the fractal structure of the shadows of these BH spacetimes was investigated. The deviations from the Kerr metric in our JP setup are parametrized by the parameter $\epsilon_3$, which we set to $\epsilon_3=1$, and corresponds to a prolate deformation of Kerr.

\begin{figure}[htbp]
\begin{center}
   \includegraphics[width=0.24\textwidth]{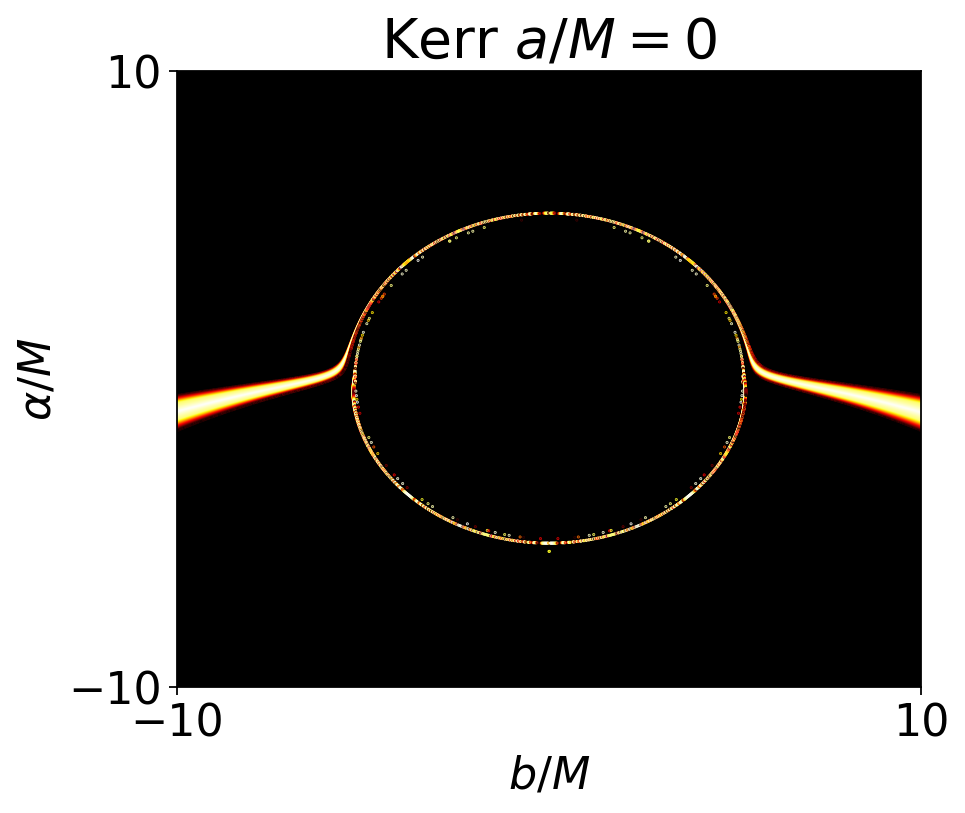}\includegraphics[width=0.24\textwidth]{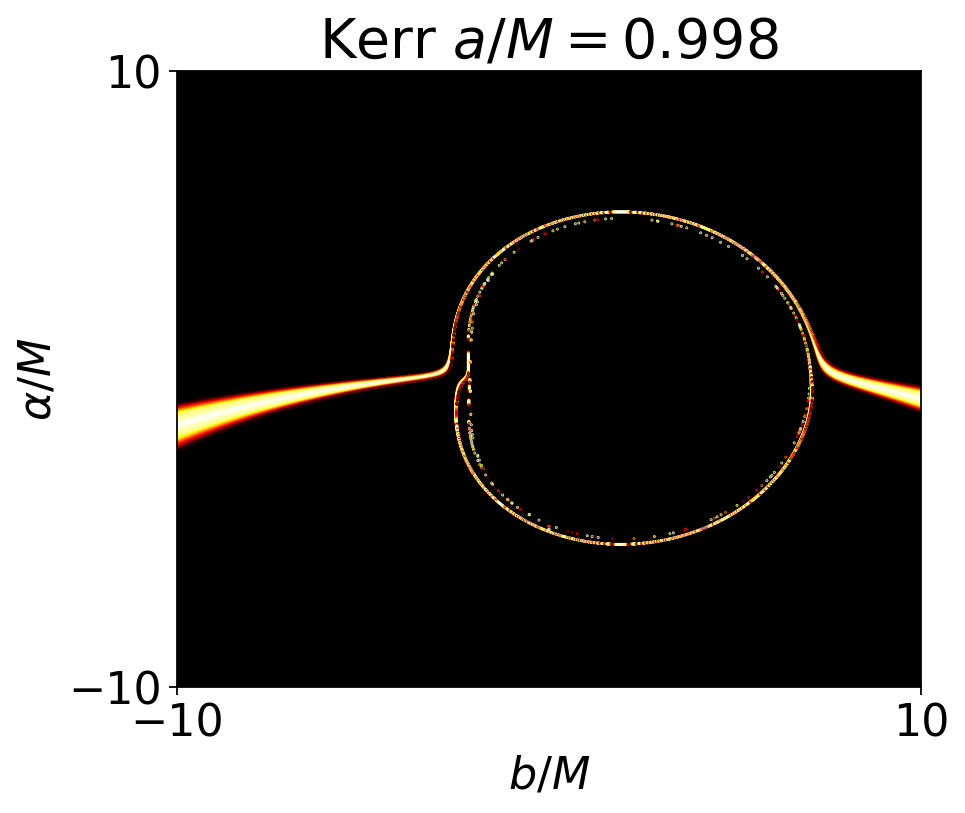}   \includegraphics[width=0.24\textwidth]{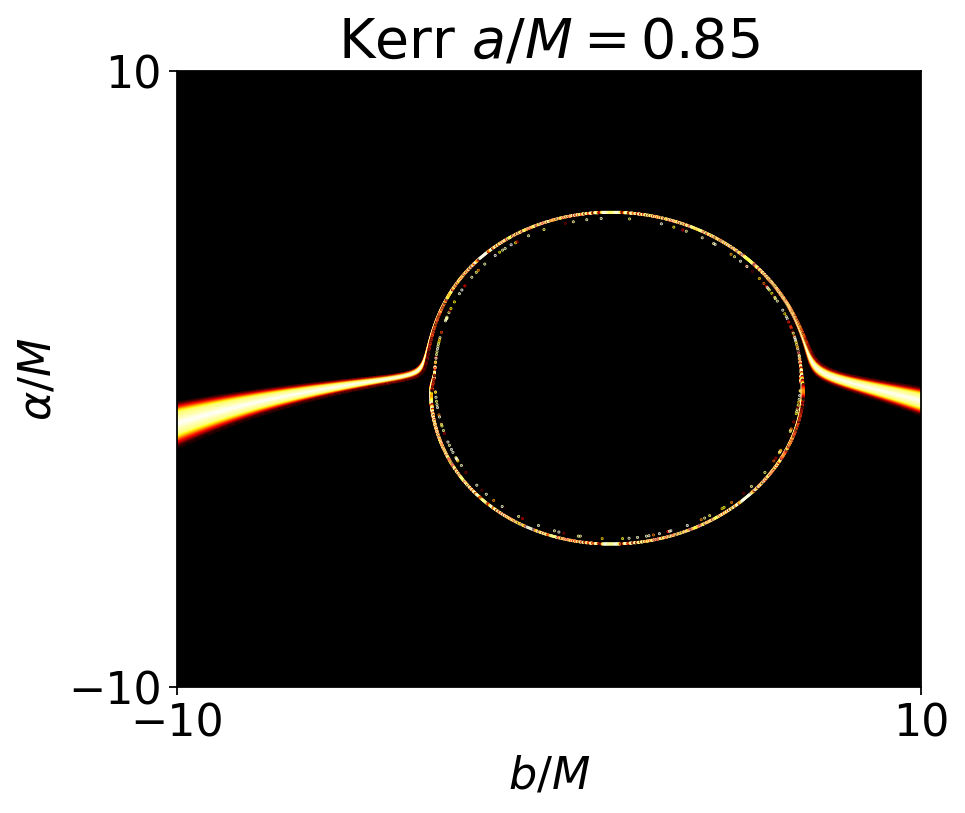}\includegraphics[width=0.24\textwidth]{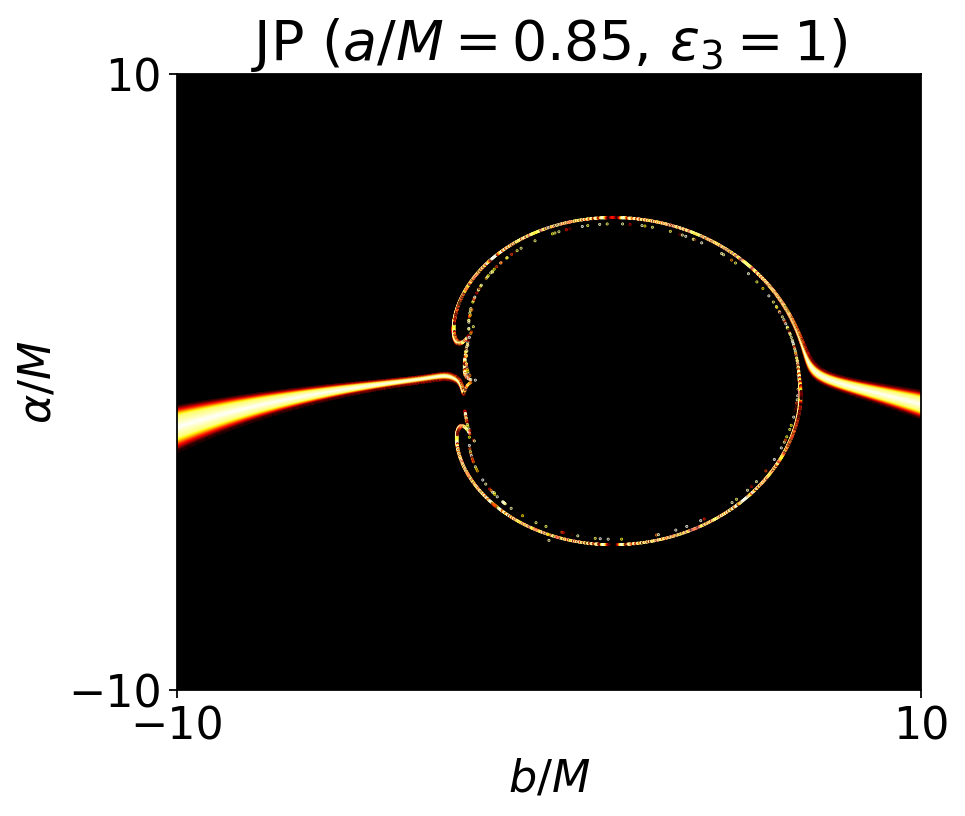}
 \caption{Zoom in image of the light-ring regions for three Kerr BHs with spins $a=0, 0.85M,$ and $0.998M$, and a JP BH with spin $a=0.85M$ and deformation $\epsilon_3=1$. The observer is at $\theta_0=86^{\circ}$ from the axis of symmetry. The image is truncated at $b=\pm10M$ and $\alpha=\pm10M$. The figures have been produced using the emissivity profile given by eq. \eqref{eq:emis2}.}
 \label{fig:nonKerr}
\end{center}
\end{figure}

At this point, we need to establish a strategy for selecting the spectral information of the light-ring only. 
To do this, we start by assuming the observer to be at such an angle with respect to the axis of symmetry of the BH and the accretion disk, so that the disk does not obscure the line of sight of the observer. We start with this angle being $\theta_0=86^{\circ}$. Fig. \ref{fig:nonKerr} shows the images of the light-rings for a Schwarzschild BH, a Kerr BH rotating with a rotation rate $a=0.85M$, a Kerr BH rotating with a rotation rate $a=0.998M$, and a JP BH with $a=0.85M$ and $\epsilon_3=1$. In order to focus on the light-ring, we have truncated the image at $b=\pm10M$ and $\alpha=\pm10M$, a region that contains the light-ring, but with  a part of the strongly lensed image of the accretion disk missing. This is the image of the far side of the disk, which at edge-on inclination becomes the lower part of the Einstein ring (see \cref{fig:E_image}). Spatially isolating the light-ring will not affect our discussion on its properties with respect to its shape.
%
%
There is also the technical issue of how one isolates the region of the light-ring from the image of the entire disk. The unique ability of cross-correlation interferometric imaging versus filled-aperture (i.e. single telescope) total power imaging lies in the ability of the former (but not the latter) to chose a special $(u,v)$-coverage that completely filters out extended emission regions, even if as bright as the direct spectral line radiation from a BLR gas disk, while retaining their high spatial-frequency (i.e. small structure) sensitivity. Such $(u,v)$-tailored observations of SMBHs, illuminated by the IR lines of the BLR disk (especially its neutral phase BLR$^{0}$), using space-borne interferometers may one day be able to select only the high spatial frequency emission as that in Fig. \ref{fig:nonKerr}.

The shape of the rotating Kerr BHs light-ring, as we can see in \cref{fig:nonKerr}, is quite similar to the shape of the non-rotating BH (even though it is a little shifted to the right), while the shape of the JP black hole has the very distinctive eyebrow-like features (discussed in \cite{Kostaros:2024vbn}) that clearly deviate from circularity. One could say that it would be relatively hard to distinguish between the Kerr BHs with different rotations only by the shape of the light-ring. Contrary to this, the JP light-ring is easily distinguishable by its shape, if it can be resolved. In addition, having an independent measurement of the mass of the BH from the Einstein ring, allows for possible tests of the non-Kerr nature of the spacetime by measuring the diameter of the light-ring, since for some non-Kerr BHs, its size depends also on the deviations from Kerr \cite{Johannsen:2013vgc}.
There is one last thing to note regarding the shape of the light-ring and observations. Deviations of the ring geometry for the JP spacetime, where the ring could have some gaps or some smaller scale features ($''$eyebrows$''$), may produce characteristic signatures  in the $(u,v)$-plane  that will be worth exploring since {\it especially in interferometric phase signals any strong $''$break$''$ from emission region symmetry can be very sensitively registered.}

\begin{figure}[htbp]
\begin{center}
   \includegraphics[width=0.47\textwidth]{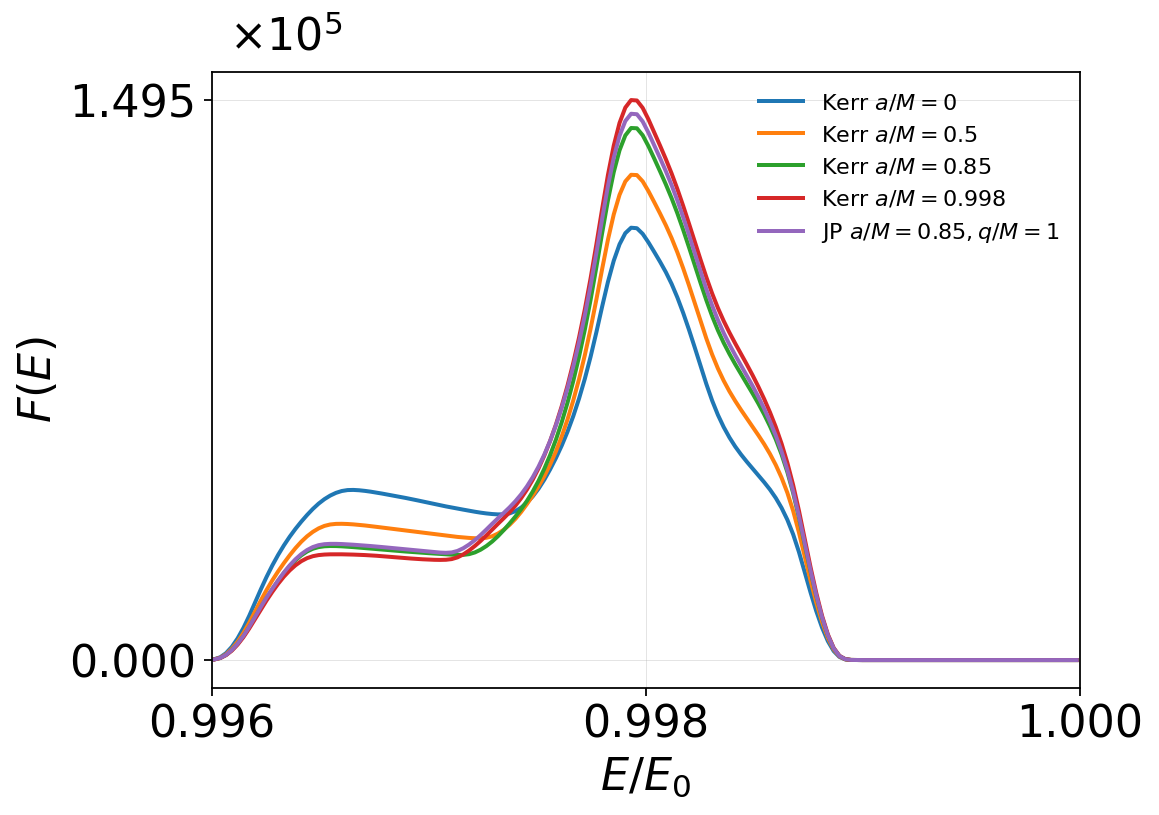} \caption{The spectra coming from the light-ring regions of \cref{fig:nonKerr}. The differences observed between the different BHs turn out to be an artifact of the artificial exclusion of parts of the disk emission.}
 \label{fig:nonKerr2}
\end{center}
\end{figure}

The next subject of our investigation is the spectrum of the light-ring.
Fig. \ref{fig:nonKerr2} shows the various spectra for the different light-rings. The figure shows the spectra for the Kerr BHs and the JP BH, with the corresponding parameters as the ones we have in \cref{fig:nonKerr}. Also, the spectra contain the light coming only from the selected region in the image plane (see \cref{fig:nonKerr}).  
%
%
A first thing to note is that the spectrum coming from the ring is relatively narrow and with some small gravitational redshift. This is due to the fact that the radiation the light-ring redirects towards the observer has been emitted mostly perpendicular to the direction of the Keplerian gas velocity field (i.e., radially) and hence suffers little Doppler effect, reflecting mostly the turbulent linewidth, while the gravitational redshift is due to the strength of the gravitational field at the position of emission, which is small for the BLR disk.

\begin{figure}[htbp]
\begin{center}
   \includegraphics[width=0.24\textwidth]{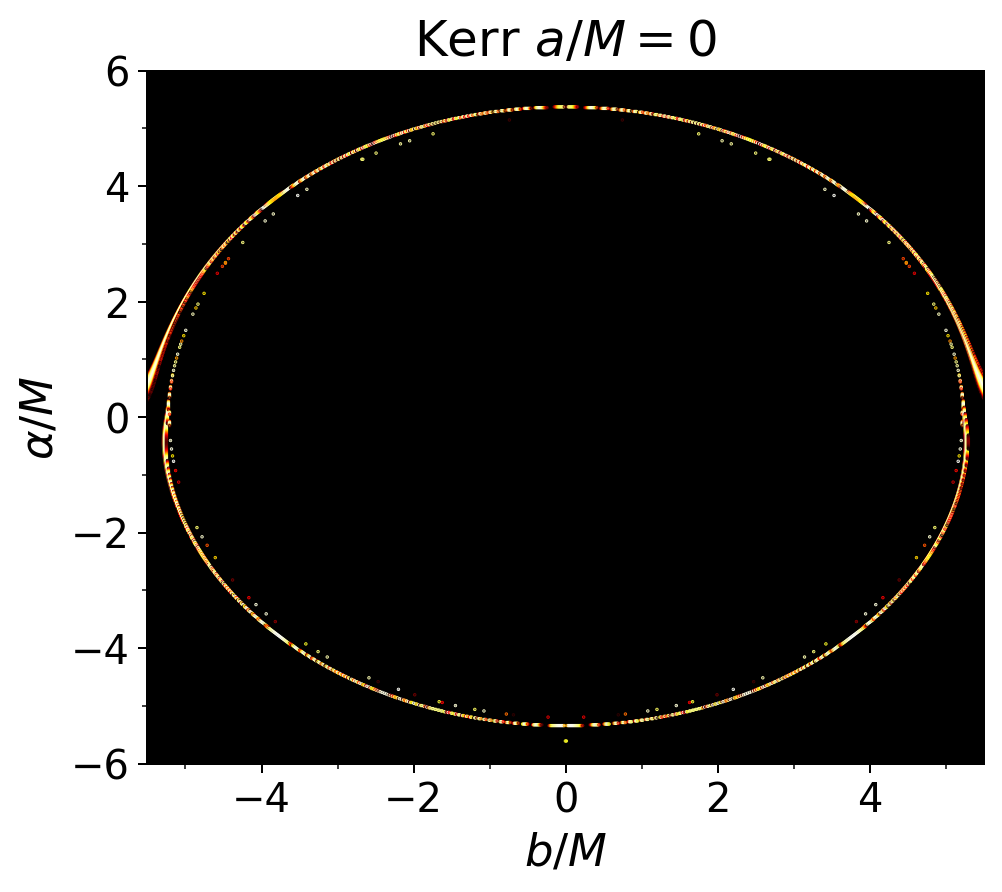}\includegraphics[width=0.24\textwidth]{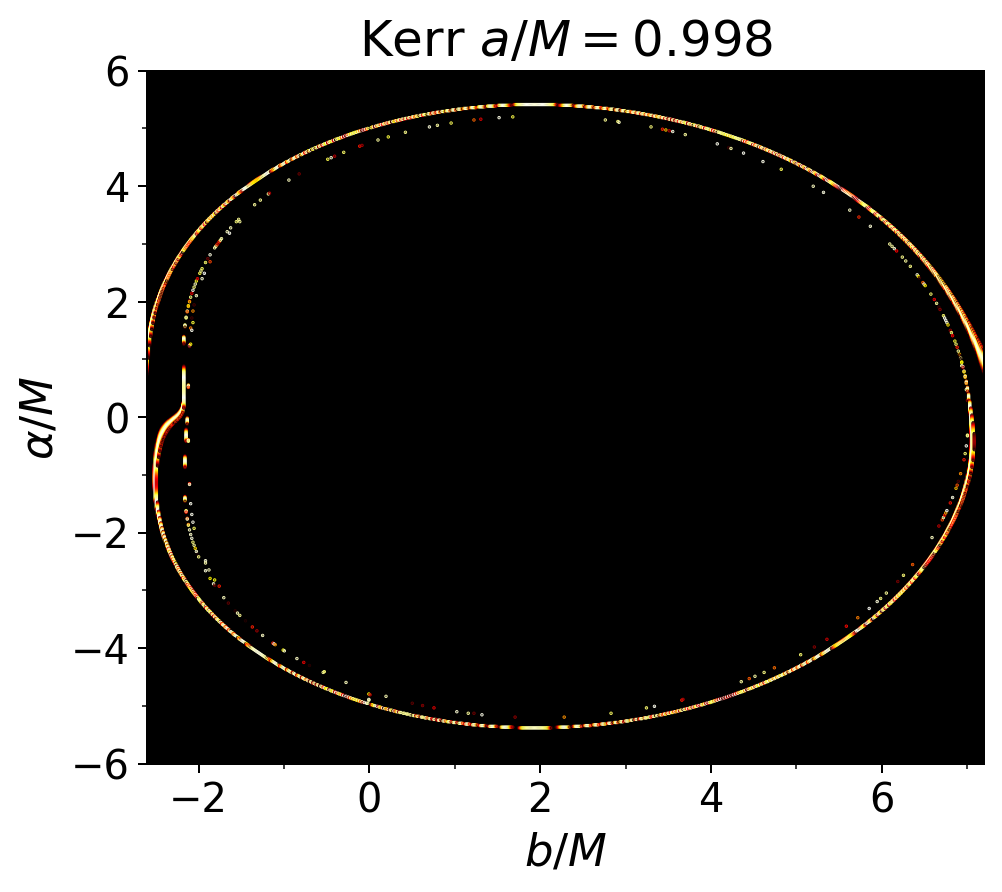}   \includegraphics[width=0.24\textwidth]{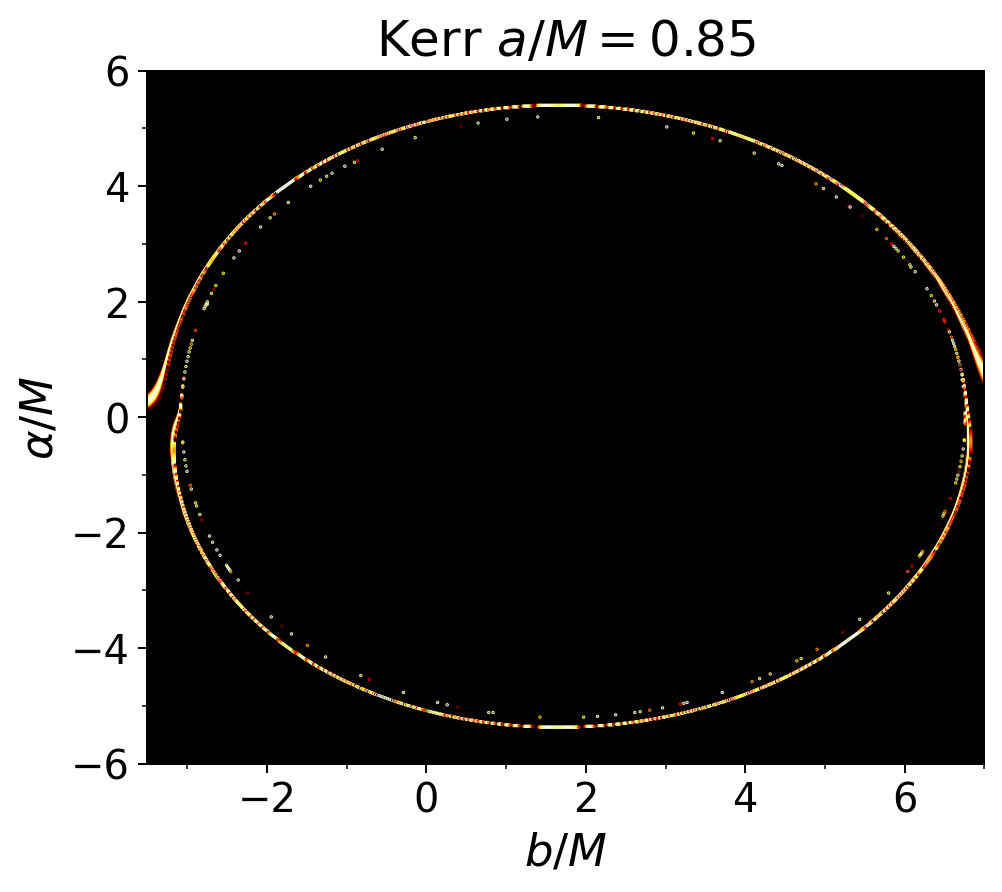}\includegraphics[width=0.24\textwidth]{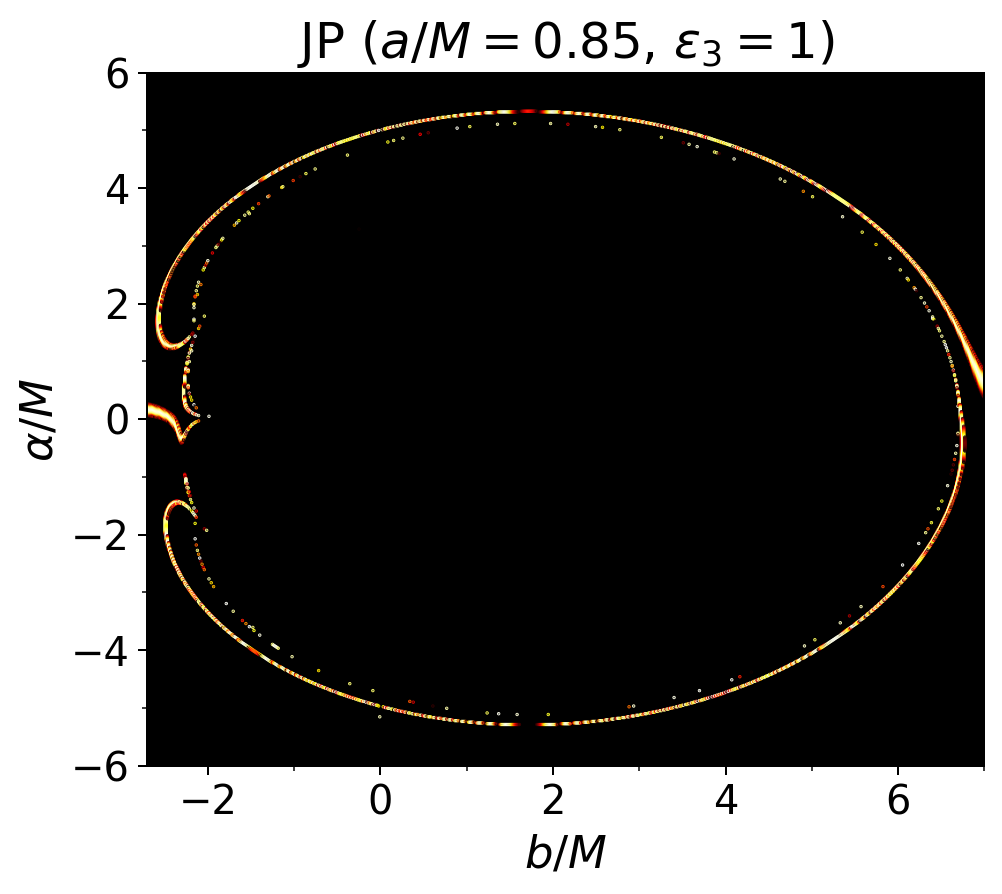}
 \caption{Zoom in image of the light-rings for the same BHs. Here, the image is truncated closer to the light-ring. The figures have been produced using the emissivity profile given by eq. \eqref{eq:emis2}.}
 \label{fig:nonKerr-2}
\end{center}
\end{figure}

There remains however an open issue regarding our image truncation strategy. The truncation was arbitrary, which means that an arbitrary part of the disk is included in the spectrum, while an equally arbitrary part is being left out. It is not a priori clear how this would affect the information in the spectrum. Fig. \ref{fig:nonKerr2} shows that the spectra from all the BHs are quite similar, with minor differences, but one cannot be certain that the differences or the similarity are not due to the truncation. Additional tests with different truncations of the image, assuming narrower regions even closer to the light-ring, but still with arbitrary cuts of the disk, such as the ones seen in \cref{fig:nonKerr-2}, resulted in arbitrary and inconsistent changes in the spectra. For this reason, we concluded that the best strategy is to have no cuts that leave out part of the disk emission distribution.    

\begin{figure}[htbp]
\begin{center}
   \includegraphics[width=0.24\textwidth]{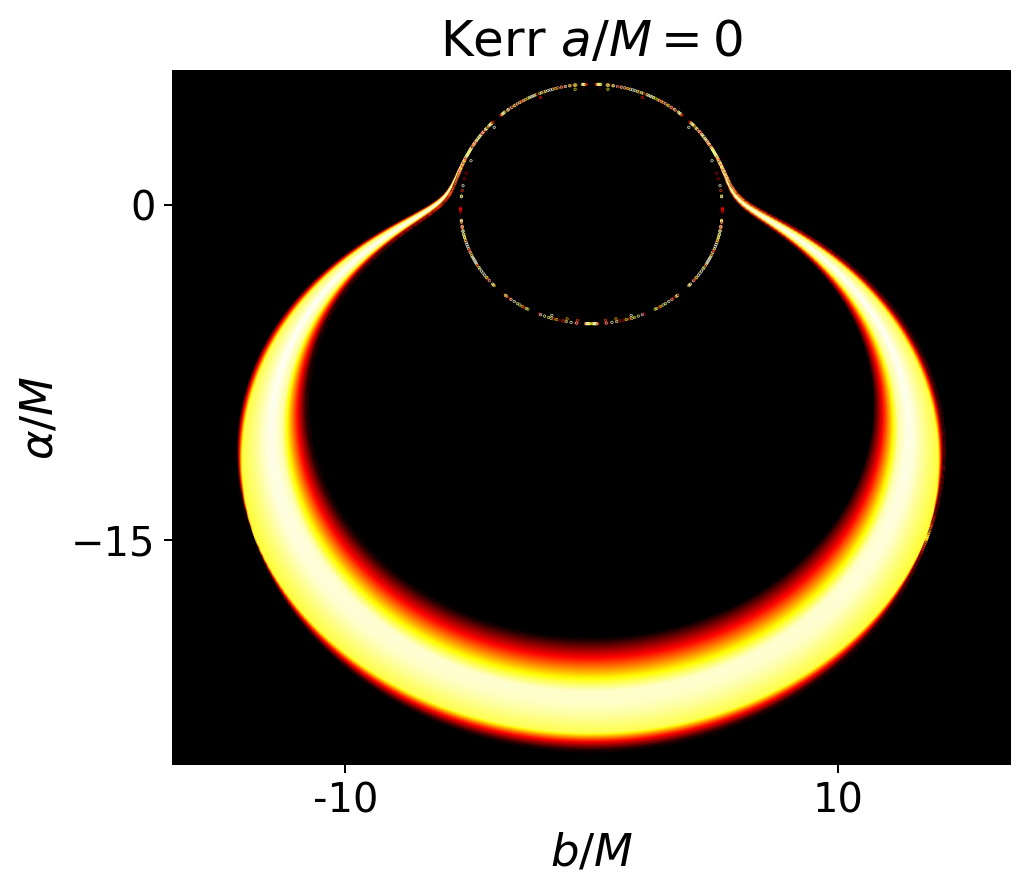}\includegraphics[width=0.24\textwidth]{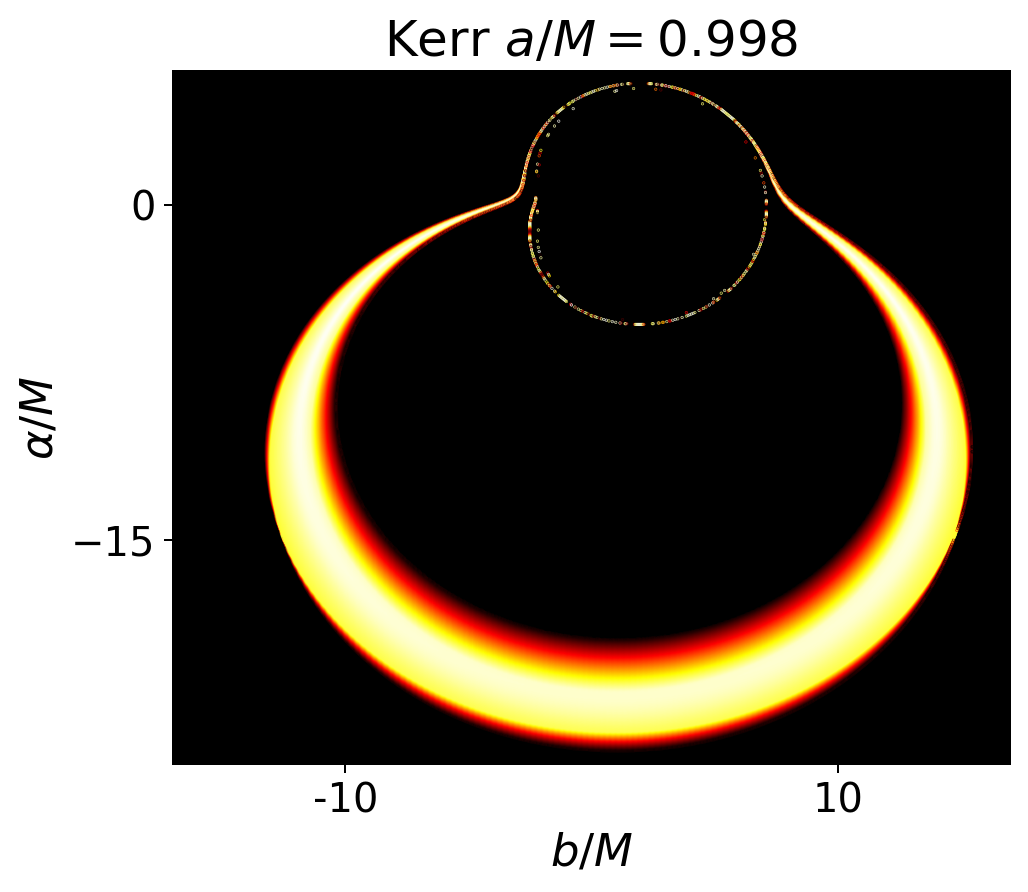}   \includegraphics[width=0.24\textwidth]{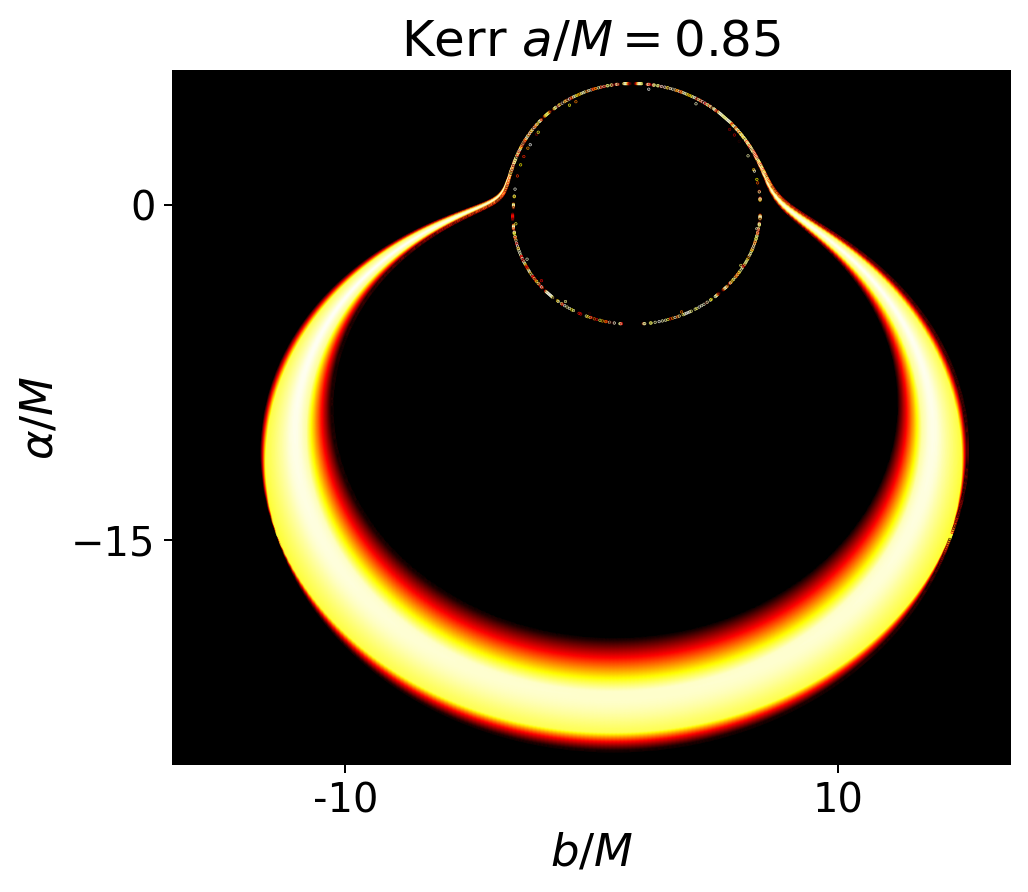}\includegraphics[width=0.24\textwidth]{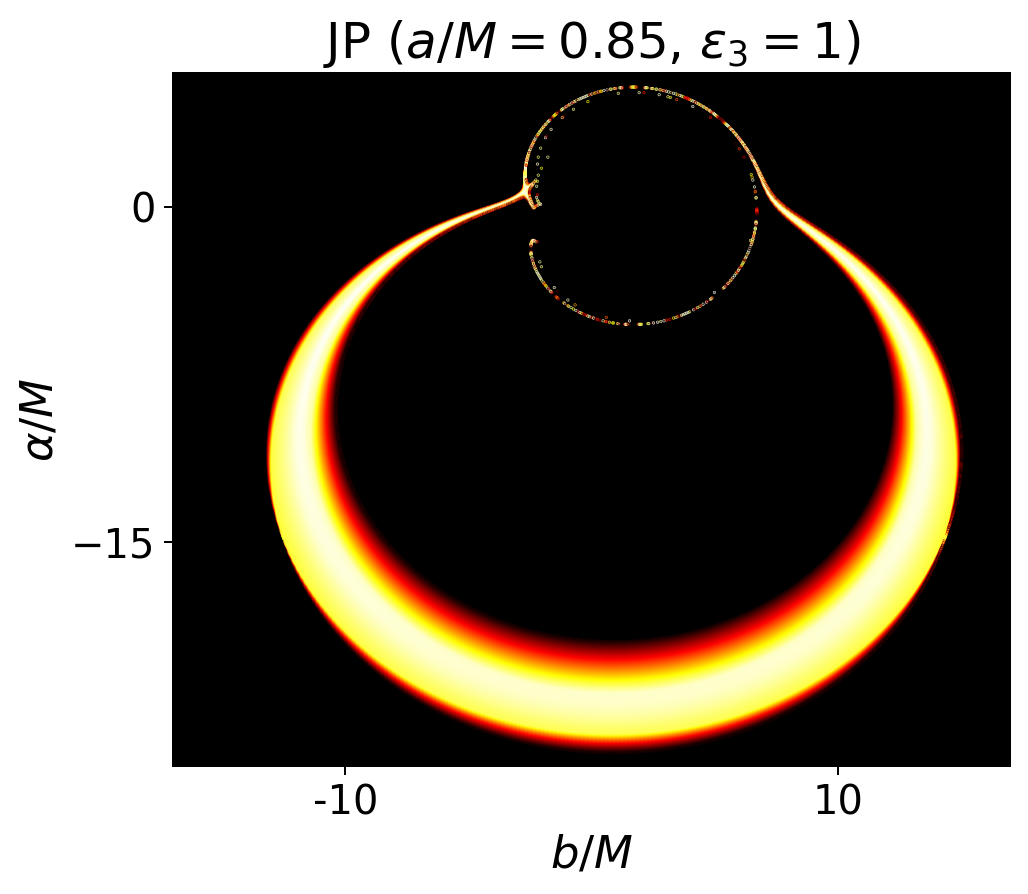}
 \caption{Zoom in image of the light-rings for the same BHs. Here, the image includes the entire disk without any parts being cut. The position of the observer in this case is at $\theta_0=80^{\circ}$ from the axis of symmetry. The figures have been produced using the emissivity profile given by eq. \eqref{eq:emis2}.}
 \label{fig:nonKerr-3}
\end{center}
\end{figure}

\begin{figure}[htbp]
\begin{center}
   \includegraphics[width=0.47\textwidth]{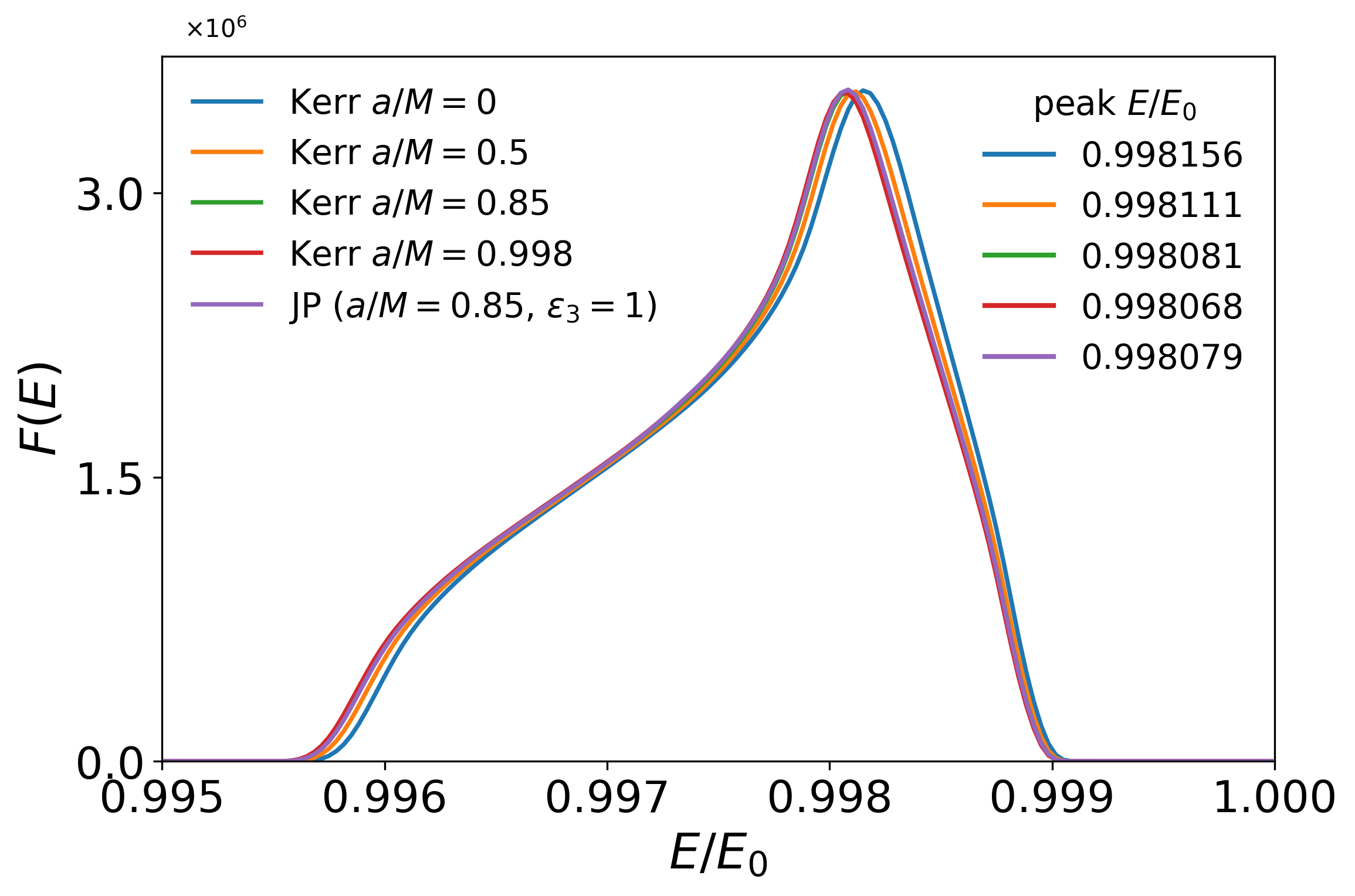} \caption{The spectra coming from the light-ring regions  of \cref{fig:nonKerr-3}. The spectra are very similar, with only a slight drift of the peak as the spin increases.}
 \label{fig:nonKerr2-2}
\end{center}
\end{figure}

To implement this, we have assumed the same BHs as before, but with the observer at $\theta_0=80^{\circ}$, so that no part of the image is affected by the direct emission of the disk between the observer and the BH. The images used for the calculation of the spectra are the ones shown in \cref{fig:nonKerr-3}, and the resulting spectra are given in \cref{fig:nonKerr2-2}, where one can immediately see the increase in the flux from the inclusion of the rest of the disk image.

Fig. \ref{fig:nonKerr2-2} shows a very small variation of the spectrum between the different BHs, with all the previous larger differences essentially eliminated. 
Returning thus to the issue of distinguishing the various BHs using the light-ring spectra, if we want to have a consistent description without introducing artificial distinctions, telling the different spacetimes apart will be challenging. At higher energies the spectra have a narrowing peak that stands out (around $E/E_0\approx 0.998$). One can see that as the rotation increases, the peak has a small drift towards lower $E/E_0$, but the effect is very small and impractical to use for obtaining the BH spin. This is especially so since the spectral lines from the actual BLRs will contain spectral features deviating from those produced by perfectly Keplerian disks with a uniform brightness and constant local turbulence. Such features will also $''$illuminate$''$ the central BH, and be superimposed on the spectra seen in Fig. \ref{fig:nonKerr2-2}, practically $''$burying$''$ the BH spin signal. Thus, it seems that only the purely geometric method described in Section \ref{sec:light-ring} remains practical for BH spin estimates, at least in this context.

\begin{figure}[htbp]
\begin{center}
   \includegraphics[width=0.24\textwidth]{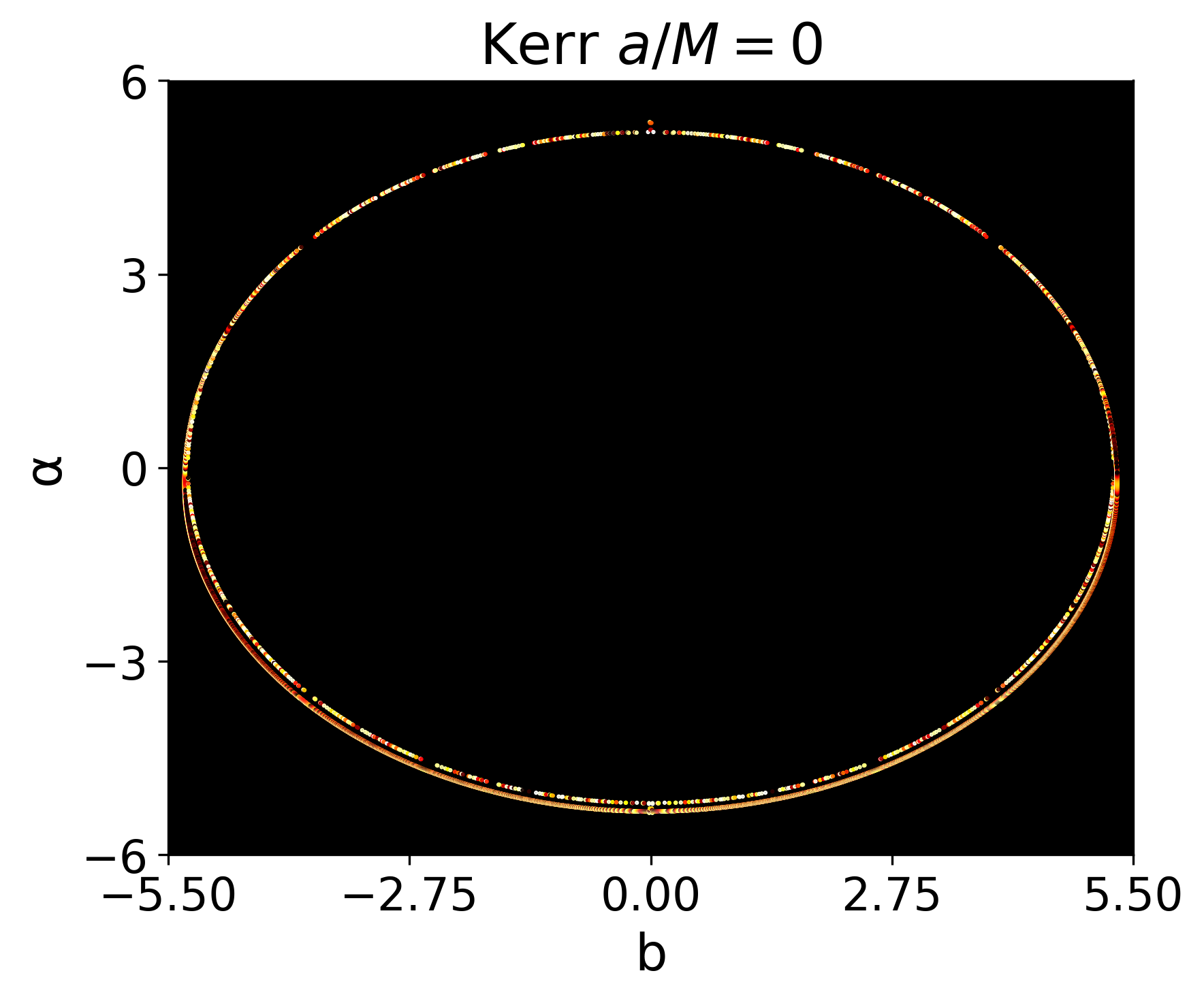}\includegraphics[width=0.24\textwidth]{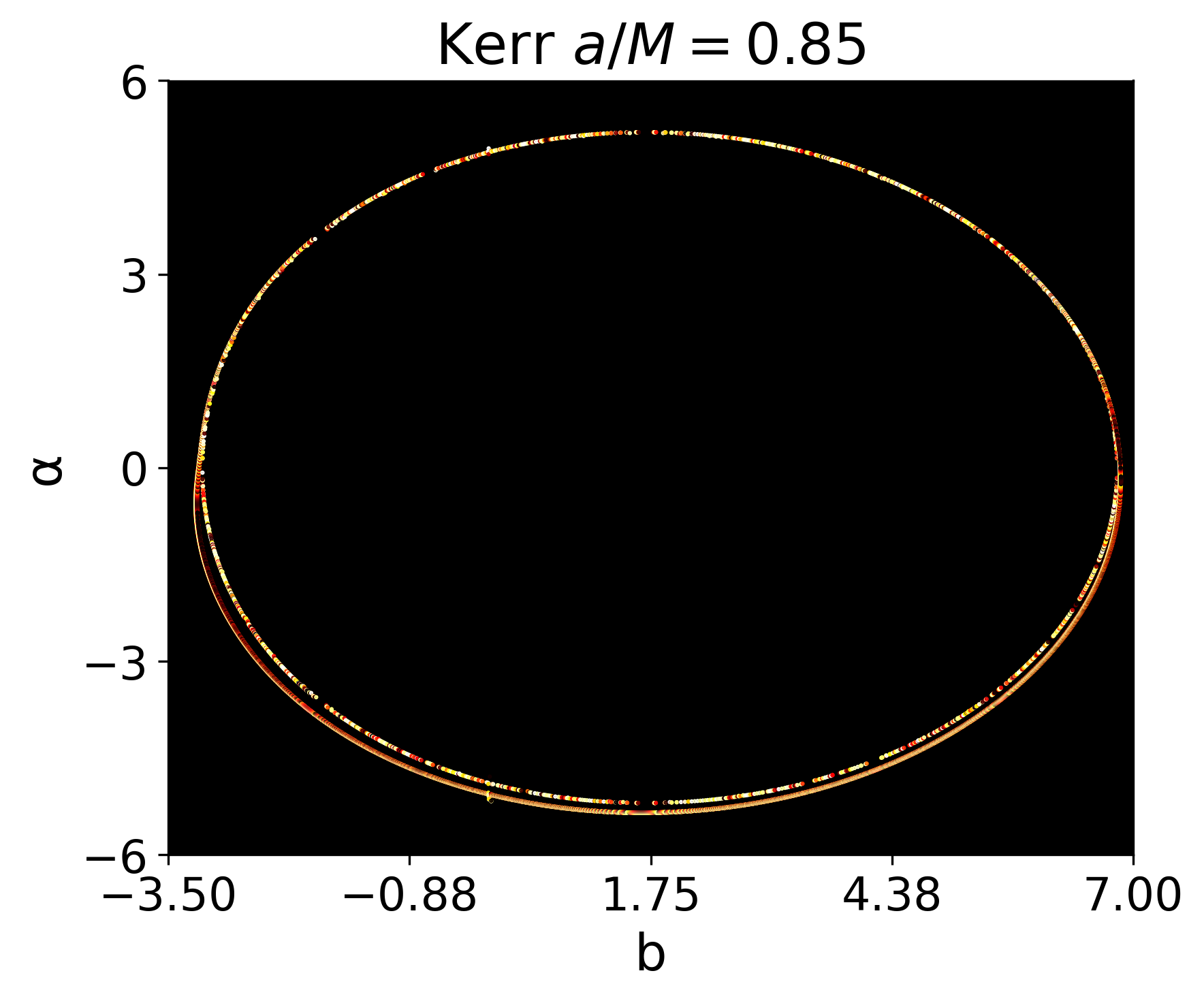}   \includegraphics[width=0.24\textwidth]{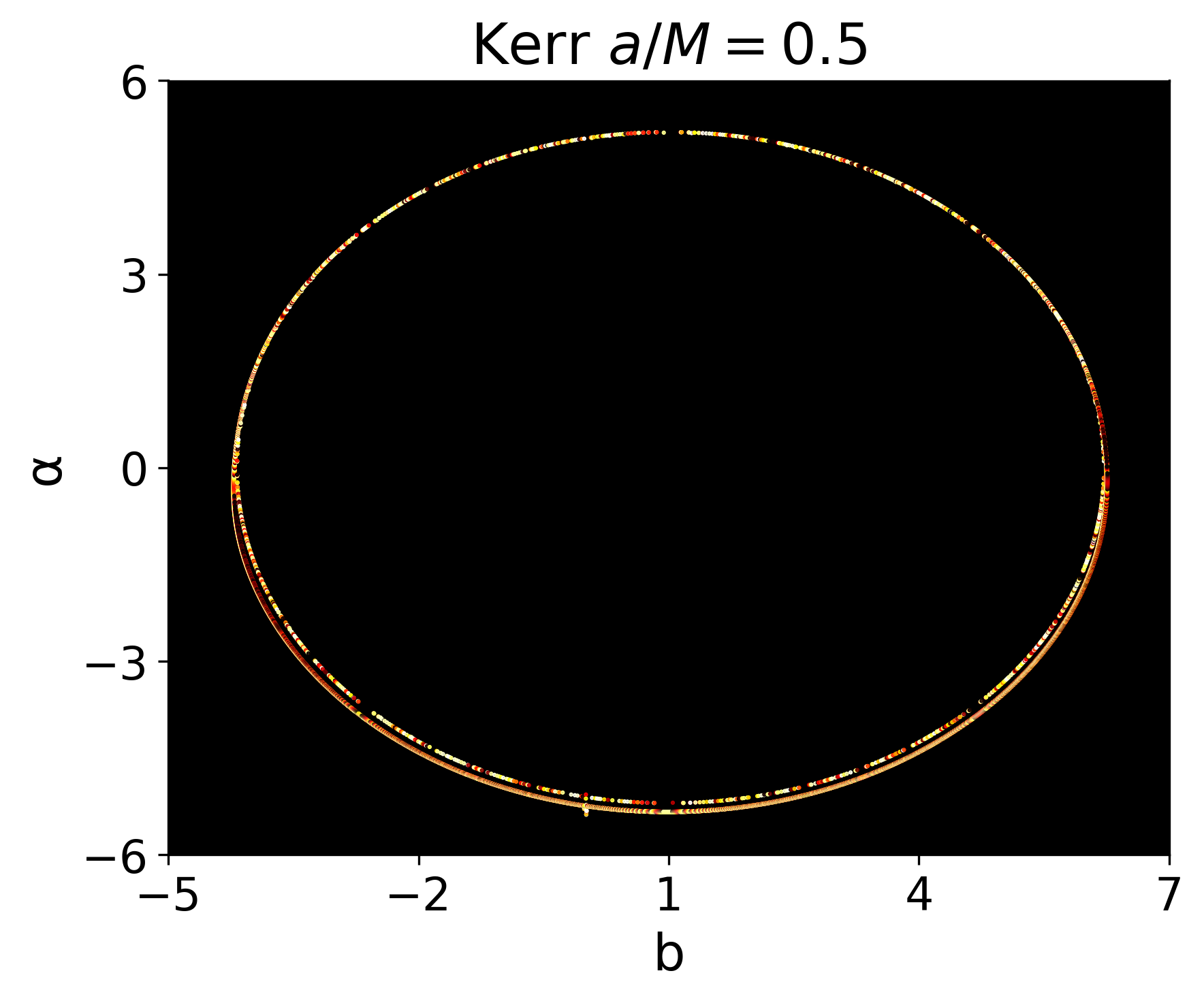}\includegraphics[width=0.24\textwidth]{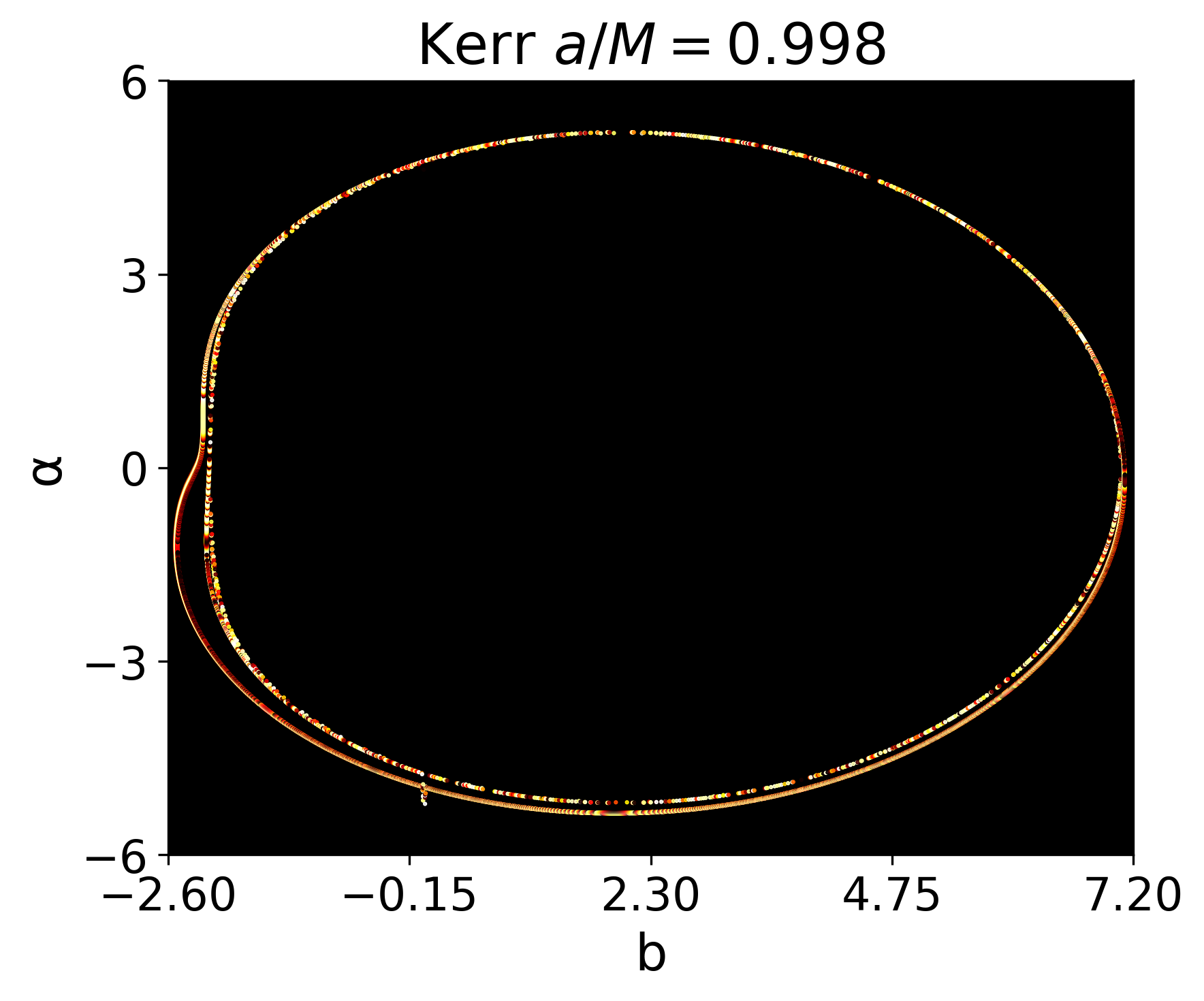}
 \caption{Zoom in image of the light-rings for Kerr BHs. Here, the image includes only a higher order image of the entire disk without any parts being cut. The position of the observer in this case is at $\theta_0=80^{\circ}$ from the axis of symmetry. The figures have been produced using the emissivity given by eq. \eqref{eq:emis2}.}
 \label{fig:nonKerr-3-2}
\end{center}
\end{figure}

\begin{figure}[htbp]
\begin{center}
   \includegraphics[width=0.47\textwidth]{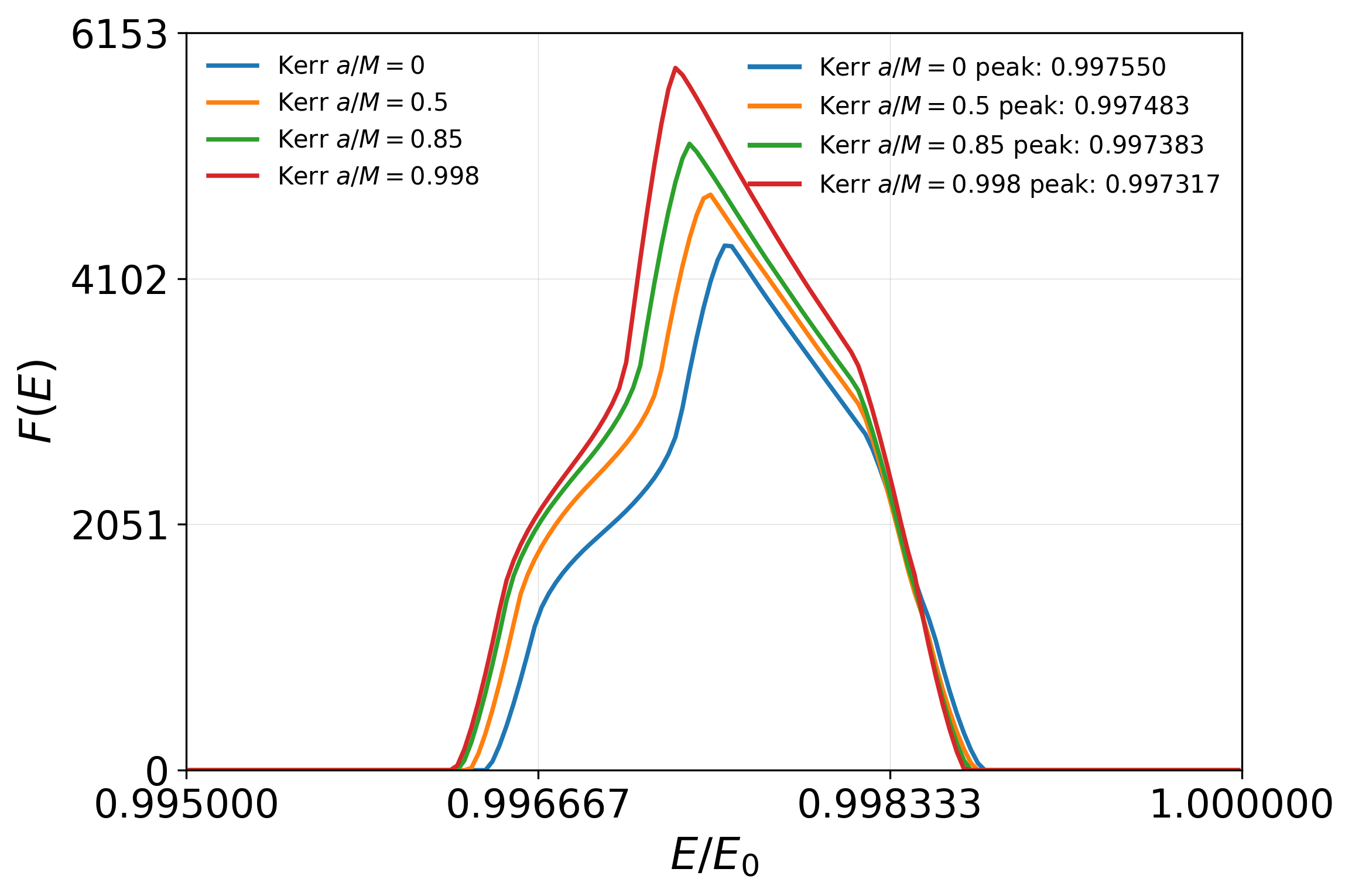} \caption{The spectra coming from the light-ring regions  of \cref{fig:nonKerr-3-2}. There is a clear frequency shift between the peaks of the non-rotating and the maximally rotating line of $\approx 0.0002.$}
 \label{fig:nonKerr2-3}
\end{center}
\end{figure}

We close this section with a brief explanation of the observed energy/frequency shift of the peak of the spectrum with BH rotation. Let's assume, for simplicity, that as the disk illuminates the light-ring, for the most part, the light rays are straight lines. The rays either hit the light-ring from the right or the left at an impact parameter which is approximately the size of the ring on the observer's screen, $\alpha_{\rm LR}$. Just as the case of the Einstein ring, this implies that the light ray goes through the accretion disk at some angle related to $\alpha_{\rm LR}$ and $D_{\rm LS}$ and therefore will have some Doppler shift due to the projection of the horizontal orbital velocity of the gas along the ray. Similar to the Einstein ring case, the frequency shift will be maximum along the equatorial plane, i.e., for $\alpha=0$ on the image plane. The shift, therefore, will be in our approximation,
\be \left(\frac{\Delta \nu}{\nu}\right)_{\rm LR}^{\pm} = \pm\frac{u_K}{c}\frac{\alpha_{\rm LR}^{\mp}}{D_{\rm LS}},
\ee
where the $(+)$ sign corresponds to blue-shift, that the rays coming from the left have, and the $(-)$ sign corresponds to the rays coming from the right, which are red-shifted (see \cref{fig:cartoon1}). For a non-rotating BH the impact parameter of the light-ring is the same on both sides, with $\alpha_{\rm LR}^{\mp}\approx 5.1M$. This introduces no shift of the spectral line. For a Kerr BH rotating at $a=0.998M$, though, as we have seen, the left side of the light-ring is at $b\approx-2.2M$, while the right side is at $b\approx +6.9M$. This means that the left side is blue-shifted by 

$$\left(\frac{\Delta \nu}{\nu}\right)_+=\left(\frac{M}{D_{\rm LS}}\right)^{1/2}\frac{2.2M}{D_{\rm LS}},$$
while the right side is red-shifted by
$$\left(\frac{\Delta \nu}{\nu}\right)_-=-\left(\frac{M}{D_{\rm LS}}\right)^{1/2}\frac{6.9M}{D_{\rm LS}},$$
where everything is expressed in geometric units. For $D_{\rm LS}=500M$, the two shifts are approximately $(\Delta\nu/\nu)_+\approx0.0002$ and $(\Delta\nu/\nu)_-\approx-0.0006$. These asymmetric shifts affect the general position of the line, which can be approximated to have shifted by the average of the two, i.e., by $\left((\Delta\nu/\nu)_+ +(\Delta\nu/\nu)_-\right)/2\approx-0.0002$.  
Therefore, rotation can introduce such a shift to the line coming from the light-ring, i.e., introduce a spin-dependent red-shift. This is approximately what we observe for the peak frequencies in \cref{fig:nonKerr2-2}. The effect can be better isolated within our ray-tracing procedure by choosing complete disk images that are even closer to the light-ring than the ones in \cref{fig:nonKerr-3}. These images are given in \cref{fig:nonKerr-3-2} (notice the differences with those in \cref{fig:nonKerr-2}). Fig. \ref{fig:nonKerr2-3} gives the corresponding spectra denoting exactly the effect we described. 
%
%
%
This effect is worth exploring further for possible use in measuring light-ring shifts, either due to rotation or deviations from Kerr.

All the previous discussion regarding the use of the various light rings, their morphology, and special spectra in order to obtain structural information on the geometry of the spacetime near BHs serves to underline the unique benefits of BH illumination by spectral line radiation emanating from gas disks at significant distances away from the ISCO region. The IR spectral lines from the BLR$^{0}$ disks uniquely $''$select$''$ these far regions, against any IR continuum (which would almost certainly have strong emission contributions, with geometries far more complicated than those of a thin disk). 
Images like the second one of the top row of \cref{fig:TT_bigscale1} that shows the full disk, or the images in  \cref{fig:nonKerr-3} that show the first strongly lensed image of the disk (together with the higher order images), and finally those in \cref{fig:nonKerr-3-2} (or rather realistically, \cref{fig:nonKerr-2}) showing the next order image of the disk scattered by the BH (along with the higher order, more tightly packed images that form the light-ring)  would be hopeless to separate in such a clear fashion if the continuum radiation of the hot, ionized, strongly non-uniform accretion disk that extends all the way down to the ISCO was used as the BH illuminator.
Spectral line illumination from the more distant BLR gas disk (and especially its  BLR$^{0}$ phase) on the other hand, coupled with $(u,v)$-tailored interferometric imaging, can, in principle, one day
deliver exquisite structure information on the light-rings, providing information on the spacetime geometry near BHs, using space-based interferometers operating at IR wavelengths where the most luminous BLR$^{0}$ lines are expected\footnote{Such interferometers would have to be space-borne as IR radiation, with the exception of a narrow window around $\lambda \sim 2\, \mu m$ is absorbed by the Earth's atmosphere.}. These lines, their luminosities, and which of them are optimal for such SMBH  illumination are the subject of upcoming work \cite{Astro_paper, Chem_paper}.

\section{Conclusions}
\label{sec:concl}
%

In this work, we propose using spectral line radiation from a more distant source rather than continuum radiation from
a near-in disk as the background radiation illuminating the SMBHs at galactic centers in order to examine the strong gravity effects near them. Such spectral line radiation fields can be provided by the BLRs, gas disks located at $r\sim (10^2-10^4) R_s$ away from the BH. The lines considered here emanate from the neutral BLR$^{0}$ disk (on which the ionized BLR (BLR$^{+}$) is only a thin outer layer) \cite{Thi:2024dny}, where special thermal conditions facilitate well-excited neutral atomic and molecular lines at IR wavelengths. The latter allows the BH spacetime geometry-related emission features to remain observable along near equatorial lines of sight (where they are the strongest),  unaffected by the interstellar extinction from dust in the host galaxy. Low gas turbulence, more uniform and less time-varying radiation fields originating from thin gas disk configurations are among the benefits of BLR$^{0}$ spectral line illumination of SMBHs (\cite{Astro_paper}). 

The major advantage of a disk-like illumination source located well beyond the ISCO region of the BH is that the disk radiation directly reaching the distant observer is no longer overlaid on the purely geometric brightness patterns expected from  BH spacetime geometry,i.e., the light-ring. This keeps the complicated accretion disk Astrophysics of the luminous, hot, ionized and strongly MHD-turbulent inner accretion disk (whose radio continuum radiation was used by the EHT to outline the BH shadows) away from becoming a necessary (and very difficult) part in the extraction of pure GR effects from the associated interferometric images.
In this regard, the advantage of using spectral line rather than continuum radiation fields as the BH illumination background is purely geometric, in the sense that if a similarly distant, isolated, thin-disk-geometry source that emits continuum instead of spectral line radiation could be found, the advantage would be exactly the same. The spectral line emission  then simply plays the role of geometrically isolating such a source (the BLR$^{0}$)  from the background of all the other $''$competing$''$ sources emitting near the BH. 
 
A geometrical configuration of the BH and a more distant illumination source then allows, on the one hand, the formation of an Einstein ring (forming for lines of sight near the BH/BLR disk equatorial plane) and, on the other hand, the clear discerning of the emission of the light-ring that is closer to the BH, without any confusion due to a luminous inner accretion disk background. 
Nevertheless, distinct advantages uniquely associated with a spectral line rather than continuum radiation field providing the BH illumination do exist, and go beyond the aforementioned favorable geometrical configuration.
The simplest one is the information on the gas disk inclination encoded in the BLR global line profile, which could be used to model uncertainties due to unknown inclination values, which impact the phenomenology of the light-ring. For example, for Kerr BHs, the rotation and the inclination have similar effects in the displacement of the light-ring with respect to the center of the image plane at $(b,\alpha)=(0,0)$. Information from the spectrum could, in principle, lift such degeneracies. 

Furthermore, as it was shown in \cite{PRL_companion}, and expanded here, the more distant illumination source, in conjunction with the spectral line radiation, produces some unique lensing phenomenology. The geometry of the disk, as we have explained, leads to the creation of an Einstein ring, which additionally leaves a very characteristic signature on the spectrum of the BLR part of the accretion disk. The combination of these two effects provides a novel way for measuring the mass of SMBHs, when the line of sight towards the system is appropriate and the Einstein ring can be resolved. We have also verified that the  SMBH mass estimates using the spectral signature and the image of the Einstein ring can provide a relatively robust result. Of course, the aforementioned analysis pertains to equatorial or near equatorial lines of sight where such GR optical effects manifest themselves. This makes the typical thermal conditions of the BLR$^{0}$  particularly fortunate as they are responsible for the expected strong IR lines \cite{Thi:2024dny, Astro_paper}, and thus for the Einstein ring and light-ring being $''$outlined$''$ by them. This then keeps these strong-gravity features observationally accessible to future IR interferometers,  unaffected by cosmic dust absorption, which will be prominent for e.g. optical or UV wavelengths along equatorial or near equatorial BH/BLR-disk lines of sight through the BH host galaxy.

Beyond BH mass measurements, one can also aim to measure the BH spin with the help of the light-ring. The dominant effect of the spin on the light-ring is its displacement with respect to the image center at $(b,\alpha)=(0,0)$. Such displacements could be measured in the future using interferometric observations at IR wavelengths, utilizing the phase space of the $(u,v)$-plane. The Einstein ring can facilitate this by providing a physical reference for framing such small light-ring positional shifts against the BH center,  marked by the center of the Einstein ring 
emission feature. Future interferometric imaging at IR wavelengths could also aim to identify deformations of the light-ring, related to the spin of the BH or even deviations from the Kerr spacetime. Again, having a gas disk and its associated spectral emission provide the BH illumination from distances well beyond the ISCO regions confers a crucial advantage in keeping the corresponding images $''$clean$''$ of disk versus light-ring emission overlays, allowing the tests described in this work to be conducted. Here, we find that the effects of spin on the spectral line profile associated solely with the BH light-ring  to be very small, and impractical in this setting as a possible spin estimator, even under the ideal spectral line illumination conditions by the BLR$^{0}$ gas disks considered here, leaving the purely geometrical light-ring center displacement method as the most viable one.

Finally, the entire discussion here assumes a stationary disk with a steady brightness distribution. There are more interesting prospects to consider in cases with variability, both spatial and temporal, in the disk emission, such as local flares or large-scale features, such as spiral configurations, that move around the disk. Such time-varying signals can produce effects like echoes in the disk images, and the light scattered by the light-ring that will encode information on the spacetime {\it and} the disk structure \cite{Yfantis:2025jtw}. Moreover, emission features related to moving structures in the disk could be used to perform tomography of the disk, and spacetime tomography, now aided by the fact of a spectral line rather than a continuum radiation BH illumination. These effects will be the subject of future work.

\section*{Acknowledgments}

This work has received no funding. For we are poor but innovative and happy. Some of us would like to thank our lucky stars for being able to work on such stuff at an advanced age, without driving our collaborators mad. PP would also like to thank the bar-tenders and numerous patrons in the Schumman's bar (Odeonplatz, Munich) and the Bulldog bar (Agia Sofia square, Thessaloniki)... who inspired him to go on, even as they never knew it... 
The ray-tracing and radiative transfer code used to produce the images and the spectra can be found at the repository [\url{https://github.com/
GPappasGR/Spectral_BH_shadows_BLR}]

\appendix
\section{Other density profile}

\begin{figure*}
\begin{center}
\includegraphics[width=0.22\textwidth]{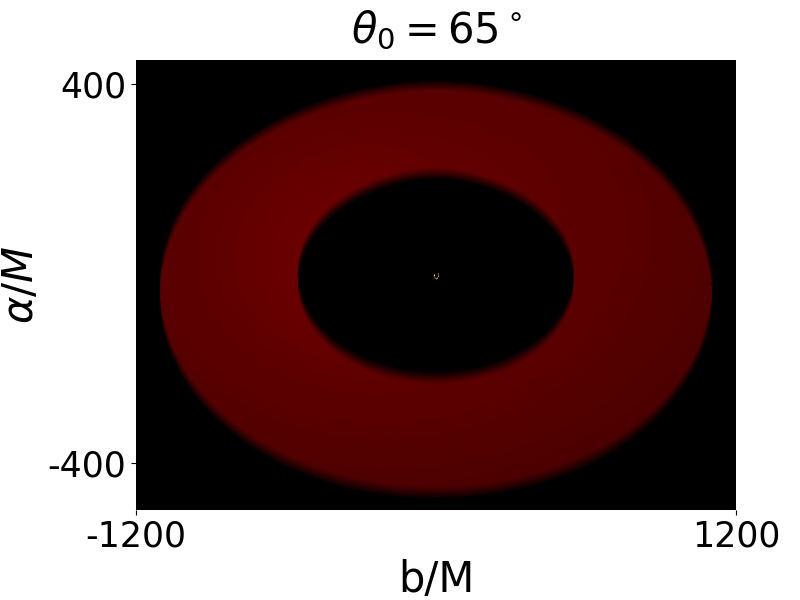}
\includegraphics[width=0.22\textwidth]{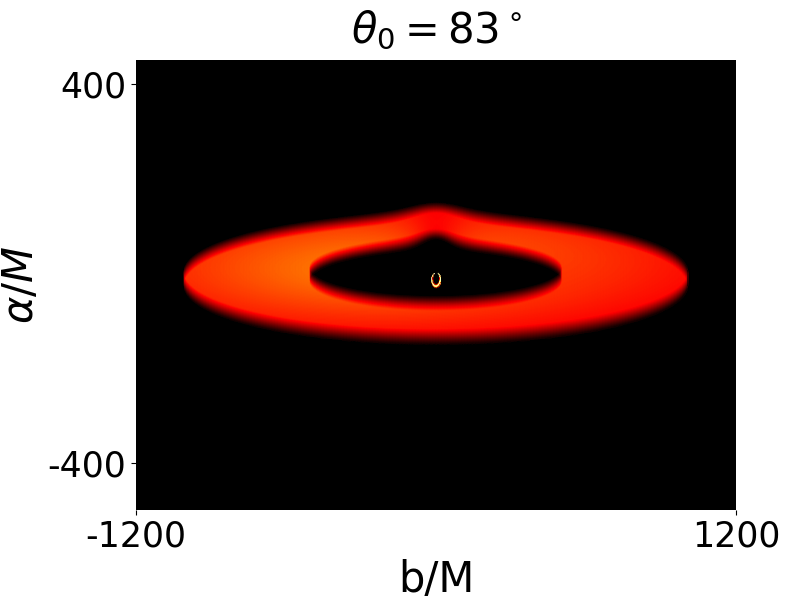}
\includegraphics[width=0.22\textwidth]{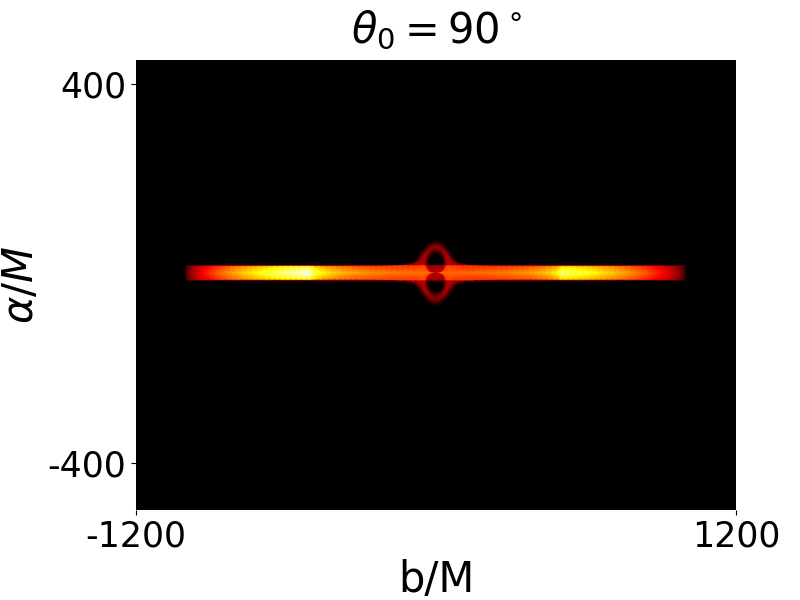}
\includegraphics[width=0.22\textwidth]{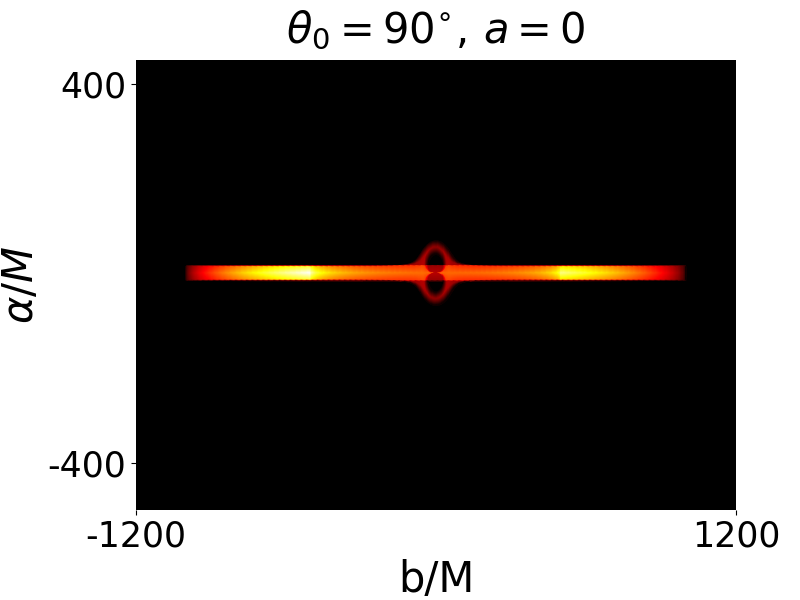}

\includegraphics[width=0.22\textwidth]{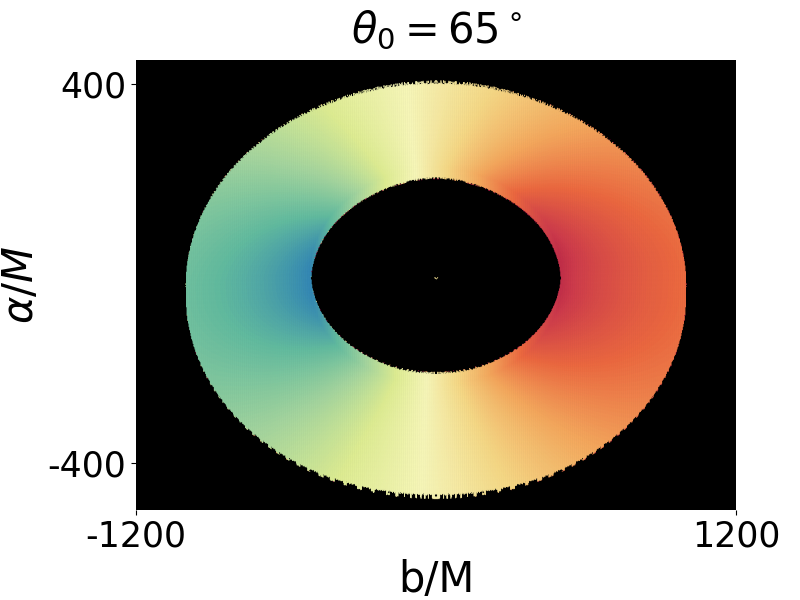}
\includegraphics[width=0.22\textwidth]{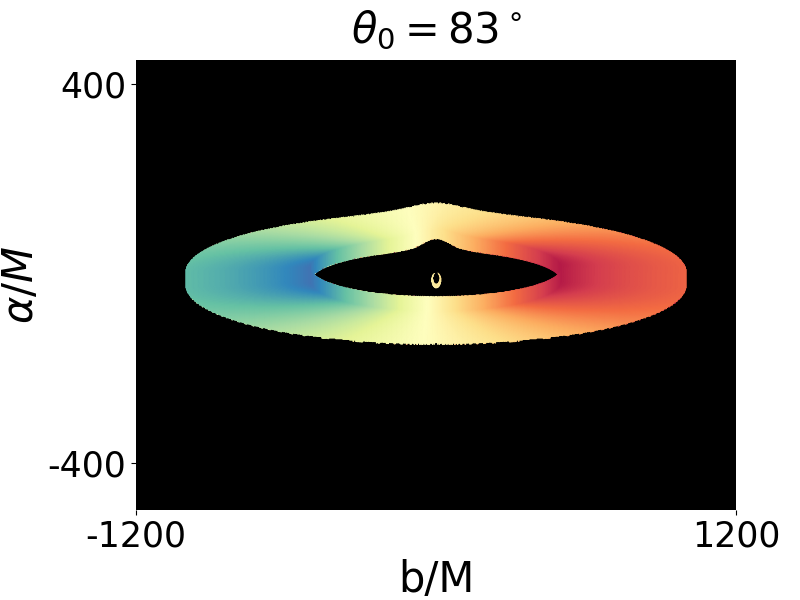}
\includegraphics[width=0.22\textwidth]{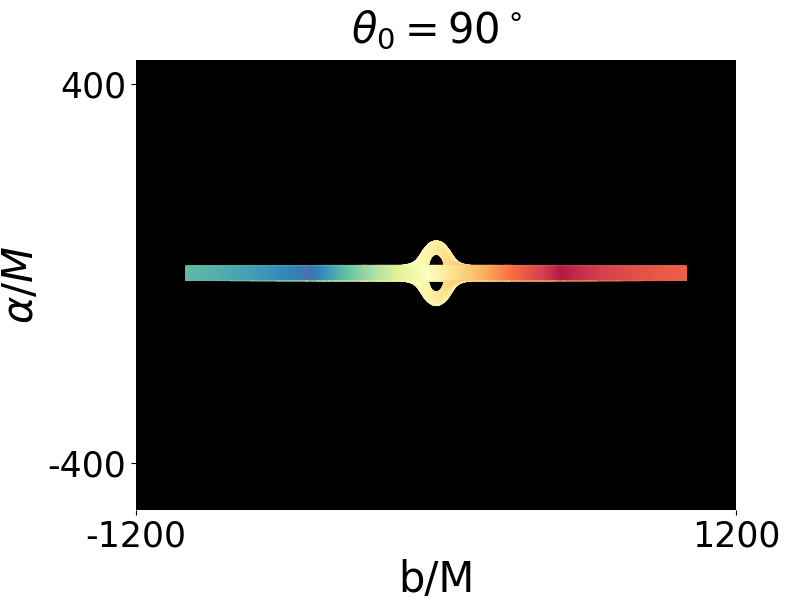}
\includegraphics[width=0.22\textwidth]{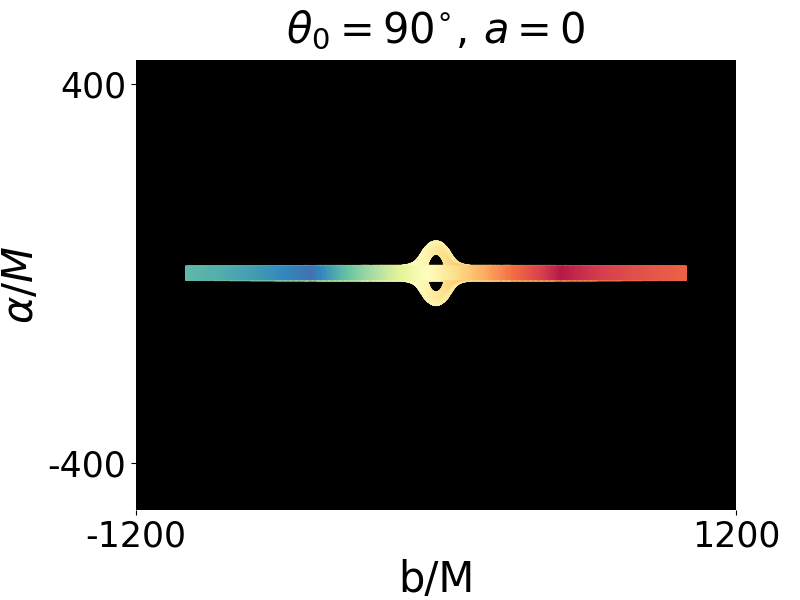}

\includegraphics[width=0.22\textwidth]{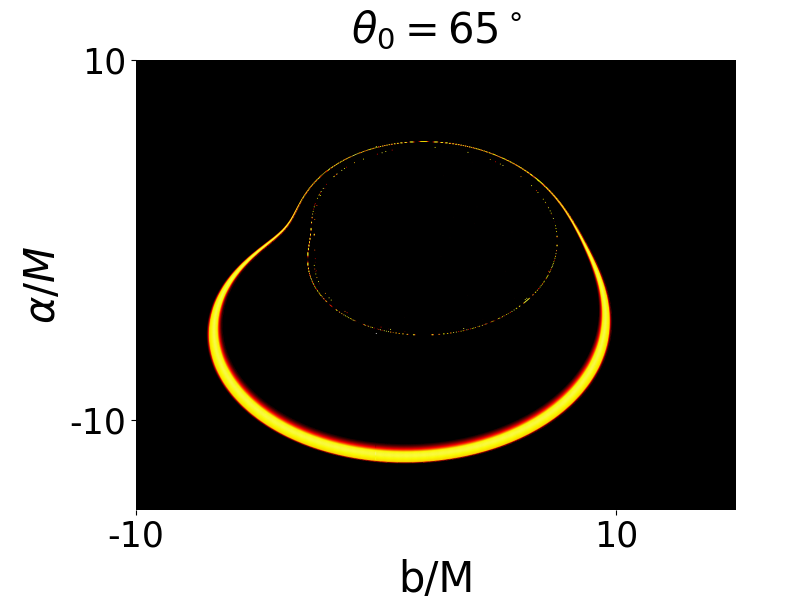}
\includegraphics[width=0.22\textwidth]{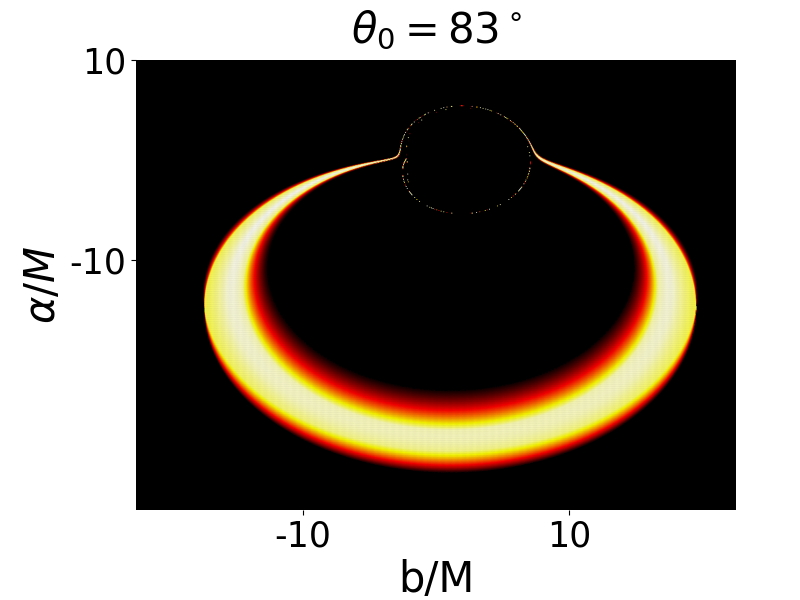}
\includegraphics[width=0.22\textwidth]{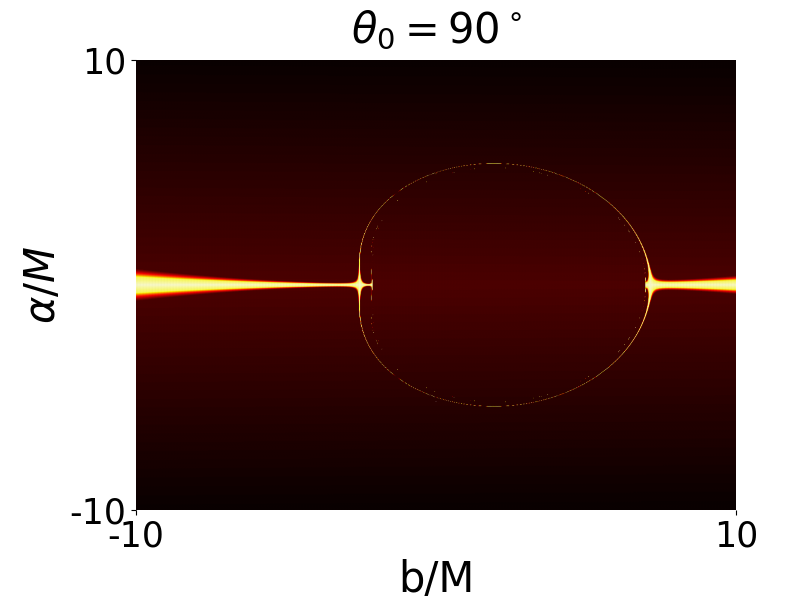}
\includegraphics[width=0.22\textwidth]{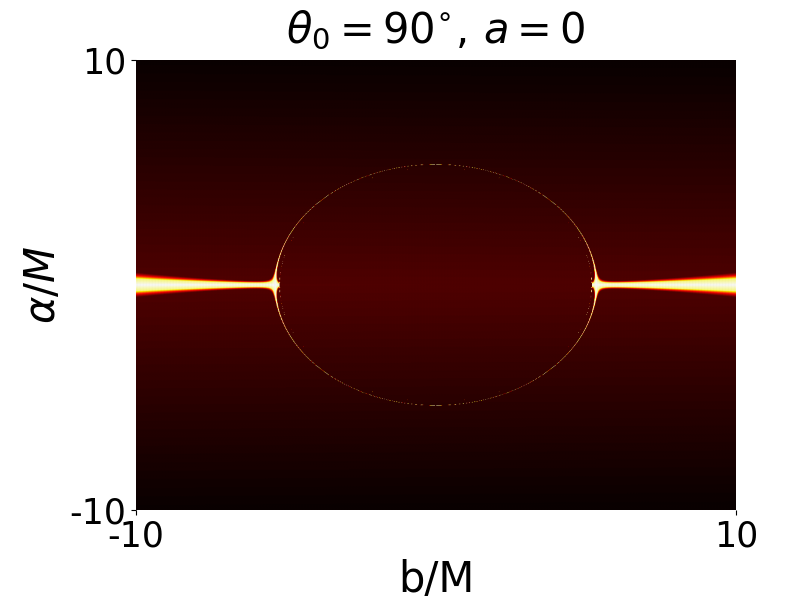}

\includegraphics[width=0.22\textwidth]{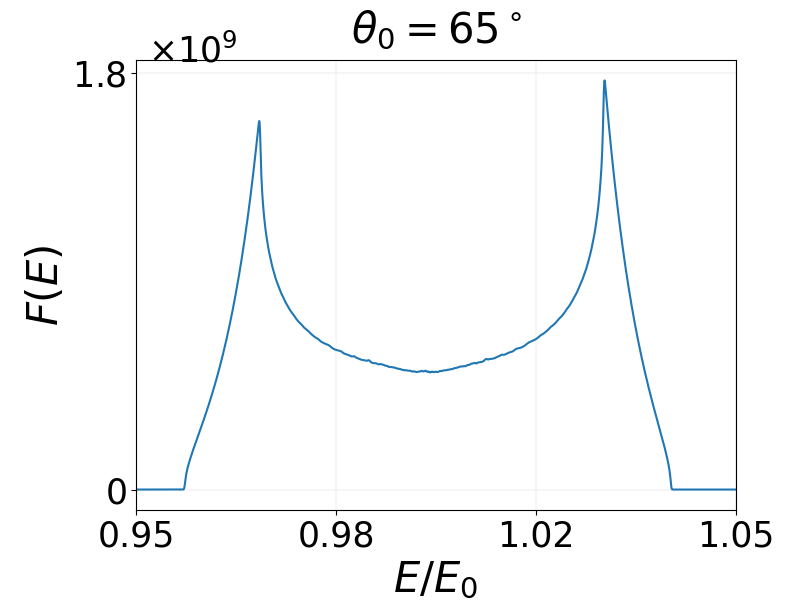}
\includegraphics[width=0.22\textwidth]{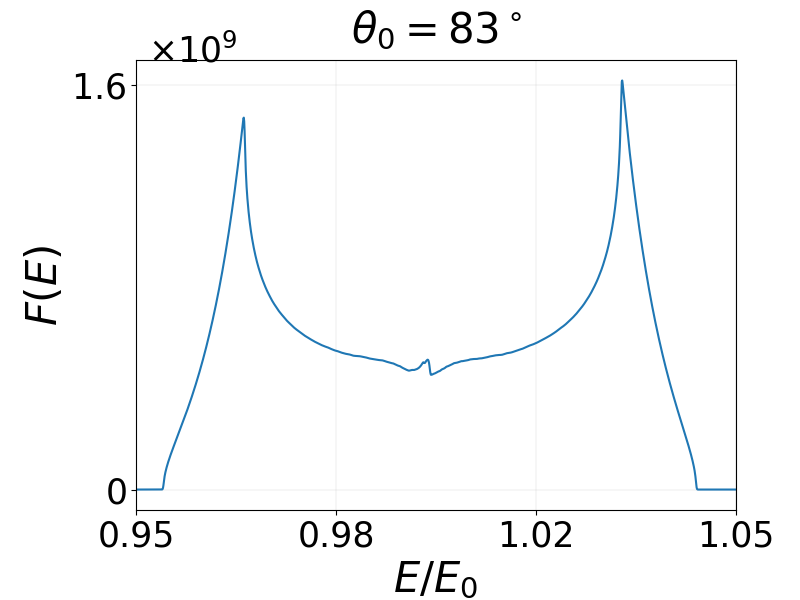}
\includegraphics[width=0.22\textwidth]{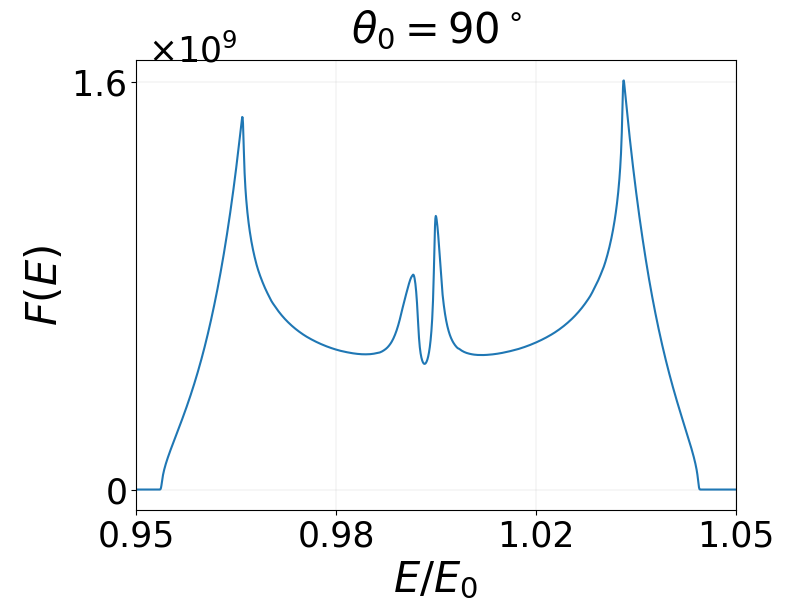}
\includegraphics[width=0.22\textwidth]{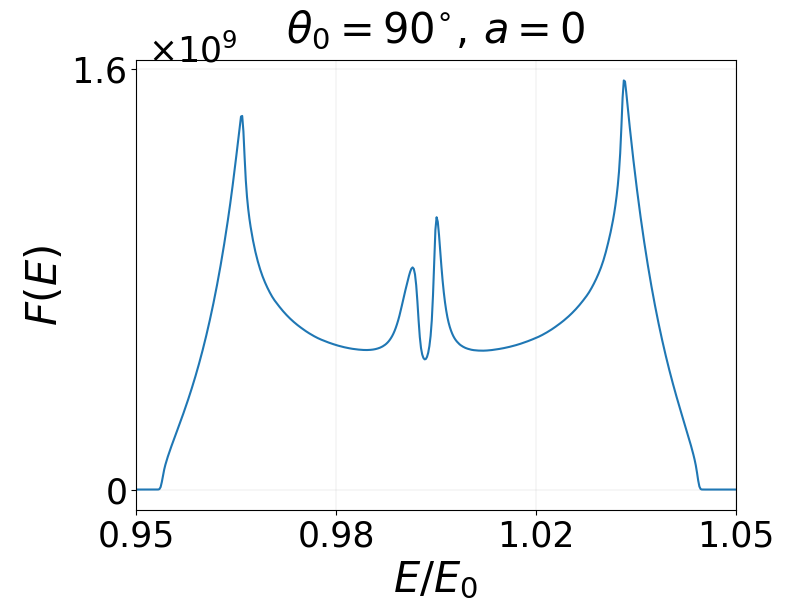}
\caption{Same figure as \cref{fig:TT_bigscale1}, where the images have been produced using the emissivity profile given by eq. \eqref{eq:emis1}.}
\label{fig:TT_bigscale}
\end{center}
\end{figure*}

In this appendix, we will provide some figures that have been produced using the alternative profile for the emissivity, given in eq. \eqref{eq:emis1}. This will provide additional support for the robustness of the results and also give an idea of how the matter distribution in the accretion disk affects the various images and spectra. We remind that the main difference between the two profiles is that the one has a more steep drop in the density of the matterial with the increasing radial distance. This has the overall effect of providing less flux in the final images and spectra. 

We first present in \cref{fig:TT_bigscale} the same figure as in \cref{fig:TT_bigscale1}, using this time the emissivity profile given by eq. \eqref{eq:emis1}. Two differences stand out. While in the case of \cref{fig:TT_bigscale1} we can see that there is a region where the emission is brighter, in \cref{fig:TT_bigscale} we can see that in general the emission is more uniformly distributed. The second difference is that as we have mentioned previously, the emitted flux is larger in this case, since there is more material available. Apart from these two differences, the phenomenology is the same for both emissivity profiles and especially with respect to the Einstein ring.

\begin{figure}
\begin{center} 
    \includegraphics[width=0.45\textwidth]{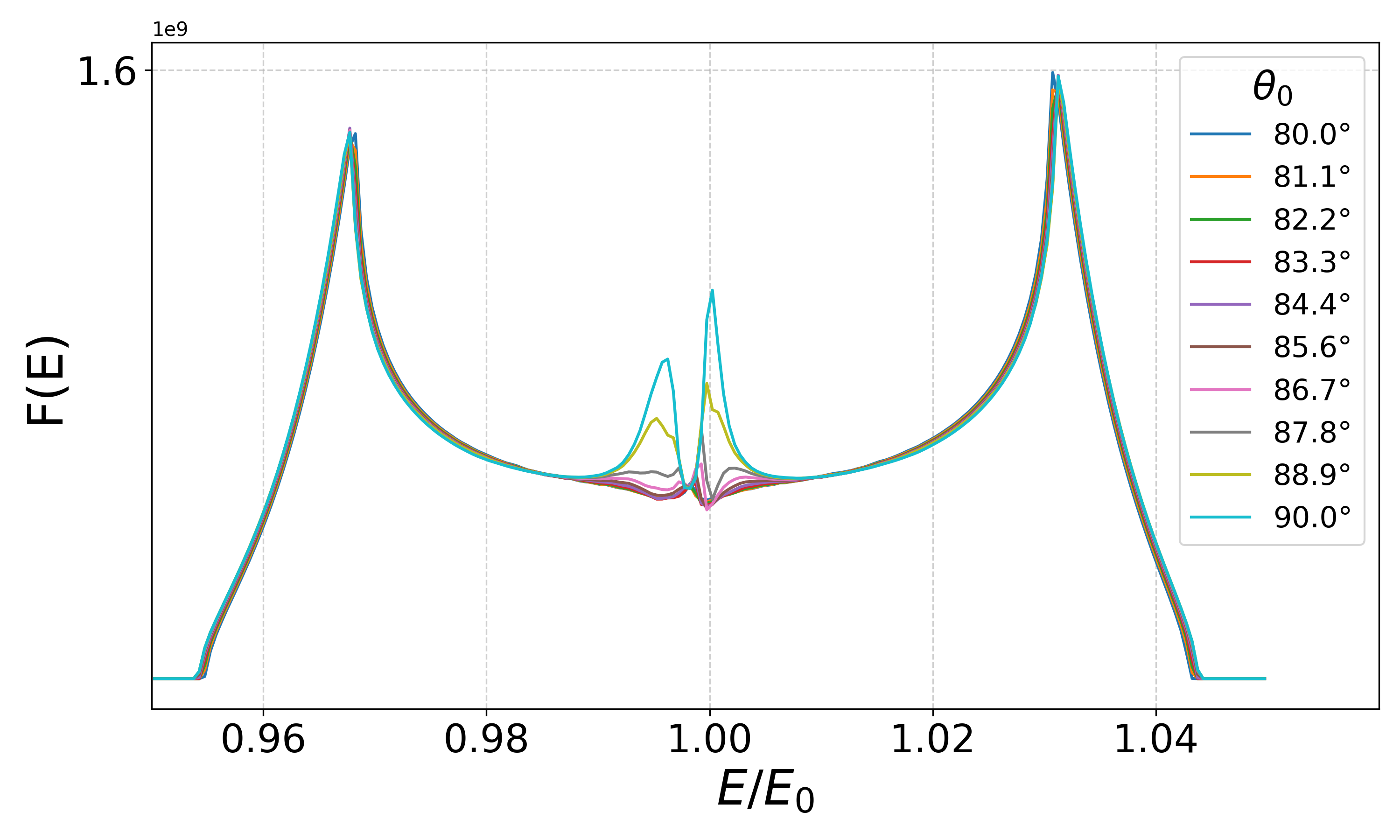}  
 \caption{Demonstration of the emergence of the central magnified double peak of the spectral line for different inclinations. The figure has been produced using the emissivity profile given by eq. \eqref{eq:emis1}.}
 \label{fig:E_spectra2}
\end{center}
\end{figure} 

The same holds for the dependence on the inclination of the emergence of the twin-peaks at the center of the spectrum, due to the Einstein ring. Fig \ref{fig:E_spectra2} shows the exact same behaviour as the one seen in \cref{fig:E_spectra}. The only differences are with respect to the total flux from the line and a minor change in the sharpness of the peaks. 

\begin{figure}
\begin{center} \includegraphics[width=0.45\textwidth]{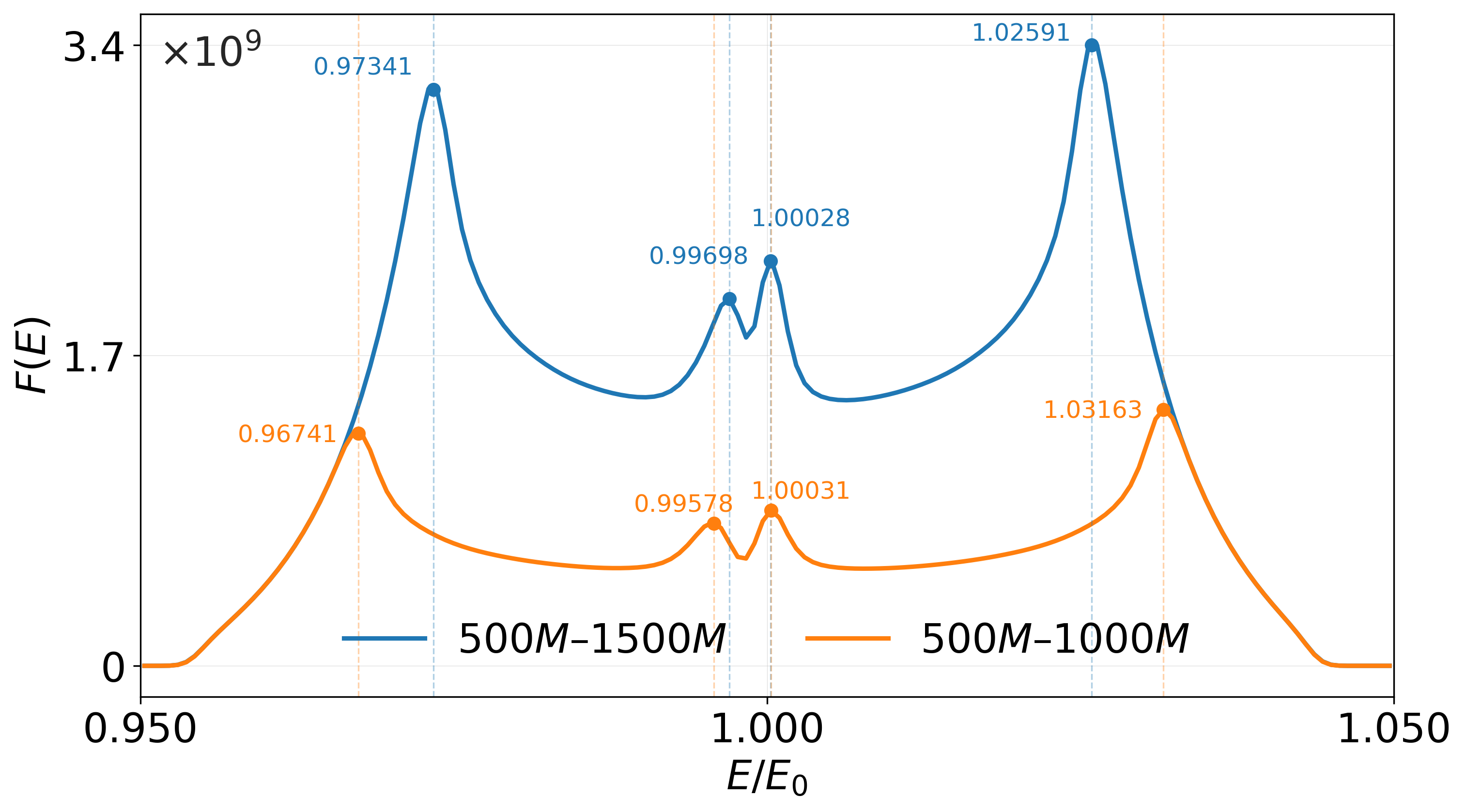}
\includegraphics[width=0.45\textwidth]{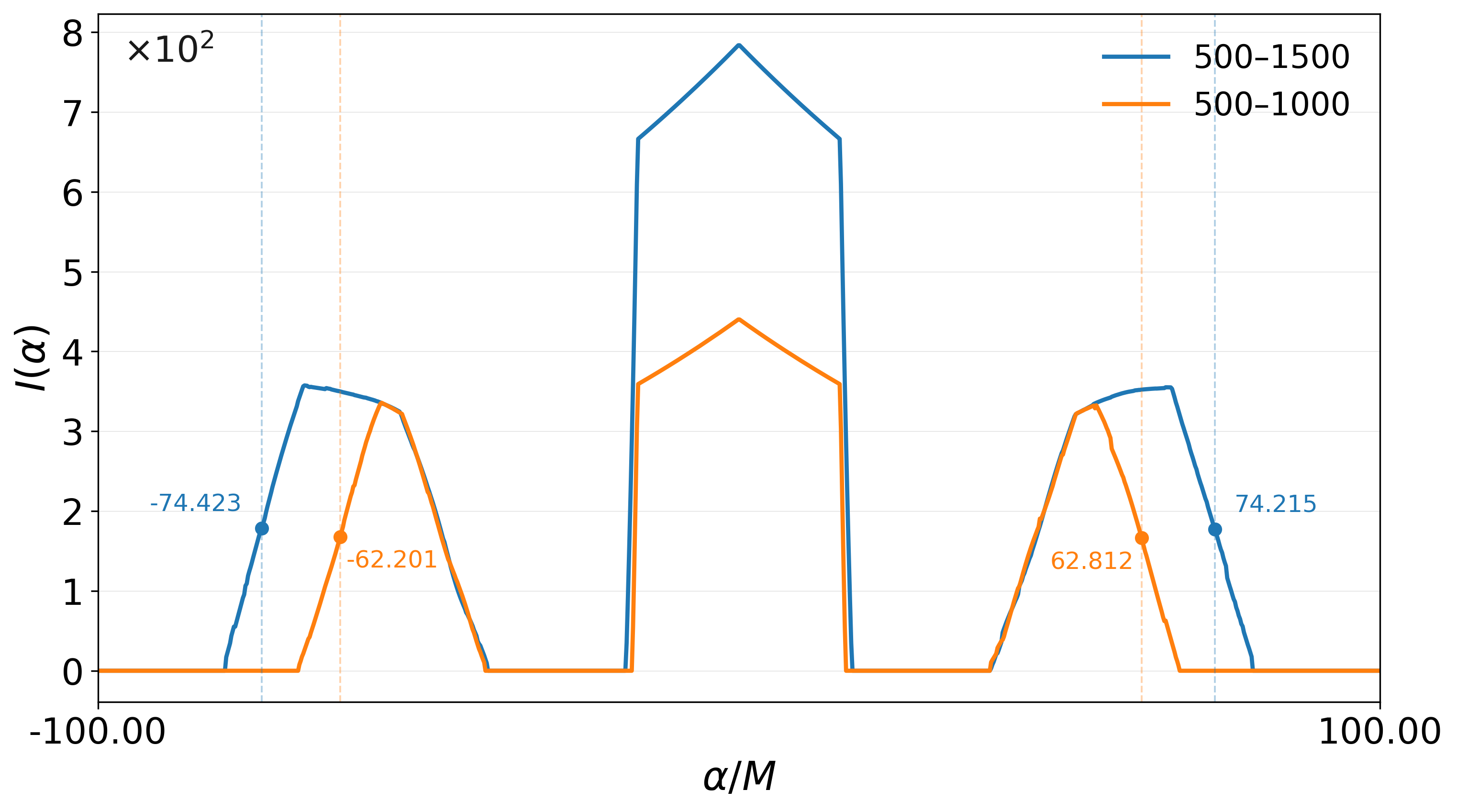}
 \caption{Effect on the spectrum and the Einstein ring of changes in the outer radius of the emission. These figures have been produced using the emissivity given in eq. \eqref{eq:emis1}.}
 \label{fig:larger_disk2}
\end{center}
\end{figure}

Our final comparison is with respect to the estimation of the mass of the BH from the frequency difference of the central twin-peaks and the size of the Einstein ring. Fig. \ref{fig:larger_disk2} shows the same plots as \cref{fig:larger_disk}, but for the emissivity of eq.\eqref{eq:emis1}. We use these plots to perform the same estimation of the mass of the BH, using again two disks that extend out to $1000M$ and $1500M$.
Using the position for which the intensity becomes half of the maximum value for estimating the outer $\alpha_E^{eff}$, we have for the first case that $\alpha_E^{eff}\simeq 62M$, while $\Delta \nu/\nu \simeq 0.0045$. This gives a mass of $M\simeq1.04$. For the second case, where the disk extends out to $1500M$, the estimate for the outer size of the Einstein ring is $\alpha_E^{eff}\simeq 74M$, while $\Delta \nu/\nu \simeq 0.0033$. This gives a mass of $M\simeq1.006$, which is in good agreement with the previous result. In general, all the results from both emissivity profiles seem to give consistent and relatively robust mass estimates for the BH.

\bibliography{bibliography.bib}

\begin{thebibliography}{75}%
\makeatletter
\providecommand \@ifxundefined [1]{%
 \@ifx{#1\undefined}
}%
\providecommand \@ifnum [1]{%
 \ifnum #1\expandafter \@firstoftwo
 \else \expandafter \@secondoftwo
 \fi
}%
\providecommand \@ifx [1]{%
 \ifx #1\expandafter \@firstoftwo
 \else \expandafter \@secondoftwo
 \fi
}%
\providecommand \natexlab [1]{#1}%
\providecommand \enquote  [1]{``#1''}%
\providecommand \bibnamefont  [1]{#1}%
\providecommand \bibfnamefont [1]{#1}%
\providecommand \citenamefont [1]{#1}%
\providecommand \href@noop [0]{\@secondoftwo}%
\providecommand \href [0]{\begingroup \@sanitize@url \@href}%
\providecommand \@href[1]{\@@startlink{#1}\@@href}%
\providecommand \@@href[1]{\endgroup#1\@@endlink}%
\providecommand \@sanitize@url [0]{\catcode `\\12\catcode `\$12\catcode
  `\&12\catcode `\#12\catcode `\^12\catcode `\_12\catcode `\%12\relax}%
\providecommand \@@startlink[1]{}%
\providecommand \@@endlink[0]{}%
\providecommand \url  [0]{\begingroup\@sanitize@url \@url }%
\providecommand \@url [1]{\endgroup\@href {#1}{\urlprefix }}%
\providecommand \urlprefix  [0]{URL }%
\providecommand \Eprint [0]{\href }%
\providecommand \doibase [0]{https://doi.org/}%
\providecommand \selectlanguage [0]{\@gobble}%
\providecommand \bibinfo  [0]{\@secondoftwo}%
\providecommand \bibfield  [0]{\@secondoftwo}%
\providecommand \translation [1]{[#1]}%
\providecommand \BibitemOpen [0]{}%
\providecommand \bibitemStop [0]{}%
\providecommand \bibitemNoStop [0]{.\EOS\space}%
\providecommand \EOS [0]{\spacefactor3000\relax}%
\providecommand \BibitemShut  [1]{\csname bibitem#1\endcsname}%
\let\auto@bib@innerbib\@empty
\bibitem [{\citenamefont {Collaboration}(2019{\natexlab{a}})}]{Akiyama_2019}%
  \BibitemOpen
  \bibfield  {author} {\bibinfo {author} {\bibfnamefont {T.~E. H.~T.}\
  \bibnamefont {Collaboration}},\ }\bibfield  {title} {\bibinfo {title} {First
  m87 event horizon telescope results. i. the shadow of the supermassive black
  hole},\ }\href {https://doi.org/10.3847/2041-8213/ab0ec7} {\bibfield
  {journal} {\bibinfo  {journal} {The Astrophysical Journal Letters}\ }\textbf
  {\bibinfo {volume} {875}},\ \bibinfo {pages} {L1} (\bibinfo {year}
  {2019}{\natexlab{a}})}\BibitemShut {NoStop}%
\bibitem [{\citenamefont {Collaboration}(2019{\natexlab{b}})}]{EHT_M87_2}%
  \BibitemOpen
  \bibfield  {author} {\bibinfo {author} {\bibfnamefont {T.~E. H.~T.}\
  \bibnamefont {Collaboration}},\ }\bibfield  {title} {\bibinfo {title} {{First
  M87 Event Horizon Telescope Results. II. Array and Instrumentation}},\ }\href
  {https://doi.org/10.3847/2041-8213/ab0c96} {\bibfield  {journal} {\bibinfo
  {journal} {The Astrophysical Journal Letters}\ }\textbf {\bibinfo {volume}
  {875}},\ \bibinfo {eid} {L2} (\bibinfo {year} {2019}{\natexlab{b}})},\
  \Eprint {https://arxiv.org/abs/1906.11239} {arXiv:1906.11239 [astro-ph.IM]}
  \BibitemShut {NoStop}%
\bibitem [{\citenamefont {Collaboration}(2019{\natexlab{c}})}]{EHT_M87_3}%
  \BibitemOpen
  \bibfield  {author} {\bibinfo {author} {\bibfnamefont {T.~E. H.~T.}\
  \bibnamefont {Collaboration}},\ }\bibfield  {title} {\bibinfo {title} {{First
  M87 Event Horizon Telescope Results. III. Data Processing and Calibration}},\
  }\href {https://doi.org/10.3847/2041-8213/ab0c57} {\bibfield  {journal}
  {\bibinfo  {journal} {The Astrophysical Journal Letters}\ }\textbf {\bibinfo
  {volume} {875}},\ \bibinfo {eid} {L3} (\bibinfo {year}
  {2019}{\natexlab{c}})},\ \Eprint {https://arxiv.org/abs/1906.11240}
  {arXiv:1906.11240 [astro-ph.GA]} \BibitemShut {NoStop}%
\bibitem [{\citenamefont {Collaboration}(2019{\natexlab{d}})}]{EHT_M87_4}%
  \BibitemOpen
  \bibfield  {author} {\bibinfo {author} {\bibfnamefont {T.~E. H.~T.}\
  \bibnamefont {Collaboration}},\ }\bibfield  {title} {\bibinfo {title} {{First
  M87 Event Horizon Telescope Results. IV. Imaging the Central Supermassive
  Black Hole}},\ }\href {https://doi.org/10.3847/2041-8213/ab0e85} {\bibfield
  {journal} {\bibinfo  {journal} {The Astrophysical Journal Letters}\ }\textbf
  {\bibinfo {volume} {875}},\ \bibinfo {eid} {L4} (\bibinfo {year}
  {2019}{\natexlab{d}})},\ \Eprint {https://arxiv.org/abs/1906.11241}
  {arXiv:1906.11241 [astro-ph.GA]} \BibitemShut {NoStop}%
\bibitem [{\citenamefont {Collaboration}(2019{\natexlab{e}})}]{EHT_M87_5}%
  \BibitemOpen
  \bibfield  {author} {\bibinfo {author} {\bibfnamefont {T.~E. H.~T.}\
  \bibnamefont {Collaboration}},\ }\bibfield  {title} {\bibinfo {title} {{First
  M87 Event Horizon Telescope Results. V. Physical Origin of the Asymmetric
  Ring}},\ }\href {https://doi.org/10.3847/2041-8213/ab0f43} {\bibfield
  {journal} {\bibinfo  {journal} {The Astrophysical Journal Letters}\ }\textbf
  {\bibinfo {volume} {875}},\ \bibinfo {eid} {L5} (\bibinfo {year}
  {2019}{\natexlab{e}})},\ \Eprint {https://arxiv.org/abs/1906.11242}
  {arXiv:1906.11242 [astro-ph.GA]} \BibitemShut {NoStop}%
\bibitem [{\citenamefont {Collaboration}(2019{\natexlab{f}})}]{EHT_M87_6}%
  \BibitemOpen
  \bibfield  {author} {\bibinfo {author} {\bibfnamefont {T.~E. H.~T.}\
  \bibnamefont {Collaboration}},\ }\bibfield  {title} {\bibinfo {title} {{First
  M87 Event Horizon Telescope Results. VI. The Shadow and Mass of the Central
  Black Hole}},\ }\href {https://doi.org/10.3847/2041-8213/ab1141} {\bibfield
  {journal} {\bibinfo  {journal} {The Astrophysical Journal Letters}\ }\textbf
  {\bibinfo {volume} {875}},\ \bibinfo {eid} {L6} (\bibinfo {year}
  {2019}{\natexlab{f}})},\ \Eprint {https://arxiv.org/abs/1906.11243}
  {arXiv:1906.11243 [astro-ph.GA]} \BibitemShut {NoStop}%
\bibitem [{\citenamefont {Collaboration}(2022{\natexlab{a}})}]{EHT_Sgr1}%
  \BibitemOpen
  \bibfield  {author} {\bibinfo {author} {\bibfnamefont {T.~E. H.~T.}\
  \bibnamefont {Collaboration}},\ }\bibfield  {title} {\bibinfo {title} {{First
  Sagittarius A* Event Horizon Telescope Results. I. The Shadow of the
  Supermassive Black Hole in the Center of the Milky Way}},\ }\href
  {https://doi.org/10.3847/2041-8213/ac6674} {\bibfield  {journal} {\bibinfo
  {journal} {The Astrophysical Journal Letters}\ }\textbf {\bibinfo {volume}
  {930}},\ \bibinfo {eid} {L12} (\bibinfo {year}
  {2022}{\natexlab{a}})}\BibitemShut {NoStop}%
\bibitem [{\citenamefont {Collaboration}(2022{\natexlab{b}})}]{EHT_Sgr2}%
  \BibitemOpen
  \bibfield  {author} {\bibinfo {author} {\bibfnamefont {T.~E. H.~T.}\
  \bibnamefont {Collaboration}},\ }\bibfield  {title} {\bibinfo {title} {{First
  Sagittarius A* Event Horizon Telescope Results. II. EHT and Multiwavelength
  Observations, Data Processing, and Calibration}},\ }\href
  {https://doi.org/10.3847/2041-8213/ac6675} {\bibfield  {journal} {\bibinfo
  {journal} {The Astrophysical Journal Letters}\ }\textbf {\bibinfo {volume}
  {930}},\ \bibinfo {eid} {L13} (\bibinfo {year}
  {2022}{\natexlab{b}})}\BibitemShut {NoStop}%
\bibitem [{\citenamefont {Collaboration}(2022{\natexlab{c}})}]{EHT_Sgr3}%
  \BibitemOpen
  \bibfield  {author} {\bibinfo {author} {\bibfnamefont {T.~E. H.~T.}\
  \bibnamefont {Collaboration}},\ }\bibfield  {title} {\bibinfo {title} {{First
  Sagittarius A* Event Horizon Telescope Results. III. Imaging of the Galactic
  Center Supermassive Black Hole}},\ }\href
  {https://doi.org/10.3847/2041-8213/ac6429} {\bibfield  {journal} {\bibinfo
  {journal} {The Astrophysical Journal Letters}\ }\textbf {\bibinfo {volume}
  {930}},\ \bibinfo {eid} {L14} (\bibinfo {year}
  {2022}{\natexlab{c}})}\BibitemShut {NoStop}%
\bibitem [{\citenamefont {Collaboration}(2022{\natexlab{d}})}]{EHT_Sgr4}%
  \BibitemOpen
  \bibfield  {author} {\bibinfo {author} {\bibfnamefont {T.~E. H.~T.}\
  \bibnamefont {Collaboration}},\ }\bibfield  {title} {\bibinfo {title} {{First
  Sagittarius A* Event Horizon Telescope Results. IV. Variability, Morphology,
  and Black Hole Mass}},\ }\href {https://doi.org/10.3847/2041-8213/ac6736}
  {\bibfield  {journal} {\bibinfo  {journal} {The Astrophysical Journal
  Letters}\ }\textbf {\bibinfo {volume} {930}},\ \bibinfo {eid} {L15} (\bibinfo
  {year} {2022}{\natexlab{d}})}\BibitemShut {NoStop}%
\bibitem [{\citenamefont {Collaboration}(2022{\natexlab{e}})}]{EHT_Sgr5}%
  \BibitemOpen
  \bibfield  {author} {\bibinfo {author} {\bibfnamefont {T.~E. H.~T.}\
  \bibnamefont {Collaboration}},\ }\bibfield  {title} {\bibinfo {title} {{First
  Sagittarius A* Event Horizon Telescope Results. V. Testing Astrophysical
  Models of the Galactic Center Black Hole}},\ }\href
  {https://doi.org/10.3847/2041-8213/ac6672} {\bibfield  {journal} {\bibinfo
  {journal} {The Astrophysical Journal Letters}\ }\textbf {\bibinfo {volume}
  {930}},\ \bibinfo {eid} {L16} (\bibinfo {year}
  {2022}{\natexlab{e}})}\BibitemShut {NoStop}%
\bibitem [{\citenamefont {Collaboration}(2022{\natexlab{f}})}]{EHT_Sgr6}%
  \BibitemOpen
  \bibfield  {author} {\bibinfo {author} {\bibfnamefont {T.~E. H.~T.}\
  \bibnamefont {Collaboration}},\ }\bibfield  {title} {\bibinfo {title} {{First
  Sagittarius A* Event Horizon Telescope Results. VI. Testing the Black Hole
  Metric}},\ }\href {https://doi.org/10.3847/2041-8213/ac6756} {\bibfield
  {journal} {\bibinfo  {journal} {The Astrophysical Journal Letters}\ }\textbf
  {\bibinfo {volume} {930}},\ \bibinfo {eid} {L17} (\bibinfo {year}
  {2022}{\natexlab{f}})}\BibitemShut {NoStop}%
\bibitem [{\citenamefont {Collaboration}(2024)}]{EHT_M87_7}%
  \BibitemOpen
  \bibfield  {author} {\bibinfo {author} {\bibfnamefont {T.~E. H.~T.}\
  \bibnamefont {Collaboration}},\ }\bibfield  {title} {\bibinfo {title} {{The
  persistent shadow of the supermassive black hole of M 87. I. Observations,
  calibration, imaging, and analysis}},\ }\href
  {https://doi.org/10.1051/0004-6361/202347932} {\bibfield  {journal} {\bibinfo
   {journal} {{A\&A}}\ }\textbf {\bibinfo {volume} {681}},\ \bibinfo {eid}
  {A79} (\bibinfo {year} {2024})}\BibitemShut {NoStop}%
\bibitem [{\citenamefont {V\"olkel}\ \emph {et~al.}(2021)\citenamefont
  {V\"olkel}, \citenamefont {Barausse}, \citenamefont {Franchini},\ and\
  \citenamefont {Broderick}}]{Volkel:2020xlc}%
  \BibitemOpen
  \bibfield  {author} {\bibinfo {author} {\bibfnamefont {S.~H.}\ \bibnamefont
  {V\"olkel}}, \bibinfo {author} {\bibfnamefont {E.}~\bibnamefont {Barausse}},
  \bibinfo {author} {\bibfnamefont {N.}~\bibnamefont {Franchini}},\ and\
  \bibinfo {author} {\bibfnamefont {A.~E.}\ \bibnamefont {Broderick}},\
  }\bibfield  {title} {\bibinfo {title} {{EHT tests of the strong-field regime
  of general relativity}},\ }\href {https://doi.org/10.1088/1361-6382/ac27ed}
  {\bibfield  {journal} {\bibinfo  {journal} {Class. Quant. Grav.}\ }\textbf
  {\bibinfo {volume} {38}},\ \bibinfo {pages} {21LT01} (\bibinfo {year}
  {2021})},\ \Eprint {https://arxiv.org/abs/2011.06812} {arXiv:2011.06812
  [gr-qc]} \BibitemShut {NoStop}%
\bibitem [{\citenamefont {Glampedakis}\ and\ \citenamefont
  {Pappas}(2021)}]{Glampedakis:2021oie}%
  \BibitemOpen
  \bibfield  {author} {\bibinfo {author} {\bibfnamefont {K.}~\bibnamefont
  {Glampedakis}}\ and\ \bibinfo {author} {\bibfnamefont {G.}~\bibnamefont
  {Pappas}},\ }\bibfield  {title} {\bibinfo {title} {{Can supermassive black
  hole shadows test the Kerr metric?}},\ }\href
  {https://doi.org/10.1103/PhysRevD.104.L081503} {\bibfield  {journal}
  {\bibinfo  {journal} {Phys. Rev. D}\ }\textbf {\bibinfo {volume} {104}},\
  \bibinfo {pages} {L081503} (\bibinfo {year} {2021})},\ \Eprint
  {https://arxiv.org/abs/2102.13573} {arXiv:2102.13573 [gr-qc]} \BibitemShut
  {NoStop}%
\bibitem [{\citenamefont {Glampedakis}\ and\ \citenamefont
  {Pappas}(2023)}]{Glampedakis:2023eek}%
  \BibitemOpen
  \bibfield  {author} {\bibinfo {author} {\bibfnamefont {K.}~\bibnamefont
  {Glampedakis}}\ and\ \bibinfo {author} {\bibfnamefont {G.}~\bibnamefont
  {Pappas}},\ }\bibfield  {title} {\bibinfo {title} {{Is a black hole shadow a
  reliable test of the no-hair theorem?}},\ }\href
  {https://doi.org/10.1103/PhysRevD.107.064001} {\bibfield  {journal} {\bibinfo
   {journal} {Phys. Rev. D}\ }\textbf {\bibinfo {volume} {107}},\ \bibinfo
  {pages} {064001} (\bibinfo {year} {2023})},\ \Eprint
  {https://arxiv.org/abs/2302.06140} {arXiv:2302.06140 [gr-qc]} \BibitemShut
  {NoStop}%
\bibitem [{\citenamefont {Gralla}\ \emph {et~al.}(2020)\citenamefont {Gralla},
  \citenamefont {Lupsasca},\ and\ \citenamefont {Marrone}}]{Gralla:2020srx}%
  \BibitemOpen
  \bibfield  {author} {\bibinfo {author} {\bibfnamefont {S.~E.}\ \bibnamefont
  {Gralla}}, \bibinfo {author} {\bibfnamefont {A.}~\bibnamefont {Lupsasca}},\
  and\ \bibinfo {author} {\bibfnamefont {D.~P.}\ \bibnamefont {Marrone}},\
  }\bibfield  {title} {\bibinfo {title} {{The shape of the black hole photon
  ring: A precise test of strong-field general relativity}},\ }\href
  {https://doi.org/10.1103/PhysRevD.102.124004} {\bibfield  {journal} {\bibinfo
   {journal} {Phys. Rev. D}\ }\textbf {\bibinfo {volume} {102}},\ \bibinfo
  {pages} {124004} (\bibinfo {year} {2020})},\ \Eprint
  {https://arxiv.org/abs/2008.03879} {arXiv:2008.03879 [gr-qc]} \BibitemShut
  {NoStop}%
\bibitem [{\citenamefont {Younsi}\ \emph {et~al.}(2023)\citenamefont {Younsi},
  \citenamefont {Psaltis},\ and\ \citenamefont {\"Ozel}}]{Younsi:2021dxe}%
  \BibitemOpen
  \bibfield  {author} {\bibinfo {author} {\bibfnamefont {Z.}~\bibnamefont
  {Younsi}}, \bibinfo {author} {\bibfnamefont {D.}~\bibnamefont {Psaltis}},\
  and\ \bibinfo {author} {\bibfnamefont {F.}~\bibnamefont {\"Ozel}},\
  }\bibfield  {title} {\bibinfo {title} {{Black Hole Images as Tests of General
  Relativity: Effects of Spacetime Geometry}},\ }\href
  {https://doi.org/10.3847/1538-4357/aca58a} {\bibfield  {journal} {\bibinfo
  {journal} {Astrophys. J.}\ }\textbf {\bibinfo {volume} {942}},\ \bibinfo
  {pages} {47} (\bibinfo {year} {2023})},\ \Eprint
  {https://arxiv.org/abs/2111.01752} {arXiv:2111.01752 [astro-ph.HE]}
  \BibitemShut {NoStop}%
\bibitem [{\citenamefont {Bauer}\ \emph {et~al.}(2022)\citenamefont {Bauer},
  \citenamefont {C\'ardenas-Avenda\~no}, \citenamefont {Gammie},\ and\
  \citenamefont {Yunes}}]{Bauer:2021atk}%
  \BibitemOpen
  \bibfield  {author} {\bibinfo {author} {\bibfnamefont {A.~M.}\ \bibnamefont
  {Bauer}}, \bibinfo {author} {\bibfnamefont {A.}~\bibnamefont
  {C\'ardenas-Avenda\~no}}, \bibinfo {author} {\bibfnamefont {C.~F.}\
  \bibnamefont {Gammie}},\ and\ \bibinfo {author} {\bibfnamefont
  {N.}~\bibnamefont {Yunes}},\ }\bibfield  {title} {\bibinfo {title}
  {{Spherical Accretion in Alternative Theories of Gravity}},\ }\href
  {https://doi.org/10.3847/1538-4357/ac3a03} {\bibfield  {journal} {\bibinfo
  {journal} {Astrophys. J.}\ }\textbf {\bibinfo {volume} {925}},\ \bibinfo
  {pages} {119} (\bibinfo {year} {2022})},\ \Eprint
  {https://arxiv.org/abs/2111.02178} {arXiv:2111.02178 [gr-qc]} \BibitemShut
  {NoStop}%
\bibitem [{\citenamefont {Gralla}(2020)}]{Gralla:2020nwp}%
  \BibitemOpen
  \bibfield  {author} {\bibinfo {author} {\bibfnamefont {S.~E.}\ \bibnamefont
  {Gralla}},\ }\bibfield  {title} {\bibinfo {title} {{Measuring the shape of a
  black hole photon ring}},\ }\href
  {https://doi.org/10.1103/PhysRevD.102.044017} {\bibfield  {journal} {\bibinfo
   {journal} {Phys. Rev. D}\ }\textbf {\bibinfo {volume} {102}},\ \bibinfo
  {pages} {044017} (\bibinfo {year} {2020})},\ \Eprint
  {https://arxiv.org/abs/2005.03856} {arXiv:2005.03856 [astro-ph.HE]}
  \BibitemShut {NoStop}%
\bibitem [{\citenamefont {Lima}\ \emph {et~al.}(2021)\citenamefont {Lima},
  \citenamefont {Crispino}, \citenamefont {Cunha},\ and\ \citenamefont
  {Herdeiro}}]{Lima:2021las}%
  \BibitemOpen
  \bibfield  {author} {\bibinfo {author} {\bibfnamefont {H.~C.~D.}\
  \bibnamefont {Lima}, \bibfnamefont {Junior.}}, \bibinfo {author}
  {\bibfnamefont {L.~C.~B.}\ \bibnamefont {Crispino}}, \bibinfo {author}
  {\bibfnamefont {P.~V.~P.}\ \bibnamefont {Cunha}},\ and\ \bibinfo {author}
  {\bibfnamefont {C.~A.~R.}\ \bibnamefont {Herdeiro}},\ }\bibfield  {title}
  {\bibinfo {title} {{Can different black holes cast the same shadow?}},\
  }\href {https://doi.org/10.1103/PhysRevD.103.084040} {\bibfield  {journal}
  {\bibinfo  {journal} {Phys. Rev. D}\ }\textbf {\bibinfo {volume} {103}},\
  \bibinfo {pages} {084040} (\bibinfo {year} {2021})},\ \Eprint
  {https://arxiv.org/abs/2102.07034} {arXiv:2102.07034 [gr-qc]} \BibitemShut
  {NoStop}%
\bibitem [{\citenamefont {Medeiros}\ \emph {et~al.}(2020)\citenamefont
  {Medeiros}, \citenamefont {Psaltis},\ and\ \citenamefont
  {\"Ozel}}]{Medeiros:2019cde}%
  \BibitemOpen
  \bibfield  {author} {\bibinfo {author} {\bibfnamefont {L.}~\bibnamefont
  {Medeiros}}, \bibinfo {author} {\bibfnamefont {D.}~\bibnamefont {Psaltis}},\
  and\ \bibinfo {author} {\bibfnamefont {F.}~\bibnamefont {\"Ozel}},\
  }\bibfield  {title} {\bibinfo {title} {{A Parametric model for the shapes of
  black-hole shadows in non-Kerr spacetimes}},\ }\href
  {https://doi.org/10.3847/1538-4357/ab8bd1} {\bibfield  {journal} {\bibinfo
  {journal} {Astrophys. J.}\ }\textbf {\bibinfo {volume} {896}},\ \bibinfo
  {pages} {7} (\bibinfo {year} {2020})}\BibitemShut {NoStop}%
\bibitem [{\citenamefont {Gralla}\ \emph {et~al.}(2019)\citenamefont {Gralla},
  \citenamefont {Holz},\ and\ \citenamefont {Wald}}]{Gralla:2019xty}%
  \BibitemOpen
  \bibfield  {author} {\bibinfo {author} {\bibfnamefont {S.~E.}\ \bibnamefont
  {Gralla}}, \bibinfo {author} {\bibfnamefont {D.~E.}\ \bibnamefont {Holz}},\
  and\ \bibinfo {author} {\bibfnamefont {R.~M.}\ \bibnamefont {Wald}},\
  }\bibfield  {title} {\bibinfo {title} {{Black Hole Shadows, Photon Rings, and
  Lensing Rings}},\ }\href {https://doi.org/10.1103/PhysRevD.100.024018}
  {\bibfield  {journal} {\bibinfo  {journal} {Phys. Rev. D}\ }\textbf {\bibinfo
  {volume} {100}},\ \bibinfo {pages} {024018} (\bibinfo {year} {2019})},\
  \Eprint {https://arxiv.org/abs/1906.00873} {arXiv:1906.00873 [astro-ph.HE]}
  \BibitemShut {NoStop}%
\bibitem [{\citenamefont {Johannsen}\ and\ \citenamefont
  {Psaltis}(2010)}]{Johannsen_2010}%
  \BibitemOpen
  \bibfield  {author} {\bibinfo {author} {\bibfnamefont {T.}~\bibnamefont
  {Johannsen}}\ and\ \bibinfo {author} {\bibfnamefont {D.}~\bibnamefont
  {Psaltis}},\ }\bibfield  {title} {\bibinfo {title} {{TESTING} {THE}
  {NO}-{HAIR} {THEOREM} {WITH} {OBSERVATIONS} {IN} {THE} {ELECTROMAGNETIC}
  {SPECTRUM}. {II}. {BLACK} {HOLE} {IMAGES}},\ }\href
  {https://doi.org/10.1088/0004-637x/718/1/446} {\bibfield  {journal} {\bibinfo
   {journal} {The Astrophysical Journal}\ }\textbf {\bibinfo {volume} {718}},\
  \bibinfo {pages} {446} (\bibinfo {year} {2010})}\BibitemShut {NoStop}%
\bibitem [{\citenamefont {Olmo}\ \emph {et~al.}(2023)\citenamefont {Olmo},
  \citenamefont {Rosa}, \citenamefont {Rubiera-Garcia},\ and\ \citenamefont
  {Saez-Chillon~Gomez}}]{Olmo:2023lil}%
  \BibitemOpen
  \bibfield  {author} {\bibinfo {author} {\bibfnamefont {G.~J.}\ \bibnamefont
  {Olmo}}, \bibinfo {author} {\bibfnamefont {J.~L.}\ \bibnamefont {Rosa}},
  \bibinfo {author} {\bibfnamefont {D.}~\bibnamefont {Rubiera-Garcia}},\ and\
  \bibinfo {author} {\bibfnamefont {D.}~\bibnamefont {Saez-Chillon~Gomez}},\
  }\bibfield  {title} {\bibinfo {title} {{Shadows and photon rings of regular
  black holes and geonic horizonless compact objects}},\ }\href
  {https://doi.org/10.1088/1361-6382/aceacd} {\bibfield  {journal} {\bibinfo
  {journal} {Class. Quant. Grav.}\ }\textbf {\bibinfo {volume} {40}},\ \bibinfo
  {pages} {174002} (\bibinfo {year} {2023})},\ \Eprint
  {https://arxiv.org/abs/2302.12064} {arXiv:2302.12064 [gr-qc]} \BibitemShut
  {NoStop}%
\bibitem [{\citenamefont {Akiyama}\ \emph
  {et~al.}(2021{\natexlab{a}})\citenamefont {Akiyama} \emph
  {et~al.}}]{EventHorizonTelescope:2021bee}%
  \BibitemOpen
  \bibfield  {author} {\bibinfo {author} {\bibfnamefont {K.}~\bibnamefont
  {Akiyama}} \emph {et~al.} (\bibinfo {collaboration} {Event Horizon
  Telescope}),\ }\bibfield  {title} {\bibinfo {title} {{First M87 Event Horizon
  Telescope Results. VII. Polarization of the Ring}},\ }\href
  {https://doi.org/10.3847/2041-8213/abe71d} {\bibfield  {journal} {\bibinfo
  {journal} {Astrophys. J. Lett.}\ }\textbf {\bibinfo {volume} {910}},\
  \bibinfo {pages} {L12} (\bibinfo {year} {2021}{\natexlab{a}})},\ \Eprint
  {https://arxiv.org/abs/2105.01169} {arXiv:2105.01169 [astro-ph.HE]}
  \BibitemShut {NoStop}%
\bibitem [{\citenamefont {Akiyama}\ \emph
  {et~al.}(2021{\natexlab{b}})\citenamefont {Akiyama} \emph
  {et~al.}}]{EventHorizonTelescope:2021srq}%
  \BibitemOpen
  \bibfield  {author} {\bibinfo {author} {\bibfnamefont {K.}~\bibnamefont
  {Akiyama}} \emph {et~al.} (\bibinfo {collaboration} {Event Horizon
  Telescope}),\ }\bibfield  {title} {\bibinfo {title} {{First M87 Event Horizon
  Telescope Results. VIII. Magnetic Field Structure near The Event Horizon}},\
  }\href {https://doi.org/10.3847/2041-8213/abe4de} {\bibfield  {journal}
  {\bibinfo  {journal} {Astrophys. J. Lett.}\ }\textbf {\bibinfo {volume}
  {910}},\ \bibinfo {pages} {L13} (\bibinfo {year} {2021}{\natexlab{b}})},\
  \Eprint {https://arxiv.org/abs/2105.01173} {arXiv:2105.01173 [astro-ph.HE]}
  \BibitemShut {NoStop}%
\bibitem [{\citenamefont {Akiyama}\ \emph {et~al.}(2023)\citenamefont {Akiyama}
  \emph {et~al.}}]{EventHorizonTelescope:2023gtd}%
  \BibitemOpen
  \bibfield  {author} {\bibinfo {author} {\bibfnamefont {K.}~\bibnamefont
  {Akiyama}} \emph {et~al.} (\bibinfo {collaboration} {Event Horizon
  Telescope}),\ }\bibfield  {title} {\bibinfo {title} {{First M87 Event Horizon
  Telescope Results. IX. Detection of Near-horizon Circular Polarization}},\
  }\href {https://doi.org/10.3847/2041-8213/acff70} {\bibfield  {journal}
  {\bibinfo  {journal} {Astrophys. J. Lett.}\ }\textbf {\bibinfo {volume}
  {957}},\ \bibinfo {pages} {L20} (\bibinfo {year} {2023})},\ \Eprint
  {https://arxiv.org/abs/2311.10976} {arXiv:2311.10976 [astro-ph.HE]}
  \BibitemShut {NoStop}%
\bibitem [{\citenamefont {Ozel}\ \emph {et~al.}(2022)\citenamefont {Ozel},
  \citenamefont {Psaltis},\ and\ \citenamefont {Younsi}}]{Ozel:2021ayr}%
  \BibitemOpen
  \bibfield  {author} {\bibinfo {author} {\bibfnamefont {F.}~\bibnamefont
  {Ozel}}, \bibinfo {author} {\bibfnamefont {D.}~\bibnamefont {Psaltis}},\ and\
  \bibinfo {author} {\bibfnamefont {Z.}~\bibnamefont {Younsi}},\ }\bibfield
  {title} {\bibinfo {title} {{Black Hole Images as Tests of General Relativity:
  Effects of Plasma Physics}},\ }\href
  {https://doi.org/10.3847/1538-4357/ac9fcb} {\bibfield  {journal} {\bibinfo
  {journal} {Astrophys. J.}\ }\textbf {\bibinfo {volume} {941}},\ \bibinfo
  {pages} {88} (\bibinfo {year} {2022})},\ \Eprint
  {https://arxiv.org/abs/2111.01123} {arXiv:2111.01123 [astro-ph.HE]}
  \BibitemShut {NoStop}%
\bibitem [{\citenamefont {Lara}\ \emph {et~al.}(2021)\citenamefont {Lara},
  \citenamefont {V\"olkel},\ and\ \citenamefont {Barausse}}]{Lara:2021zth}%
  \BibitemOpen
  \bibfield  {author} {\bibinfo {author} {\bibfnamefont {G.}~\bibnamefont
  {Lara}}, \bibinfo {author} {\bibfnamefont {S.~H.}\ \bibnamefont {V\"olkel}},\
  and\ \bibinfo {author} {\bibfnamefont {E.}~\bibnamefont {Barausse}},\
  }\bibfield  {title} {\bibinfo {title} {{Separating astrophysics and geometry
  in black hole images}},\ }\href {https://doi.org/10.1103/PhysRevD.104.124041}
  {\bibfield  {journal} {\bibinfo  {journal} {Phys. Rev. D}\ }\textbf {\bibinfo
  {volume} {104}},\ \bibinfo {pages} {124041} (\bibinfo {year} {2021})},\
  \Eprint {https://arxiv.org/abs/2110.00026} {arXiv:2110.00026 [gr-qc]}
  \BibitemShut {NoStop}%
\bibitem [{\citenamefont {Fernandes}\ \emph {et~al.}(2024)\citenamefont
  {Fernandes}, \citenamefont {Burrage}, \citenamefont {Eichhorn},\ and\
  \citenamefont {Sotiriou}}]{Fernandes:2024ztk}%
  \BibitemOpen
  \bibfield  {author} {\bibinfo {author} {\bibfnamefont {P.~G.~S.}\
  \bibnamefont {Fernandes}}, \bibinfo {author} {\bibfnamefont {C.}~\bibnamefont
  {Burrage}}, \bibinfo {author} {\bibfnamefont {A.}~\bibnamefont {Eichhorn}},\
  and\ \bibinfo {author} {\bibfnamefont {T.~P.}\ \bibnamefont {Sotiriou}},\
  }\bibfield  {title} {\bibinfo {title} {{Shadows and properties of
  spin-induced scalarized black holes with and without a Ricci coupling}},\
  }\href {https://doi.org/10.1103/PhysRevD.109.104033} {\bibfield  {journal}
  {\bibinfo  {journal} {Phys. Rev. D}\ }\textbf {\bibinfo {volume} {109}},\
  \bibinfo {pages} {104033} (\bibinfo {year} {2024})},\ \Eprint
  {https://arxiv.org/abs/2403.14596} {arXiv:2403.14596 [gr-qc]} \BibitemShut
  {NoStop}%
\bibitem [{\citenamefont {Gyulchev}\ \emph {et~al.}(2024)\citenamefont
  {Gyulchev}, \citenamefont {Roy}, \citenamefont {Collodel}, \citenamefont
  {Nedkova}, \citenamefont {Yazadjiev},\ and\ \citenamefont
  {Doneva}}]{Gyulchev:2024iel}%
  \BibitemOpen
  \bibfield  {author} {\bibinfo {author} {\bibfnamefont {G.~N.}\ \bibnamefont
  {Gyulchev}}, \bibinfo {author} {\bibfnamefont {A.}~\bibnamefont {Roy}},
  \bibinfo {author} {\bibfnamefont {L.~G.}\ \bibnamefont {Collodel}}, \bibinfo
  {author} {\bibfnamefont {P.~G.}\ \bibnamefont {Nedkova}}, \bibinfo {author}
  {\bibfnamefont {S.~S.}\ \bibnamefont {Yazadjiev}},\ and\ \bibinfo {author}
  {\bibfnamefont {D.~D.}\ \bibnamefont {Doneva}},\ }\bibfield  {title}
  {\bibinfo {title} {{Shadows of rotating hairy Kerr black holes coupled to
  time periodic scalar fields with a nonflat target space}},\ }\href
  {https://doi.org/10.1103/PhysRevD.109.104051} {\bibfield  {journal} {\bibinfo
   {journal} {Phys. Rev. D}\ }\textbf {\bibinfo {volume} {109}},\ \bibinfo
  {pages} {104051} (\bibinfo {year} {2024})},\ \Eprint
  {https://arxiv.org/abs/2402.08469} {arXiv:2402.08469 [gr-qc]} \BibitemShut
  {NoStop}%
\bibitem [{\citenamefont {Deliyski}\ \emph {et~al.}(2025)\citenamefont
  {Deliyski}, \citenamefont {Gyulchev}, \citenamefont {Nedkova},\ and\
  \citenamefont {Yazadjiev}}]{DeliyskiPhysRevD.111.064068}%
  \BibitemOpen
  \bibfield  {author} {\bibinfo {author} {\bibfnamefont {V.}~\bibnamefont
  {Deliyski}}, \bibinfo {author} {\bibfnamefont {G.}~\bibnamefont {Gyulchev}},
  \bibinfo {author} {\bibfnamefont {P.}~\bibnamefont {Nedkova}},\ and\ \bibinfo
  {author} {\bibfnamefont {S.}~\bibnamefont {Yazadjiev}},\ }\bibfield  {title}
  {\bibinfo {title} {Observing naked singularities with the present and
  next-generation event horizon telescope},\ }\href
  {https://doi.org/10.1103/PhysRevD.111.064068} {\bibfield  {journal} {\bibinfo
   {journal} {Phys. Rev. D}\ }\textbf {\bibinfo {volume} {111}},\ \bibinfo
  {pages} {064068} (\bibinfo {year} {2025})}\BibitemShut {NoStop}%
\bibitem [{\citenamefont {Ben~Achour}\ \emph {et~al.}(2025)\citenamefont
  {Ben~Achour}, \citenamefont {Gourgoulhon},\ and\ \citenamefont
  {Roussille}}]{BenAchour:2025uzp}%
  \BibitemOpen
  \bibfield  {author} {\bibinfo {author} {\bibfnamefont {J.}~\bibnamefont
  {Ben~Achour}}, \bibinfo {author} {\bibfnamefont {E.}~\bibnamefont
  {Gourgoulhon}},\ and\ \bibinfo {author} {\bibfnamefont {H.}~\bibnamefont
  {Roussille}},\ }\bibfield  {title} {\bibinfo {title} {{Black hole photon ring
  beyond General Relativity: an integrable parametrization}},\ }\href@noop {}
  {\  (\bibinfo {year} {2025})},\ \Eprint {https://arxiv.org/abs/2506.09882}
  {arXiv:2506.09882 [gr-qc]} \BibitemShut {NoStop}%
\bibitem [{\citenamefont {Kostaros}\ and\ \citenamefont
  {Pappas}(2022)}]{Kostaros_2022}%
  \BibitemOpen
  \bibfield  {author} {\bibinfo {author} {\bibfnamefont {K.}~\bibnamefont
  {Kostaros}}\ and\ \bibinfo {author} {\bibfnamefont {G.}~\bibnamefont
  {Pappas}},\ }\bibfield  {title} {\bibinfo {title} {Chaotic photon orbits and
  shadows of a non-kerr object described by the hartle–thorne spacetime},\
  }\href {https://doi.org/10.1088/1361-6382/ac7028} {\bibfield  {journal}
  {\bibinfo  {journal} {Classical and Quantum Gravity}\ }\textbf {\bibinfo
  {volume} {39}},\ \bibinfo {pages} {134001} (\bibinfo {year}
  {2022})}\BibitemShut {NoStop}%
\bibitem [{\citenamefont {Kostaros}\ \emph {et~al.}(2024)\citenamefont
  {Kostaros}, \citenamefont {Papadopoulos},\ and\ \citenamefont
  {Pappas}}]{Kostaros:2024vbn}%
  \BibitemOpen
  \bibfield  {author} {\bibinfo {author} {\bibfnamefont {K.}~\bibnamefont
  {Kostaros}}, \bibinfo {author} {\bibfnamefont {P.}~\bibnamefont
  {Papadopoulos}},\ and\ \bibinfo {author} {\bibfnamefont {G.}~\bibnamefont
  {Pappas}},\ }\bibfield  {title} {\bibinfo {title} {{Fractal signatures of
  non-Kerr spacetimes in the shadow of light-ring bifurcations}},\ }\href
  {https://doi.org/10.1103/PhysRevD.110.024001} {\bibfield  {journal} {\bibinfo
   {journal} {Phys. Rev. D}\ }\textbf {\bibinfo {volume} {110}},\ \bibinfo
  {pages} {024001} (\bibinfo {year} {2024})},\ \Eprint
  {https://arxiv.org/abs/2405.09653} {arXiv:2405.09653 [gr-qc]} \BibitemShut
  {NoStop}%
\bibitem [{\citenamefont {Papadopoulos}\ \emph {et~al.}()\citenamefont
  {Papadopoulos}, \citenamefont {Kostaros}, \citenamefont {Pappas},\ and\
  \citenamefont {Thi}}]{Astro_paper}%
  \BibitemOpen
  \bibfield  {author} {\bibinfo {author} {\bibfnamefont {P.}~\bibnamefont
  {Papadopoulos}}, \bibinfo {author} {\bibfnamefont {K.}~\bibnamefont
  {Kostaros}}, \bibinfo {author} {\bibfnamefont {G.}~\bibnamefont {Pappas}},\
  and\ \bibinfo {author} {\bibfnamefont {W.-F.}\ \bibnamefont {Thi}},\
  }\bibfield  {title} {\bibinfo {title} {{Spectral line illumination of Black
  Holes: a new class of observational tests for Kerr Spacetime signatures}},\
  }\href@noop {} {\ }\Eprint {https://arxiv.org/abs/in preparation} {in
  preparation} \BibitemShut {NoStop}%
\bibitem [{\citenamefont {Thi}\ and\ \citenamefont
  {Papadopoulos}(2024)}]{Thi:2024dny}%
  \BibitemOpen
  \bibfield  {author} {\bibinfo {author} {\bibfnamefont {W.-F.}\ \bibnamefont
  {Thi}}\ and\ \bibinfo {author} {\bibfnamefont {P.}~\bibnamefont
  {Papadopoulos}},\ }\bibfield  {title} {\bibinfo {title} {{The neutral gas
  phase nearest to supermassive black holes - Massive neutral-atom- and
  molecule-rich broad line regions in active galactic nuclei}},\ }\href
  {https://doi.org/10.1051/0004-6361/202449905} {\bibfield  {journal} {\bibinfo
   {journal} {Astron. Astrophys.}\ }\textbf {\bibinfo {volume} {688}},\
  \bibinfo {pages} {L20} (\bibinfo {year} {2024})},\ \Eprint
  {https://arxiv.org/abs/2407.07574} {arXiv:2407.07574 [astro-ph.GA]}
  \BibitemShut {NoStop}%
\bibitem [{\citenamefont {Thi}\ \emph {et~al.}()\citenamefont {Thi},
  \citenamefont {Papadopoulos}, \citenamefont {Kostaros},\ and\ \citenamefont
  {Pappas}}]{Chem_paper}%
  \BibitemOpen
  \bibfield  {author} {\bibinfo {author} {\bibfnamefont {W.-F.}\ \bibnamefont
  {Thi}}, \bibinfo {author} {\bibfnamefont {P.}~\bibnamefont {Papadopoulos}},
  \bibinfo {author} {\bibfnamefont {K.}~\bibnamefont {Kostaros}},\ and\
  \bibinfo {author} {\bibfnamefont {G.}~\bibnamefont {Pappas}},\ }\href@noop {}
  {\ }\Eprint {https://arxiv.org/abs/in preparation} {in preparation}
  \BibitemShut {NoStop}%
\bibitem [{\citenamefont {{Seyfert, C. K.}}(1943)}]{Seyf43}%
  \BibitemOpen
  \bibfield  {author} {\bibinfo {author} {\bibnamefont {{Seyfert, C. K.}}},\
  }\bibfield  {title} {\bibinfo {title} {{Nuclear Emission in Spiral
  Nebulae}},\ }\href@noop {} {\bibfield  {journal} {\bibinfo  {journal} {The
  Astrophysical Journal}\ }\textbf {\bibinfo {volume} {97}},\ \bibinfo {pages}
  {28} (\bibinfo {year} {1943})}\BibitemShut {NoStop}%
\bibitem [{\citenamefont {{Baldwin, J. A., Ferland, G. J., Korista, K. T.,
  Hamann, F. \& Dietrich, M.}}(2003)}]{Bald03}%
  \BibitemOpen
  \bibfield  {author} {\bibinfo {author} {\bibnamefont {{Baldwin, J. A.,
  Ferland, G. J., Korista, K. T., Hamann, F. \& Dietrich, M.}}},\ }\bibfield
  {title} {\bibinfo {title} {{The mass of Quasar Broad Emission line
  regions}},\ }\href@noop {} {\bibfield  {journal} {\bibinfo  {journal} {\apj}\
  }\textbf {\bibinfo {volume} {582}},\ \bibinfo {pages} {590} (\bibinfo {year}
  {2003})}\BibitemShut {NoStop}%
\bibitem [{\citenamefont {{Peterson, B. M.}}(2006)}]{Pet06}%
  \BibitemOpen
  \bibfield  {author} {\bibinfo {author} {\bibnamefont {{Peterson, B. M.}}},\
  }\href@noop {} {\emph {\bibinfo {title} {{Physics of Active Galactic Nuclei
  at all scales: The Broad Line Region in Active Galactic Nuclei}}}}\ (\bibinfo
   {publisher} {Springer Berlin, Heidelberg},\ \bibinfo {year} {2006})\ pp.\
  \bibinfo {pages} {77--100}\BibitemShut {NoStop}%
\bibitem [{\citenamefont {Gaskell}(2009)}]{Gas09}%
  \BibitemOpen
  \bibfield  {author} {\bibinfo {author} {\bibfnamefont {C.~M.}\ \bibnamefont
  {Gaskell}},\ }\bibfield  {title} {\bibinfo {title} {{What broad emission
  lines tell us about how active galactic nuclei work}},\ }\href
  {https://doi.org/https://doi.org/10.1016/j.newar.2009.09.006} {\bibfield
  {journal} {\bibinfo  {journal} {New Astronomy Reviews}\ }\textbf {\bibinfo
  {volume} {53}},\ \bibinfo {pages} {140} (\bibinfo {year} {2009})}\BibitemShut
  {NoStop}%
\bibitem [{\citenamefont {{Czerny, B., \& Hryniewicz, K.}}(2011)}]{Czer11}%
  \BibitemOpen
  \bibfield  {author} {\bibinfo {author} {\bibnamefont {{Czerny, B., \&
  Hryniewicz, K.}}},\ }\bibfield  {title} {\bibinfo {title} {{The origin of the
  broad line region in active galactic nuclei}},\ }\href@noop {} {\bibfield
  {journal} {\bibinfo  {journal} {Astronomy \& Astrophysics Letters}\ }\textbf
  {\bibinfo {volume} {525}},\ \bibinfo {pages} {L8} (\bibinfo {year}
  {2011})}\BibitemShut {NoStop}%
\bibitem [{\citenamefont {{Czerny, B.}}(2019)}]{Czer19}%
  \BibitemOpen
  \bibfield  {author} {\bibinfo {author} {\bibnamefont {{Czerny, B.}}},\
  }\bibfield  {title} {\bibinfo {title} {{Modelling broad emission lines in
  active galactic nuclei}},\ }\href
  {https://doi.org/https://doi.org/10.1515/astro-2019-0018} {\bibfield
  {journal} {\bibinfo  {journal} {Open Astronomy}\ }\textbf {\bibinfo {volume}
  {28}},\ \bibinfo {pages} {200} (\bibinfo {year} {2019})}\BibitemShut
  {NoStop}%
\bibitem [{\citenamefont {{Baskin, A., Laor, A., \& Stern, J.
  }}(2014)}]{Bask14}%
  \BibitemOpen
  \bibfield  {author} {\bibinfo {author} {\bibnamefont {{Baskin, A., Laor, A.,
  \& Stern, J. }}},\ }\bibfield  {title} {\bibinfo {title} {{Radiative pressure
  confinment - II. Application to the broad line region in active galactic
  nuclei}},\ }\href@noop {} {\bibfield  {journal} {\bibinfo  {journal} {Monthly
  Notices of the Royal Astronomical Society}\ }\textbf {\bibinfo {volume}
  {438}},\ \bibinfo {pages} {604} (\bibinfo {year} {2014})}\BibitemShut
  {NoStop}%
\bibitem [{\citenamefont {{Baskin, A., \& Laor, A.}}(2018)}]{Bask18}%
  \BibitemOpen
  \bibfield  {author} {\bibinfo {author} {\bibnamefont {{Baskin, A., \& Laor,
  A.}}},\ }\bibfield  {title} {\bibinfo {title} {{Dust inflated accretion disk
  as the origin of the broad line region in active galactic nuclei}},\
  }\href@noop {} {\bibfield  {journal} {\bibinfo  {journal} {Monthly Notices of
  the Royal Astronomical Society}\ }\textbf {\bibinfo {volume} {474}},\
  \bibinfo {pages} {1970} (\bibinfo {year} {2018})}\BibitemShut {NoStop}%
\bibitem [{\citenamefont {{Ferland, G. J., Done, C., Jin, C., Landt, H., \&
  Ward, M. J.}}(2020)}]{Fer20}%
  \BibitemOpen
  \bibfield  {author} {\bibinfo {author} {\bibnamefont {{Ferland, G. J., Done,
  C., Jin, C., Landt, H., \& Ward, M. J.}}},\ }\bibfield  {title} {\bibinfo
  {title} {{State of the art AGN SEDs for photoinization models: BLR
  predictions confront the observations}},\ }\href@noop {} {\bibfield
  {journal} {\bibinfo  {journal} {Monthly Notices of the Royal Astronomical
  Society}\ }\textbf {\bibinfo {volume} {494}},\ \bibinfo {pages} {5917}
  (\bibinfo {year} {2020})}\BibitemShut {NoStop}%
\bibitem [{\citenamefont {{Czerny, B., Cao, S., Vikram, K. J. et al.
  }}(2023)}]{Czer23}%
  \BibitemOpen
  \bibfield  {author} {\bibinfo {author} {\bibnamefont {{Czerny, B., Cao, S.,
  Vikram, K. J. et al. }}},\ }\bibfield  {title} {\bibinfo {title} {{Accretion
  disks, quasars and Cosmology: meandering towards understanding}},\
  }\href@noop {} {\bibfield  {journal} {\bibinfo  {journal} {Astrophysics and
  Space Science}\ }\textbf {\bibinfo {volume} {368}},\ \bibinfo {pages} {8}
  (\bibinfo {year} {2023})}\BibitemShut {NoStop}%
\bibitem [{\citenamefont {Kostaros}\ \emph {et~al.}(2025)\citenamefont
  {Kostaros}, \citenamefont {Pappas}, \citenamefont {Papadopoulos},\ and\
  \citenamefont {Thi}}]{PRL_companion}%
  \BibitemOpen
  \bibfield  {author} {\bibinfo {author} {\bibfnamefont {K.}~\bibnamefont
  {Kostaros}}, \bibinfo {author} {\bibfnamefont {G.}~\bibnamefont {Pappas}},
  \bibinfo {author} {\bibfnamefont {P.}~\bibnamefont {Papadopoulos}},\ and\
  \bibinfo {author} {\bibfnamefont {W.-F.}\ \bibnamefont {Thi}},\ }\bibfield
  {title} {\bibinfo {title} {{An Einstein ring fingerprint around SMBHs
  illuminated by BLR spectral lines}},\ }\href@noop {} {\  (\bibinfo {year}
  {2025})},\ \Eprint {https://arxiv.org/abs/2509.08690} {arXiv:2509.08690
  [gr-qc]} \BibitemShut {NoStop}%
\bibitem [{\citenamefont {Storchi-Bergmann}\ \emph {et~al.}(2017)\citenamefont
  {Storchi-Bergmann}, \citenamefont {Schimoia}, \citenamefont {Peterson},
  \citenamefont {Elvis}, \citenamefont {Denney}, \citenamefont {Eracleous},\
  and\ \citenamefont {Nemmen}}]{Storchi_Bergmann_2017}%
  \BibitemOpen
  \bibfield  {author} {\bibinfo {author} {\bibfnamefont {T.}~\bibnamefont
  {Storchi-Bergmann}}, \bibinfo {author} {\bibfnamefont {J.~S.}\ \bibnamefont
  {Schimoia}}, \bibinfo {author} {\bibfnamefont {B.~M.}\ \bibnamefont
  {Peterson}}, \bibinfo {author} {\bibfnamefont {M.}~\bibnamefont {Elvis}},
  \bibinfo {author} {\bibfnamefont {K.~D.}\ \bibnamefont {Denney}}, \bibinfo
  {author} {\bibfnamefont {M.}~\bibnamefont {Eracleous}},\ and\ \bibinfo
  {author} {\bibfnamefont {R.~S.}\ \bibnamefont {Nemmen}},\ }\bibfield  {title}
  {\bibinfo {title} {Double-peaked profiles: Ubiquitous signatures of disks in
  the broad emission lines of active galactic nuclei},\ }\href
  {https://doi.org/10.3847/1538-4357/835/2/236} {\bibfield  {journal} {\bibinfo
   {journal} {The Astrophysical Journal}\ }\textbf {\bibinfo {volume} {835}},\
  \bibinfo {pages} {236} (\bibinfo {year} {2017})}\BibitemShut {NoStop}%
\bibitem [{\citenamefont {{Younsi, Z.}}\ \emph {et~al.}(2012)\citenamefont
  {{Younsi, Z.}}, \citenamefont {{Wu, K.}},\ and\ \citenamefont {{Fuerst, S.
  V.}}}]{Younsi2012}%
  \BibitemOpen
  \bibfield  {author} {\bibinfo {author} {\bibnamefont {{Younsi, Z.}}},
  \bibinfo {author} {\bibnamefont {{Wu, K.}}},\ and\ \bibinfo {author}
  {\bibnamefont {{Fuerst, S. V.}}},\ }\bibfield  {title} {\bibinfo {title}
  {General relativistic radiative transfer: formulation and emission from
  structured tori around black holes},\ }\href
  {https://doi.org/10.1051/0004-6361/201219599} {\bibfield  {journal} {\bibinfo
   {journal} {A\&A}\ }\textbf {\bibinfo {volume} {545}},\ \bibinfo {pages}
  {A13} (\bibinfo {year} {2012})}\BibitemShut {NoStop}%
\bibitem [{\citenamefont {Fuerst}\ and\ \citenamefont
  {Wu}(2004)}]{Fuerst_2004}%
  \BibitemOpen
  \bibfield  {author} {\bibinfo {author} {\bibfnamefont {S.~V.}\ \bibnamefont
  {Fuerst}}\ and\ \bibinfo {author} {\bibfnamefont {K.}~\bibnamefont {Wu}},\
  }\bibfield  {title} {\bibinfo {title} {Radiation transfer of emission lines
  in curved space-time},\ }\href {https://doi.org/10.1051/0004-6361:20035814}
  {\bibfield  {journal} {\bibinfo  {journal} {Astronomy {\&} Astrophysics}\
  }\textbf {\bibinfo {volume} {424}},\ \bibinfo {pages} {733} (\bibinfo {year}
  {2004})}\BibitemShut {NoStop}%
\bibitem [{\citenamefont {Fuerst}\ and\ \citenamefont
  {Wu}(2007)}]{Fuerst_2007}%
  \BibitemOpen
  \bibfield  {author} {\bibinfo {author} {\bibfnamefont {S.~V.}\ \bibnamefont
  {Fuerst}}\ and\ \bibinfo {author} {\bibfnamefont {K.}~\bibnamefont {Wu}},\
  }\bibfield  {title} {\bibinfo {title} {Line emission from optically thick
  relativistic accretion tori},\ }\href
  {https://doi.org/10.1051/0004-6361:20066008} {\bibfield  {journal} {\bibinfo
  {journal} {Astronomy {\&} Astrophysics}\ }\textbf {\bibinfo {volume} {474}},\
  \bibinfo {pages} {55} (\bibinfo {year} {2007})}\BibitemShut {NoStop}%
\bibitem [{\citenamefont {Bardeen}(1973)}]{Bardeen:1973tla}%
  \BibitemOpen
  \bibfield  {author} {\bibinfo {author} {\bibfnamefont {J.~M.}\ \bibnamefont
  {Bardeen}},\ }\bibfield  {title} {\bibinfo {title} {{Timelike and null
  geodesics in the Kerr metric}},\ }\href@noop {} {\bibfield  {journal}
  {\bibinfo  {journal} {Proceedings, Ecole d'Et\'e de Physique Th\'eorique: Les
  Astres Occlus : Les Houches, France, August, 1972, 215-240}\ ,\ \bibinfo
  {pages} {215}} (\bibinfo {year} {1973})}\BibitemShut {NoStop}%
\bibitem [{\citenamefont {{Thorne}}(1974)}]{1974ApJThorne}%
  \BibitemOpen
  \bibfield  {author} {\bibinfo {author} {\bibfnamefont {K.~S.}\ \bibnamefont
  {{Thorne}}},\ }\bibfield  {title} {\bibinfo {title} {{Disk-Accretion onto a
  Black Hole. II. Evolution of the Hole}},\ }\href
  {https://doi.org/10.1086/152991} {\bibfield  {journal} {\bibinfo  {journal}
  {\apj}\ }\textbf {\bibinfo {volume} {191}},\ \bibinfo {pages} {507} (\bibinfo
  {year} {1974})}\BibitemShut {NoStop}%
\bibitem [{\citenamefont {Johannsen}\ and\ \citenamefont
  {Psaltis}(2011)}]{Johannsen:2011dh}%
  \BibitemOpen
  \bibfield  {author} {\bibinfo {author} {\bibfnamefont {T.}~\bibnamefont
  {Johannsen}}\ and\ \bibinfo {author} {\bibfnamefont {D.}~\bibnamefont
  {Psaltis}},\ }\bibfield  {title} {\bibinfo {title} {{A Metric for Rapidly
  Spinning Black Holes Suitable for Strong-Field Tests of the No-Hair
  Theorem}},\ }\href {https://doi.org/10.1103/PhysRevD.83.124015} {\bibfield
  {journal} {\bibinfo  {journal} {Phys. Rev. D}\ }\textbf {\bibinfo {volume}
  {83}},\ \bibinfo {pages} {124015} (\bibinfo {year} {2011})},\ \Eprint
  {https://arxiv.org/abs/1105.3191} {arXiv:1105.3191 [gr-qc]} \BibitemShut
  {NoStop}%
\bibitem [{\citenamefont {{Congdon}}\ and\ \citenamefont
  {{Keeton}}(2018)}]{2018bookCongdon}%
  \BibitemOpen
  \bibfield  {author} {\bibinfo {author} {\bibfnamefont {A.~B.}\ \bibnamefont
  {{Congdon}}}\ and\ \bibinfo {author} {\bibfnamefont {C.~R.}\ \bibnamefont
  {{Keeton}}},\ }\href {https://doi.org/10.1007/978-3-030-02122-1} {\emph
  {\bibinfo {title} {{Principles of Gravitational Lensing: Light Deflection as
  a Probe of Astrophysics and Cosmology}}}}\ (\bibinfo  {publisher} {Springer
  Praxis Books},\ \bibinfo {year} {2018})\BibitemShut {NoStop}%
\bibitem [{\citenamefont {Takahashi}(2004)}]{Takahashi_2004}%
  \BibitemOpen
  \bibfield  {author} {\bibinfo {author} {\bibfnamefont {R.}~\bibnamefont
  {Takahashi}},\ }\bibfield  {title} {\bibinfo {title} {Shapes and positions of
  black hole shadows in accretion disks and spin parameters of black holes},\
  }\href {https://doi.org/10.1086/422403} {\bibfield  {journal} {\bibinfo
  {journal} {The Astrophysical Journal}\ }\textbf {\bibinfo {volume} {611}},\
  \bibinfo {pages} {996} (\bibinfo {year} {2004})}\BibitemShut {NoStop}%
\bibitem [{\citenamefont {Johannsen}(2013)}]{Johannsen:2013vgc}%
  \BibitemOpen
  \bibfield  {author} {\bibinfo {author} {\bibfnamefont {T.}~\bibnamefont
  {Johannsen}},\ }\bibfield  {title} {\bibinfo {title} {{Photon Rings around
  Kerr and Kerr-like Black Holes}},\ }\href
  {https://doi.org/10.1088/0004-637X/777/2/170} {\bibfield  {journal} {\bibinfo
   {journal} {Astrophys. J.}\ }\textbf {\bibinfo {volume} {777}},\ \bibinfo
  {pages} {170} (\bibinfo {year} {2013})},\ \Eprint
  {https://arxiv.org/abs/1501.02814} {arXiv:1501.02814 [astro-ph.HE]}
  \BibitemShut {NoStop}%
\bibitem [{\citenamefont {Broderick}\ \emph {et~al.}(2014)\citenamefont
  {Broderick}, \citenamefont {Johannsen}, \citenamefont {Loeb},\ and\
  \citenamefont {Psaltis}}]{Broderick:2013rlq}%
  \BibitemOpen
  \bibfield  {author} {\bibinfo {author} {\bibfnamefont {A.~E.}\ \bibnamefont
  {Broderick}}, \bibinfo {author} {\bibfnamefont {T.}~\bibnamefont
  {Johannsen}}, \bibinfo {author} {\bibfnamefont {A.}~\bibnamefont {Loeb}},\
  and\ \bibinfo {author} {\bibfnamefont {D.}~\bibnamefont {Psaltis}},\
  }\bibfield  {title} {\bibinfo {title} {{Testing the No-Hair Theorem with
  Event Horizon Telescope Observations of Sagittarius A*}},\ }\href
  {https://doi.org/10.1088/0004-637X/784/1/7} {\bibfield  {journal} {\bibinfo
  {journal} {Astrophys. J.}\ }\textbf {\bibinfo {volume} {784}},\ \bibinfo
  {pages} {7} (\bibinfo {year} {2014})},\ \Eprint
  {https://arxiv.org/abs/1311.5564} {arXiv:1311.5564 [astro-ph.HE]}
  \BibitemShut {NoStop}%
\bibitem [{\citenamefont {Johannsen}(2016)}]{Johannsen:2015mdd}%
  \BibitemOpen
  \bibfield  {author} {\bibinfo {author} {\bibfnamefont {T.}~\bibnamefont
  {Johannsen}},\ }\bibfield  {title} {\bibinfo {title} {{Sgr A* and General
  Relativity}},\ }\href {https://doi.org/10.1088/0264-9381/33/11/113001}
  {\bibfield  {journal} {\bibinfo  {journal} {Class. Quant. Grav.}\ }\textbf
  {\bibinfo {volume} {33}},\ \bibinfo {pages} {113001} (\bibinfo {year}
  {2016})},\ \Eprint {https://arxiv.org/abs/1512.03818} {arXiv:1512.03818
  [astro-ph.GA]} \BibitemShut {NoStop}%
\bibitem [{\citenamefont {Tsupko}(2017)}]{Tsupko2017PhysRevD}%
  \BibitemOpen
  \bibfield  {author} {\bibinfo {author} {\bibfnamefont {O.~Y.}\ \bibnamefont
  {Tsupko}},\ }\bibfield  {title} {\bibinfo {title} {Analytical calculation of
  black hole spin using deformation of the shadow},\ }\href
  {https://doi.org/10.1103/PhysRevD.95.104058} {\bibfield  {journal} {\bibinfo
  {journal} {Phys. Rev. D}\ }\textbf {\bibinfo {volume} {95}},\ \bibinfo
  {pages} {104058} (\bibinfo {year} {2017})}\BibitemShut {NoStop}%
\bibitem [{\citenamefont {Johnson}\ \emph {et~al.}(2020)\citenamefont {Johnson}
  \emph {et~al.}}]{Johnson:2019ljv}%
  \BibitemOpen
  \bibfield  {author} {\bibinfo {author} {\bibfnamefont {M.~D.}\ \bibnamefont
  {Johnson}} \emph {et~al.},\ }\bibfield  {title} {\bibinfo {title} {{Universal
  interferometric signatures of a black hole{\textquoteright}s photon ring}},\
  }\href {https://doi.org/10.1126/sciadv.aaz1310} {\bibfield  {journal}
  {\bibinfo  {journal} {Sci. Adv.}\ }\textbf {\bibinfo {volume} {6}},\ \bibinfo
  {pages} {eaaz1310} (\bibinfo {year} {2020})},\ \Eprint
  {https://arxiv.org/abs/1907.04329} {arXiv:1907.04329 [astro-ph.IM]}
  \BibitemShut {NoStop}%
\bibitem [{\citenamefont {C{\'a}rdenas-Avenda{\~n}o}\ and\ \citenamefont
  {Lupsasca}(2023)}]{Cardenas-Avendano:2023dzo}%
  \BibitemOpen
  \bibfield  {author} {\bibinfo {author} {\bibfnamefont {A.}~\bibnamefont
  {C{\'a}rdenas-Avenda{\~n}o}}\ and\ \bibinfo {author} {\bibfnamefont
  {A.}~\bibnamefont {Lupsasca}},\ }\bibfield  {title} {\bibinfo {title}
  {{Prediction for the interferometric shape of the first black hole photon
  ring}},\ }\href {https://doi.org/10.1103/PhysRevD.108.064043} {\bibfield
  {journal} {\bibinfo  {journal} {Phys. Rev. D}\ }\textbf {\bibinfo {volume}
  {108}},\ \bibinfo {pages} {064043} (\bibinfo {year} {2023})},\ \Eprint
  {https://arxiv.org/abs/2305.12956} {arXiv:2305.12956 [gr-qc]} \BibitemShut
  {NoStop}%
\bibitem [{\citenamefont {Kumar~Walia}\ \emph {et~al.}(2025)\citenamefont
  {Kumar~Walia}, \citenamefont {Kocherlakota}, \citenamefont {Chang},\ and\
  \citenamefont {Salehi}}]{KumarWalia:2024omf}%
  \BibitemOpen
  \bibfield  {author} {\bibinfo {author} {\bibfnamefont {R.}~\bibnamefont
  {Kumar~Walia}}, \bibinfo {author} {\bibfnamefont {P.}~\bibnamefont
  {Kocherlakota}}, \bibinfo {author} {\bibfnamefont {D.~O.}\ \bibnamefont
  {Chang}},\ and\ \bibinfo {author} {\bibfnamefont {K.}~\bibnamefont
  {Salehi}},\ }\bibfield  {title} {\bibinfo {title} {{Spacetime measurements
  with the photon ring}},\ }\href {https://doi.org/10.1103/PhysRevD.111.104074}
  {\bibfield  {journal} {\bibinfo  {journal} {Phys. Rev. D}\ }\textbf {\bibinfo
  {volume} {111}},\ \bibinfo {pages} {104074} (\bibinfo {year} {2025})},\
  \Eprint {https://arxiv.org/abs/2411.15119} {arXiv:2411.15119 [gr-qc]}
  \BibitemShut {NoStop}%
\bibitem [{\citenamefont {{Koutsantoniou}}(2022)}]{Koutsantoniou:2022fev}%
  \BibitemOpen
  \bibfield  {author} {\bibinfo {author} {\bibfnamefont {L.~E.}\ \bibnamefont
  {{Koutsantoniou}}},\ }\bibfield  {title} {\bibinfo {title} {{Algorithms and
  radiation dynamics for the vicinity of black holes. I. Methods and codes}},\
  }\href {https://doi.org/10.1051/0004-6361/202140682} {\bibfield  {journal}
  {\bibinfo  {journal} {A\&A}\ }\textbf {\bibinfo {volume} {657}},\ \bibinfo
  {eid} {A32} (\bibinfo {year} {2022})},\ \Eprint
  {https://arxiv.org/abs/2212.01532} {arXiv:2212.01532 [astro-ph.IM]}
  \BibitemShut {NoStop}%
\bibitem [{\citenamefont {{Koutsantoniou}}(2023)}]{Koutsantoniou:2023anr}%
  \BibitemOpen
  \bibfield  {author} {\bibinfo {author} {\bibfnamefont {L.~E.}\ \bibnamefont
  {{Koutsantoniou}}},\ }\bibfield  {title} {\bibinfo {title} {{Algorithms and
  radiation dynamics for the vicinity of black holes. II. Results}},\ }\href
  {https://doi.org/10.1051/0004-6361/202244319} {\bibfield  {journal} {\bibinfo
   {journal} {A\&A}\ }\textbf {\bibinfo {volume} {671}},\ \bibinfo {eid} {A131}
  (\bibinfo {year} {2023})}\BibitemShut {NoStop}%
\bibitem [{\citenamefont {Tim~Cornwell}\ and\ \citenamefont
  {Briggs}(1999{\natexlab{a}})}]{Corn99_N1}%
  \BibitemOpen
  \bibfield  {author} {\bibinfo {author} {\bibfnamefont {R.~B.}\ \bibnamefont
  {Tim~Cornwell}}\ and\ \bibinfo {author} {\bibfnamefont {D.~S.}\ \bibnamefont
  {Briggs}},\ }\href@noop {} {\emph {\bibinfo {title} {Synthesis imaging in
  Radio Astronomy II}}}\ (\bibinfo  {publisher} {Astronomical Society of the
  Pacific},\ \bibinfo {year} {1999})\ pp.\ \bibinfo {pages}
  {151--170}\BibitemShut {NoStop}%
\bibitem [{\citenamefont {Tim~Cornwell}\ and\ \citenamefont
  {Briggs}(1999{\natexlab{b}})}]{Corn99_N2}%
  \BibitemOpen
  \bibfield  {author} {\bibinfo {author} {\bibfnamefont {R.~B.}\ \bibnamefont
  {Tim~Cornwell}}\ and\ \bibinfo {author} {\bibfnamefont {D.~S.}\ \bibnamefont
  {Briggs}},\ }\href@noop {} {\emph {\bibinfo {title} {Synthesis imaging in
  Radio Astronomy II}}}\ (\bibinfo  {publisher} {Astronomical Society of the
  Pacific},\ \bibinfo {year} {1999})\ pp.\ \bibinfo {pages}
  {187--199}\BibitemShut {NoStop}%
\bibitem [{\citenamefont {Crane}\ and\ \citenamefont {Napier}(1989)}]{Cra89}%
  \BibitemOpen
  \bibfield  {author} {\bibinfo {author} {\bibfnamefont {P.~C.}\ \bibnamefont
  {Crane}}\ and\ \bibinfo {author} {\bibfnamefont {P.~J.}\ \bibnamefont
  {Napier}},\ }\href@noop {} {\emph {\bibinfo {title} {Synthesis imaging in
  Radio Astronomy}}}\ (\bibinfo  {publisher} {Astronomical Society of the
  Pacific},\ \bibinfo {year} {1989})\ pp.\ \bibinfo {pages}
  {139--165}\BibitemShut {NoStop}%
\bibitem [{\citenamefont {Zineb}\ \emph {et~al.}(2024)\citenamefont {Zineb},
  \citenamefont {Ozel},\ and\ \citenamefont {Psaltis}}]{Zineb:2024gwx}%
  \BibitemOpen
  \bibfield  {author} {\bibinfo {author} {\bibfnamefont {Y.~B.}\ \bibnamefont
  {Zineb}}, \bibinfo {author} {\bibfnamefont {F.}~\bibnamefont {Ozel}},\ and\
  \bibinfo {author} {\bibfnamefont {D.}~\bibnamefont {Psaltis}},\ }\bibfield
  {title} {\bibinfo {title} {{Advancing Black Hole Imaging with Space-Based
  Interferometry}},\ }\href@noop {} {\  (\bibinfo {year} {2024})},\ \Eprint
  {https://arxiv.org/abs/2412.01904} {arXiv:2412.01904 [astro-ph.IM]}
  \BibitemShut {NoStop}%
\bibitem [{\citenamefont {Fromm}\ \emph {et~al.}(2021)\citenamefont {Fromm},
  \citenamefont {Mizuno}, \citenamefont {Younsi}, \citenamefont {Olivares},
  \citenamefont {Porth}, \citenamefont {De~Laurentis}, \citenamefont {Falcke},
  \citenamefont {Kramer},\ and\ \citenamefont {Rezzolla}}]{Fromm:2021flr}%
  \BibitemOpen
  \bibfield  {author} {\bibinfo {author} {\bibfnamefont {C.~M.}\ \bibnamefont
  {Fromm}}, \bibinfo {author} {\bibfnamefont {Y.}~\bibnamefont {Mizuno}},
  \bibinfo {author} {\bibfnamefont {Z.}~\bibnamefont {Younsi}}, \bibinfo
  {author} {\bibfnamefont {H.}~\bibnamefont {Olivares}}, \bibinfo {author}
  {\bibfnamefont {O.}~\bibnamefont {Porth}}, \bibinfo {author} {\bibfnamefont
  {M.}~\bibnamefont {De~Laurentis}}, \bibinfo {author} {\bibfnamefont
  {H.}~\bibnamefont {Falcke}}, \bibinfo {author} {\bibfnamefont
  {M.}~\bibnamefont {Kramer}},\ and\ \bibinfo {author} {\bibfnamefont
  {L.}~\bibnamefont {Rezzolla}},\ }\bibfield  {title} {\bibinfo {title} {{Using
  space-VLBI to probe gravity around Sgr A*}},\ }\href
  {https://doi.org/10.1051/0004-6361/201937335} {\bibfield  {journal} {\bibinfo
   {journal} {Astron. Astrophys.}\ }\textbf {\bibinfo {volume} {649}},\
  \bibinfo {pages} {A116} (\bibinfo {year} {2021})},\ \Eprint
  {https://arxiv.org/abs/2101.08618} {arXiv:2101.08618 [astro-ph.HE]}
  \BibitemShut {NoStop}%
\bibitem [{\citenamefont {Gurvits}\ \emph {et~al.}(2021)\citenamefont {Gurvits}
  \emph {et~al.}}]{Gurvits:2019ioq}%
  \BibitemOpen
  \bibfield  {author} {\bibinfo {author} {\bibfnamefont {L.~I.}\ \bibnamefont
  {Gurvits}} \emph {et~al.},\ }\bibfield  {title} {\bibinfo {title} {{THEZA:
  TeraHertz Exploration and Zooming-in for Astrophysics: An ESA Voyage 2050
  White Paper}},\ }\href {https://doi.org/10.1007/s10686-021-09714-y}
  {\bibfield  {journal} {\bibinfo  {journal} {Exper. Astron.}\ }\textbf
  {\bibinfo {volume} {51}},\ \bibinfo {pages} {559} (\bibinfo {year} {2021})},\
  \Eprint {https://arxiv.org/abs/1908.10767} {arXiv:1908.10767 [astro-ph.IM]}
  \BibitemShut {NoStop}%
\bibitem [{\citenamefont {Yfantis}\ \emph {et~al.}(2025)\citenamefont
  {Yfantis}, \citenamefont {Palumbo},\ and\ \citenamefont
  {Mo{\'s}cibrodzka}}]{Yfantis:2025jtw}%
  \BibitemOpen
  \bibfield  {author} {\bibinfo {author} {\bibfnamefont {A.~I.}\ \bibnamefont
  {Yfantis}}, \bibinfo {author} {\bibfnamefont {D.~C.~M.}\ \bibnamefont
  {Palumbo}},\ and\ \bibinfo {author} {\bibfnamefont {M.}~\bibnamefont
  {Mo{\'s}cibrodzka}},\ }\bibfield  {title} {\bibinfo {title} {{Lensing of hot
  spots in Kerr spacetime: An empirical relation for black hole spin
  estimation}},\ }\href@noop {} {\  (\bibinfo {year} {2025})},\ \Eprint
  {https://arxiv.org/abs/2504.16218} {arXiv:2504.16218 [astro-ph.HE]}
  \BibitemShut {NoStop}%
\end{thebibliography}%

\end{document}